\documentclass[twocolumn]{aastex631}

\hypersetup{linkcolor=blue,citecolor=blue,filecolor=blue,urlcolor=blue}

\shorttitle{}
\shortauthors{}
\begin{document}

\title{X-ray Cavity Dynamics and their Role in the Gas Precipitation in Planck Sunyaev–Zeldovich (SZ) Selected Clusters}

\correspondingauthor{Valeria Olivares}
\email{valeria.olivares@uky.edu}

\author[0000-0001-6638-4324]{V.~Olivares}\affiliation{Department of Physics and Astronomy, University of Kentucky, 505 Rose Street, Lexington, KY 40506, USA.}

\author[0000-0002-3886-1258]{Y.~Su}
\affiliation{Department of Physics and Astronomy, University of Kentucky, 505 Rose Street, Lexington, KY 40506, USA.}

\author{W.~Forman}
\affiliation{Center for Astrophysics \text{\textbar} Harvard \& Smithsonian, 60 Garden Street, Cambridge, MA 02138, USA}

\author[0000-0003-2754-9258]{M.~Gaspari}
\affiliation{Department of Astrophysical Sciences, Princeton University, 4 Ivy Lane, Princeton, NJ 08544-1001, USA}

\author[0000-0002-8144-9285]{F. Andrade-Santos}
\affiliation{Department of Liberal Arts and Sciences, Berklee College of Music, 7 Haviland Street, Boston, MA 02215, USA}
\affiliation{Center for Astrophysics \text{\textbar} Harvard \& Smithsonian, 60 Garden Street, Cambridge, MA 02138, USA}

\author{P.~Salome}
\affiliation{Observatoire de Paris, LERMA, PSL University, Sorbonne University, Paris, France}

\author{P.~Nulsen}
\affiliation{Center for Astrophysics \text{\textbar} Harvard \& Smithsonian, 60 Garden Street, Cambridge, MA 02138, USA}
\author{A.~Edge}
\affiliation{Department of Physics, Durham University, Durham DH1 3LE, UK}

\author{F.~Combes}
\affiliation{Observatoire de Paris, LERMA, Collège de France, CNRS, PSL University, Sorbonne University, Paris, France}

\author{C.~Jones}
\affiliation{Center for Astrophysics \text{\textbar} Harvard \& Smithsonian, 60 Garden Street, Cambridge, MA 02138, USA}




\begin{abstract}
We study active galactic nucleus (AGN) feedback in nearby ($z<0.35$) galaxy clusters from the Planck Sunyaev–Zeldovich (SZ) sample using Chandra observations. This nearly unbiased mass-selected sample includes both relaxed and disturbed clusters and may reflect the entire AGN feedback cycle. We find that relaxed clusters better follow the one-to-one relation of cavity power versus cooling luminosity, while disturbed clusters display higher cavity power for a given cooling luminosity, likely reflecting a difference in cooling and feedback efficiency. Disturbed clusters are also found to contain asymmetric cavities when compared to relaxed clusters, hinting toward the influence of the intracluster medium (ICM) ``weather'' on the distribution and morphology of the cavities. Disturbed clusters do not have {fewer} cavities than relaxed clusters, suggesting that cavities are difficult to disrupt. Thus, multiple cavities are a natural outcome of recurrent AGN outbursts. As in previous studies, we confirm that clusters with short central cooling times, $t_{\rm cool}$, and low central entropy values, $K_{\rm 0}$, contain warm ionized (10,000~K) or cold molecular ($<$100~K) gas, consistent with ICM cooling and a precipitation/chaotic cold accretion (CCA) scenario. We analyzed archival MUSE observations that are available for 18 clusters. In 11/18 of the cases, the projected optical line emission filaments appear to be located beneath or around the cavity rims, indicating that AGN feedback plays an important role in forming the warm filaments by likely enhancing turbulence or uplift. In the remaining cases (7/18), the clusters either lack cavities or their association of filaments with cavities is vague, suggesting alternative turbulence-driven mechanisms (sloshing/mergers) or physical time delays are involved.

\end{abstract}


\section{Introduction}\label{sec:intro}
Observational evidence for mechanical AGN feedback in the nearby Universe comes from observations of massive elliptical galaxies residing at cluster centers (also known as brightest cluster galaxies, BCGs). 
The finding was made possible by joining ROSAT X-ray and Very Large Array (VLA) radio observations \citep{boehringer93,churazov00}. Later further studies with new state-of-art X-ray \textit{XMM–Newton} and \textit{Chandra}, and radio Jansky Very Large Array (JVLA) instruments, with superb sensitivity and high spatial resolution reveal surface brightness depressions in the hot intracluster medium (ICM) known as X-ray cavities, created by radio jets, are often filled with relativistic plasma whose emission is seen in radio (\citealt{birzan20,mcnamara00} and references therein, see also \citealt{fabian12} for a review). These observations revealed that mechanical AGN feedback could alleviate the so-called cooling flow problem \citep[e.g.,][]{rafferty06}, as the amount of mechanical energy input by the AGN is sufficient to offset the cooling losses of the ICM in cool-core clusters.

There has been a growing number of known local cooling flow clusters hosting large multi-phase filamentary structures with extensions of several tens of kpc over the central regions \citep{Crawford99,cowie77,cowie80,heckman89,
hatch07,fabian03,fabian08,mcdonald10,mcdonald12,fabian16,tremblay18,hamer19,olivares19,russell19}. In the same vein, it has been found that the presence of star formation and nebular emission in BCGs occurs in clusters with low central entropy and short central cooling times \citep[e.g.,][]{rafferty06,cavagnolo08}, which shows the close connection between the thermodynamic state of the hot ICM and the properties of the BCG. Early theoretical studies and simulations argued that multiphase filaments form ICM condensations via thermal instabilities (TI), when the ratio of the cooling time to dynamical time is less than 10 \citep[e.g.,][]{gaspari12,sharma12,mccourt12,li14}. Recent simulations have shown that this ratio may increase up to 20--30, given the intrinsic variance of the precipitation-driven feedback \citep[e.g.,][]{prasad18,Voit_2017,voit19,beckmann19}.

Observational studies have shown a morphological association of the bubbles with the cold molecular filaments in a few cooling flow clusters \citep{salome06,olivares19,russell19}. One scenario suggests that low entropy hot gas from the core can be uplifted by buoyantly rising bubbles, which becomes thermally unstable at higher altitude and condense to form multiphase gas \citep{Revaz_2008,pope10,li14,mcnamara16,qiu20}. Nonetheless, it is still debated whether gas precipitation necessarily requires the presence of bubbles for the cooling to occur, as AGN feedback in dynamically disturbed clusters has remained poorly studied.

There are alternative mechanisms for the condensation of multiphase filaments via turbulence-induced condensation, but also the ``sloshing'' that might be able to uplift the hot gas to altitudes where it becomes thermally unstable. \citet{Gaspari_2018} proposed that the ICM condenses through turbulent non-linear instabilities (triggered mainly by AGN feedback but also mergers), when the ratio of cooling time to the eddy-turnover time is close to unity, in a process known as Chaotic Cold Accretion (CCA). Such CCA ``rain out'' of the turbulent ICM ``weather'' is the key to trigger recurrent AGN outflows/bubbles with flicker-noise variability (\citealt{gaspari20} for a review).

Many cool-core clusters are dynamically disturbed, with clear signatures of sloshing, and some \textit{galaxy interactions/mergers} \citep{markevitch03}. An interesting question is how AGN feedback operates in these cases and how the dynamical state of the clusters affects the AGN feedback mechanism. Hydrodynamical simulations suggest that bulk and chaotic motions of the above ICM ``weather'' could disturb the AGN bubbles \citep[e.g.,][]{mendygral12,yang19,wittor20}. Recently, \citet{fabian21} suggested that outer bubbles might be destroyed due to the interaction with the cold front produced by sloshing motions.

Samples used in previous studies of central cluster galaxies were selected from X-ray surveys and, thus, contain a higher proportion of bright and relaxed cool-core clusters \citep[e.g.,][]{mcdonald10,mcdonald12,hamer16,andrade-santos17}. On the other hand, cluster selection through the Sunyaev-Zel'dovich (SZ) effect is more reflective of the total mass of the cluster as it relates to the inverse Compton interaction between the cosmic microwave background (CMB) photons and the electrons of the hot cluster plasma (see \citealt{Aghanim08}). {The Planck sample, observed with \textit{Chandra}, is an SZ selected sample of galaxy clusters at moderate redshift ($z<0.35$). Additionally, it is a representative sample of galaxy clusters, as it includes both dynamically both relaxed and disturbed systems
\citep{lovisari17,andrade-santos17}.} Since it is a nearly unbiased, mass-selected sample, it could reflects the entire AGN feedback cycle at the centers of galaxy clusters, providing a complete view of the cold gas ``rain'' and gives observational constraints on numerical simulations. 

{To date, almost all the existing studies on AGN feedback in galaxy clusters are X-ray selected and strongly biased towards relaxed systems. SZ selected samples are nearly mass limited, more representative of the true distribution of galaxy clusters. They provide a contrast to previously explored X-ray samples by including more dynamically disturbed systems that are experiencing mergers or sloshing. With this sample, we are able to assess how AGN feedback operates in disturbed systems and what is the role in the feedback loop played by mergers, sloshing, and turbulence.}



In \citet{olivares22b} (Paper I), we determined that 30 of the 164 Planck SZ selected clusters show cavities using archival \textit{Chandra} observations. Our analysis suggested no evolution of the AGN feedback across almost 8~Gyr. This paper presents the effect of the ICM ``weather'' on the AGN feedback and explores if such ``weather" is related to the formation of multiphase condensation. In Section~\ref{sec:sample}, we describe the Planck SZ sample. Section~\ref{sec:obs} presents the \textit{Chandra} X-ray and MUSE optical observations. {Section~\ref{sec:analysis} present the analysis of the observations. Sections~\ref{sec:results} and \ref{sec:discussion} present and discus the results. Section~\ref{sec:limitations} presents the limitations of the present study}. Finally, Section~\ref{sec:conclusions} summarizes our findings.

Throughout this paper, we adopted a {flat} cosmology with H$_{\rm 0}$=70\,km s$^{-1}$\,Mpc$^{-1}$ and $\Omega_{\rm m}$=0.3.

\section{Planck sample}\label{sec:sample}

The sample consists of 164 nearby ($z\leq0.35$) clusters drawn from the first Planck mission release from early 2011 \citep{planck_collaboration11} selected based on the SZ effect. The Planck clusters were observed with \textit{Chandra} over the almost full sky (and $|$b$|> 15 \degr$) by two different programs \citep{jones12}, the XVP (X-ray Visionary Program, PI: Jones) and HRC Guaranteed Time Observations (PI: Murray). All clusters observed with \textit{Chandra} have exposures sufficient to collect at least 10,000 source counts from each cluster.

The Planck sample is a nearly mass-limited sample, and covers a range of mass, $M_{\rm 500}$\footnote{The mass enclosed within a sphere of radius $r_{\rm 500}$ within which the enclosed density is 500 times the critical density at that redshift.}, from 7$\times$10$^{13}$M$_{\odot}$ to 2$\times$10$^{15}$M$_{\odot}$.

As described in \citet{andrade-santos17}, {we classify each cluster as cool-core (CC) or non-cool-core (NCC), based on their high or low central density, respectively.} Clusters with a central electron density exceeding $n_{\rm core} = 1.5\times$10$^{-2}$~cm$^{-3}$, measured at 0.01~R$_{500}$, were classified as CC, while those with lower central densities were classified as NCC, resulting in 63 classified as CC and 101 as NCC clusters.

Compared to X-ray selected samples, SZ selected samples contain a larger fraction of NCC clusters (72\% versus 56\%, \citealt{andrade-santos17}).

\section{Observations and Data Reduction}\label{sec:obs}

This paper synthesizes archival observations of the Planck clusters, from \textit{Chandra} and MUSE datasets (when available) that comprise the bulk of our analysis.

\subsection{\textit{Chandra} X-ray Observations}
All \textit{Chandra} observations used in this paper are part of the same suite as those described in \citet{olivares22b}, and listed in Table~\ref{tab:sample}. In the following, we briefly summarize the observations, but we refer the reader to \citet{olivares22b} for further details. 

The data reduction and calibration of \textit{Chandra} observations were carried out using \textit{Chandra} Interactive Analysis of Observations software (CIAO) 4.12, and the \textit{Chandra} Calibration Database (CALDB) 4.9.2.1. The observations were reprocessed using the {\tt chandra\_repro} tool of CIAO starting from the level one event files. Light curves were created using {\tt deflare} CIAO tool in the 0.5-8.0~keV and 9.0-12.0~keV energy bands. We subtracted background flares higher than 3$\sigma$ using the light curve filtering script {\tt lc\_sigma\_clip}. {Standard blank sky background} and readout artifacts were also subtracted. Point sources were detected with the {\tt wavdetect} tool of CIAO in the 0.5-8.0~keV energy band.

Images were produced in the 0.5-2.0~keV energy band. We created monochromatic exposure maps defined at the central energy of the band. Each image was normalized with an exposure time map. A background image was generated using the blank-sky fields. 
We used {\tt dmfilth} CIAO tool to replace point sources with pixel values interpolated from the surrounding background regions. {Point sources were detected in the 0.5-8.0 keV energy band, then masked before performing the spectral and imaging analysis of the clusters. Exposure corrected images were produced in the 0.5–2.0~keV band energy.}

\subsection{Presence of Cold or Warm Gas}
Cool-core clusters are expected to develop a multiphase nebula associated with ICM cooling. We indicate in Table~\ref{tab:sample_properties} when these clusters harbor either warm ionized ($\sim 10^4$ K) gas or cold molecular ($\sim$\,100~K) emission-line gas. For that purpose, we analyzed, when available, Multi-Unit Spectroscopic Explorer (MUSE) data\footnote{http://archive.eso.org/}, and also examined various optical and millimeter emission line surveys of galaxy clusters \citep[e.g.,][]{hamer16,Crawford99,donahue10,castignani20}. We find that 40\% of Planck CC clusters have warm or cold gas, while 32\% lack emission-line gas, and 28\% of the CC clusters have no available information.

\subsubsection{MUSE Observations}
MUSE is an optical image slicing integral field unit (IFU) with a wide field-of-view (FOV) of 1$\arcmin\times$1$\arcmin$, mounted on UT4 of the European Southern Observation Very Large Telescope (ESO-VLT). MUSE has a spectral resolution of R$\sim$3000 (FWHM=100~km~s$^{-1}$ at 7000~\AA), allowing us to fit individual optical emission lines. The MUSE data were reduced using the latest MUSE pipeline 2.8.5 \citep{weilbacher14} and the EsoRex command-line tool. We also performed an additional sky subtraction to the one in the pipeline using Zurich Atmosphere Purge 2.1 (ZAP; \citealt{soto16}). When optical emission lines were present in the data cube, we fitted the data following the same method described by \citet{olivares19}. The data were also corrected for Galactic foreground extinction, estimated from the \citet{schlafly11} recalibration of the \citet{schlegel98} Milky Way dust map, assuming $R_{\rm V} = 3.1$.

Table~\ref{tab:MUSE} lists the archival MUSE observations of Planck clusters used in this paper, including the program ID, average DIMM seeing, exposure time, whether or not the central BCG reveals optical emission lines, and a brief description of the distribution of the optical line emitting gas when it was detected. It is worth noting that most of the clusters selected for the MUSE observations used in this paper were selected from X-ray samples, so no purely Planck-selected clusters have been observed {(see Section~\ref{sec:limitations} for more details)}.

\section{Analysis}\label{sec:analysis}
\subsection{Density and Temperature profiles}
{We used deprojected density and temperature profiles from \citet{andrade-santos17}, that were produced following the method described in \citet{vikhlinin06}. Below, we provide some details on the procedure to obtain these profiles. Surface brightness profiles were extracted in narrow concentric annuli centered on the X-ray halo peak and computed using the \textit{Chandra} area-averaged effective area for each annulus. We assumed spherical symmetry to compute the emission measure and temperature profiles. We converted the \textit{Chandra} count rate into the emission integral, $EI = \int n_{\rm e} n_{\rm p} dV$, within each cylindrical shell. Then, we fit the emission measure profile assuming the gas density profile described in \citet{vikhlinin06},}

\begin{equation}
    n_{e} n_{p} = n_{0}^{2} \frac{(r/r_{c})^{-\alpha}}{(1+r^{2}/r^{2}_{c2})^{3\beta}} \frac{1}{(1+r^{\gamma})/r_{s}^{\gamma})^{\epsilon / \gamma}} + \frac{n_{02}^{2}}{(1+r^{2}/r^{2}_{c2})^{3\beta_{2}}},
    \label{eq:ne}
\end{equation}
where the parameters $n_{0}$ and $n_{02}$ determine the normalization of both additive components, {and $\alpha$, $\beta$, $\beta_{2}$, $\epsilon$ are fit indexes.} In this fit, all the parameters are free to vary. For examples of the emissivity profiles and gas density profiles, we refer the reader to Figures~3 and 4 of \citet{andrade-santos17}.

{The deprojected 3D temperature profile was obtained following \cite{vikhlinin06} equation,}
\begin{equation}
    T_{3D} = T_{0} \times \frac{x+T_{min}/T_{0}}{x+1} \times \frac{(r/r_{t}^{-a})}{(1+(r/r_{t})^{b})^{c/b}},
    \label{eq:T3}
\end{equation}
{where $x = (r/r_{cool})^{a_{cool}}$, $r_{t}$ and $r_{cool}$ are the transition and CC radii, respectively. $T_{min}$ is the central temperature, and $a$, $b$, $c$, and $a_{cool}$ are the indexes that determine the slopes of the temperature profile in different radial ranges \citep{andrade-santos17}.
The 3D temperature was derived by projecting a model to compare with the observed profile, in which the 3D temperature model, $T_{3D}$, is weighted by the density squared according to the spectroscopic temperature \citep{mazzotta04}. We used the following equation, }

\begin{equation}
    T_{2D} = T_{spec} = \frac{\int n_{e}^{2} T_{3D}^{1/4} dz}{\int n_{e}T_{3D}^{-3/4}dz},
\end{equation}
where $n_{e}$ is the electron density, given by Equation~\ref{eq:ne}, and $T_{3D}$ is the deprojected 3D temperature, given by Equation~\ref{eq:T3}. The spectroscopic-like temperature profiles were constructed by extracting spectra in annuli and fitting them with an absorbed {\sc APEC} model in XSPEC, leaving the metallicity free (see \citealt{andrade-santos17} for more details).

\subsection{Concentration and Centroid shift}

We measured the centroid shift, $w$, and concentration, $c$, to examine the dynamical state of each cluster. The centroid shift, $w$, is defined as the fractional scatter in the separation between the X-ray peak and the centroid obtained within ten different apertures of radius $=$ i $\times$ 0.1~R$_{500}$, with i$=$1,2,3 up to 10 or the maximum radius that fits entirely within the FOV, {$R_{max}$},
\begin{equation}
    w = \Big[ \frac{1}{N-1} \sum_{i} (\Delta_{i} - \bar{\Delta})^{2}\Big]^{1/2} \frac{1}{R_{500}}
\end{equation}
where $\Delta_{i}$ is the distance between the weighted centroid within the $i$th aperture and the X-ray peak, $\bar{\Delta}$ is the average centroid, and $N$ is the number of apertures.

The concentration, $c$, quantifies how centrally concentrated the X-ray emission of a given cluster is. \citet{santos08} shows that is a good indicator of the presence of a cool core (also see \citealt{Su2020}). It is defined as the ratio of the {flux} within two circular apertures. Following \citet{santos08}, it is calculated as,
\begin{equation}
    c_{SB40} = \frac{\int_{0}^{40}  SB(<40\,{\rm kpc})\,dr}{\int_{0}^{400}  SB(<400\,{\rm kpc})\,dr}
\end{equation}
where $\int SB(<r)$ is the flux (integrated surface brightness) within the circle of radius $r$.

Following \citet{lovisari17} concentration classification, we also calculate the concentration in the 0.1–-1.0 $R_{500}$ range, or $R_{max}$. It is defined as the ratio of the flux within two circular apertures, one that encloses the core emission ($\int SB (<0.1R_{500})$) and another the overall emission ($\int SB (<R_{500})$). We used the following expression:

\begin{equation}
    c = \frac{\int_{0}^{0.1R_{500}} SB(<0.1R_{500})\,dr}{\int_{0}^{R_{500}} SB(<R_{500})\,dr}
\end{equation}

Based on the centroid shift and concentration described in \citet{lovisari17}, we classify each cluster as disturbed ($w>0.012$ and $c<0.2$), relaxed ($w<0.012$ and $c>0.2$), or mixed. {In Table~\ref{tab:dynamical_state} we list the concentration and centroid shift of each cluster.} 

\startlongtable
\movetableright=-5cm
\begin{deluxetable}{lcccc}
\setlength{\tabcolsep}{3pt}
\tabletypesize{\scriptsize} 
\centering
\tablecaption{Dynamical State Properties of the Sample} \label{tab:dynamical_state}
\tablehead{
{Cluster} & {$R_{\rm 500}$} & {$w$}  & {$C_{\rm SB40}$} & {$C_{\rm SB0.1R500}$}}
\decimalcolnumbers
\startdata
G006.47+50.54 &    1364 & 0.0034$\pm$ 0.0003 &      0.1759 $\pm$ 0.0005 &          0.5160  $\pm$ 0.0008 \\
G008.44$-$56.35 &     779 & 0.0054$\pm$ 0.0005 &      0.0843 $\pm$ 0.0047 &          0.1686  $\pm$ 0.0060 \\
G021.09+33.25 &     699 & 0.0025$\pm$ 0.0002 &      0.3165 $\pm$ 0.0014 &          0.4117  $\pm$ 0.0015 \\
G033.46$-$48.43 &     713 & 0.0466$\pm$ 0.0047 &      0.1483 $\pm$ 0.0028 &          0.1863  $\pm$ 0.0026 \\
G033.78+77.16 &     728 & 0.0078$\pm$ 0.0008 &      0.2180 $\pm$ 0.0010 &          0.3853  $\pm$ 0.0013 \\
G036.72+14.92 &     885 & 0.0357$\pm$ 0.0036 &      0.0770 $\pm$ 0.0040 &          0.1648  $\pm$ 0.0053 \\
G042.82+56.61 &     659 & 0.0280$\pm$ 0.0028 &      0.0852 $\pm$ 0.0010 &          0.1324  $\pm$ 0.0011 \\
G044.22+48.68 &     774 & 0.0114$\pm$ 0.0011 &      0.0870 $\pm$ 0.0003 &          0.1996  $\pm$ 0.0004 \\
G049.20+30.86 &    1226 & 0.0058$\pm$ 0.0006 &      0.2233 $\pm$ 0.0021 &          0.4609  $\pm$ 0.0029 \\
G049.66$-$49.50 &     821 & 0.0075$\pm$ 0.0008 &      0.0951 $\pm$ 0.0036 &          0.1996  $\pm$ 0.0047 \\
G055.60+31.86 &    1328 & 0.0149$\pm$ 0.0015 &      0.1154 $\pm$ 0.0014 &          0.3040  $\pm$ 0.0019 \\
G056.81+36.31 &     599 & 0.0041$\pm$ 0.0004 &      0.1051 $\pm$ 0.0008 &          0.1718  $\pm$ 0.0010 \\
G057.92+27.64 &     495 & 0.0036$\pm$ 0.0004 &      0.2262 $\pm$ 0.0015 &          0.2705  $\pm$ 0.0016 \\
G062.42$-$46.41 &     289 & 0.0669$\pm$ 0.0067 &      0.0679 $\pm$ 0.0027 &          0.0686  $\pm$ 0.0033 \\
G062.92+43.70 &     347 & 0.0044$\pm$ 0.0004 &      0.2307 $\pm$ 0.0004 &          0.1961  $\pm$ 0.0004 \\
G067.23+67.46 &    1225 & 0.0170$\pm$ 0.0017 &      0.0635 $\pm$ 0.0008 &          0.2508  $\pm$ 0.0014 \\
G072.80$-$18.72 &    1121 & 0.0223$\pm$ 0.0022 &      0.0713 $\pm$ 0.0038 &          0.1748  $\pm$ 0.0047 \\
G073.96$-$27.82 &    1387 & 0.0074$\pm$ 0.0007 &      0.1170 $\pm$ 0.0010 &          0.3123  $\pm$ 0.0014 \\
G080.99$-$50.90 &    1228 & 0.0022$\pm$ 0.0002 &      0.0623 $\pm$ 0.0029 &          0.1921  $\pm$ 0.0040 \\
G086.45+15.29 &    1269 & 0.0053$\pm$ 0.0005 &      0.0823 $\pm$ 0.0044 &          0.2983  $\pm$ 0.0075 \\
G094.01+27.42 &    1079 & 0.0117$\pm$ 0.0012 &      0.2064 $\pm$ 0.0014 &          0.3084  $\pm$ 0.0015 \\
G096.85+52.46 &    1152 & 0.0070$\pm$ 0.0007 &      0.0448 $\pm$ 0.0012 &          0.1803  $\pm$ 0.0022 \\
G098.95+24.86 &     570 & 0.0439$\pm$ 0.0044 &      0.0742 $\pm$ 0.0037 &          0.1063  $\pm$ 0.0041 \\
G115.16$-$72.09 &     524 & 0.0259$\pm$ 0.0026 &      0.1535 $\pm$ 0.0005 &          0.1888  $\pm$ 0.0005 \\
G115.71+17.52 &     984 & 0.0121$\pm$ 0.0012 &      0.1358 $\pm$ 0.0052 &          0.2636  $\pm$ 0.0062 \\
G124.21$-$36.48 &    1219 & 0.0724$\pm$ 0.0072 &      0.1755 $\pm$ 0.0015 &          0.1873  $\pm$ 0.0010 \\
G125.70+53.85 &    1124 & 0.0144$\pm$ 0.0014 &      0.0616 $\pm$ 0.0032 &          0.1750  $\pm$ 0.0043 \\
G139.59+24.18 &    1210 & 0.0067$\pm$ 0.0007 &      0.1533 $\pm$ 0.0048 &          0.3162  $\pm$ 0.0062 \\
G146.33$-$15.59 &     122 & 0.0148$\pm$ 0.0015 &      0.1813 $\pm$ 0.0003 &          0.0615  $\pm$ 0.0003 \\
G164.61+46.38 &    1139 & 0.0315$\pm$ 0.0032 &      0.0823 $\pm$ 0.0047 &          0.1912  $\pm$ 0.0059 \\
G166.13+43.39 &    1189 & 0.0251$\pm$ 0.0025 &      0.0416 $\pm$ 0.0016 &          0.1538  $\pm$ 0.0025 \\
G176.28$-$35.05 &     260 & 0.0127$\pm$ 0.0013 &      0.4081 $\pm$ 0.0008 &          0.2735  $\pm$ 0.0006 \\
G180.62+76.65 &    1056 & 0.0061$\pm$ 0.0006 &      0.1141 $\pm$ 0.0037 &          0.2194  $\pm$ 0.0042 \\
G182.44$-$28.29 &     425 & 0.0020$\pm$ 0.0002 &      0.1757 $\pm$ 0.0008 &          0.1849  $\pm$ 0.0009 \\
G182.63+55.82 &     826 & 0.0061$\pm$ 0.0006 &      0.0983 $\pm$ 0.0018 &          0.2066  $\pm$ 0.0023 \\
G209.56$-$36.49 &     285 & 0.0087$\pm$ 0.0009 &      0.2747 $\pm$ 0.0006 &          0.2066  $\pm$ 0.0006 \\
G226.24+76.76 &    1204 & 0.0027$\pm$ 0.0003 &      0.0990 $\pm$ 0.0011 &          0.3178  $\pm$ 0.0018 \\
G228.49+53.12 &     824 & 0.0027$\pm$ 0.0003 &      0.1899 $\pm$ 0.0064 &          0.3390  $\pm$ 0.0083 \\
G241.74$-$30.88 &    1115 & 0.0042$\pm$ 0.0004 &      0.0728 $\pm$ 0.0036 &          0.2472  $\pm$ 0.0060 \\
G241.77$-$24.00 &     597 & 0.0046$\pm$ 0.0005 &      0.2256 $\pm$ 0.0064 &          0.2827  $\pm$ 0.0069 \\
G244.34$-$32.13 &    1279 & 0.0180$\pm$ 0.0018 &      0.1069 $\pm$ 0.0021 &          0.2462  $\pm$ 0.0026 \\
G252.96$-$56.05 &     785 & 0.0102$\pm$ 0.0010 &      0.2933 $\pm$ 0.0011 &          0.4586  $\pm$ 0.0014 \\
G253.47$-$33.72 &     962 & 0.0055$\pm$ 0.0006 &      0.0711 $\pm$ 0.0043 &          0.1857  $\pm$ 0.0061 \\
G256.45$-$65.71 &    1149 & 0.0189$\pm$ 0.0019 &      0.1225 $\pm$ 0.0046 &          0.2445  $\pm$ 0.0051 \\
G257.34$-$22.18 &    1004 & 0.0587$\pm$ 0.0059 &      0.0745 $\pm$ 0.0044 &          0.1181  $\pm$ 0.0040 \\
G260.03$-$63.44 &    1187 & 0.0213$\pm$ 0.0021 &      0.1489 $\pm$ 0.0050 &          0.2886  $\pm$ 0.0061 \\
G263.16$-$23.41 &    1264 & 0.0064$\pm$ 0.0006 &      0.1055 $\pm$ 0.0015 &          0.2889  $\pm$ 0.0023 \\
G263.66$-$22.53 &     903 & 0.0154$\pm$ 0.0015 &      0.0691 $\pm$ 0.0036 &          0.1704  $\pm$ 0.0049 \\
G264.41+19.48 &    1166 & 0.0046$\pm$ 0.0005 &      0.0534 $\pm$ 0.0037 &          0.1435  $\pm$ 0.0047 \\
G266.84+25.07 &    1070 & 0.0038$\pm$ 0.0004 &      0.2152 $\pm$ 0.0028 &          0.3993  $\pm$ 0.0035 \\
G286.58$-$31.25 &     926 & 0.0267$\pm$ 0.0027 &      0.0344 $\pm$ 0.0032 &          0.0932  $\pm$ 0.0042 \\
G303.75+33.65 &     859 & 0.0438$\pm$ 0.0044 &      0.1785 $\pm$ 0.0047 &          0.1743  $\pm$ 0.0033 \\
G304.89+45.45 &     589 & 0.0563$\pm$ 0.0056 &      0.0974 $\pm$ 0.0010 &          0.1161  $\pm$ 0.0010 \\
G306.68+61.06 &    1106 & 0.0027$\pm$ 0.0003 &      0.1135 $\pm$ 0.0005 &          0.3182  $\pm$ 0.0009 \\
G306.80+58.60 &     684 & 0.0061$\pm$ 0.0006 &      0.0860 $\pm$ 0.0022 &          0.1620  $\pm$ 0.0029 \\
G313.36+61.11 &    1378 & 0.0034$\pm$ 0.0003 &      0.1228 $\pm$ 0.0008 &          0.3986  $\pm$ 0.0014 \\
G313.87$-$17.10 &    1136 & 0.0066$\pm$ 0.0007 &      0.0861 $\pm$ 0.0029 &          0.3060  $\pm$ 0.0053 \\
G316.34+28.54 &     316 & 0.0429$\pm$ 0.0043 &      0.0955 $\pm$ 0.0005 &          0.0679  $\pm$ 0.0005 \\
G318.13$-$29.57 &     866 & 0.0177$\pm$ 0.0018 &      0.1259 $\pm$ 0.0050 &          0.2225  $\pm$ 0.0059 \\
G324.49$-$44.97 &     633 & 0.0064$\pm$ 0.0006 &      0.1184 $\pm$ 0.0053 &          0.1538  $\pm$ 0.0055 \\
G337.09$-$25.97 &     921 & 0.0066$\pm$ 0.0007 &      0.1640 $\pm$ 0.0046 &          0.3021  $\pm$ 0.0056 \\
G340.95+35.11 &     918 & 0.0054$\pm$ 0.0005 &      0.2540 $\pm$ 0.0034 &          0.3492  $\pm$ 0.0035 \\
G349.46$-$59.94 &    1470 & 0.0112$\pm$ 0.0011 &      0.0588 $\pm$ 0.0011 &          0.2908  $\pm$ 0.0022
\enddata
\tablecomments{(1) Column lists the cluster name (the prefix PLCKESZ is omitted). (2) $R_{\rm 500}$ or $R_{\rm max}$ in kpc. (3) Centroid shift (w) and uncertanties.(4) and (5) concentration $C_{\rm SB40}$ and $C_{\rm SB0.1R500}$, and one sigma uncertanties.}
\end{deluxetable}

\tabletypesize{\footnotesize}
\setlength{\tabcolsep}{3pt}
\begin{deluxetable*}{llcllccccc}
\caption{Summary of MUSE Observations}\label{tab:MUSE}
\tablehead{
\colhead{Cluster}
& \colhead{Alt name}
& \colhead{Lines?}
& \colhead{Prop. ID}
& \colhead{Exp. time}
& \colhead{Seeing}
& \colhead{H$\alpha$ distribution}
& \colhead{PI}
& \colhead{IFU References}\\
\colhead{(1)}
& \colhead{(2)}
& \colhead{(3)}
& \colhead{(4)}
& \colhead{(5)}
& \colhead{(6)}
& \colhead{(7)}
& \colhead{(8)}
& \colhead{(9)}
}
\startdata
G006.47+50.54  & A2029   & no &  0101.B-0808(A) & 540$\times$8 & 1.33 &.. & F.~Buitrago & .. \\ 
G021.09+33.25  & A2204   & yes & 60.A$-$9100(G)  & 2660  & 0.71 & Filaments &MUSE Team & (8) \\
G033.78+77.16  & A1795   & yes & 094.A$-$0859(A) & 900$\times$3 &  2.31 & Filaments & S.~Hamer & (7, 1) \\
G073.96$-$27.82 & A2390 & yes & 094.A$-$0115(B) &  1800$\times$4 &  0.75 & Filaments & J.~Richard& (0, 2)\\
G080.99$-$50.90 & A2552 & no & 0103.A-0777(B) &  970$\times$3 & 0.51 & .. & A.~Edge& .. \\
G115.16$-$72.09 & A85   & yes &  099.B$-$0193(A)  &  1260$\times$4  & 0.62 & Filaments & J.~Thomas & (3) \\ 
G176.28$-$35.05 & 2A0335+096 & yes & 094.A$-$0859(A)  &   900$\times$3 & 1.32 & Filaments &  S.~Hamer & (1) \\
G182.44$-$28.29 & A478  & yes  & 094.A$-$0859(A) &  900$\times$3 &  1.05 &  Compact filaments &  S.~Hamer & (7) \\
G209.56$-$36.49 & A496  & yes  & 095.B$-$0127(A) &  690$\times$4 & 1.24 & Filaments & E.~Emsellem & (7)\\
G241.74$-$30.88 & RXCJ0532.9-3701 & no & 0104.A-0801(B) & 970$\times$5 &  0.79 & .. &A.~Edge& .. \\ 
G241.77$-$24.00 & A3378 & yes & 0104.A$-$0801(B) &  970$\times$6 & 1.32 & Filaments & A.~Edge & (7)\\
G244.34$-$32.13 & RXCJ0528.9-3927 & yes & 0104.A-0801(A)  &  970$\times$3 & 1.41 & Filaments & A.~Edge & ..\\
G252.96$-$56.05 & A3112 & yes & 0109.A-0709(A)  &  970$\times$2 & 0.94 & Filaments & A.~Edge & (7)\\
G256.45$-$65.71 & A3017 & yes & 0104.A$-$0801(A) & 970$\times$3 & 0.59 & Filaments & A.~Edge& (7) \\
G260.03$-$63.44 & RXCJ0232.2-4420 & yes &0104.A$-$0801(B) & 970$\times$3 & 0.81 & Compact &A.~Edge & ..\\ 
G263.16$-$23.41 & AS0592 & no & 098.A-0502(A) &   1000$\times$3 & 0.32& ..& J.~Richard & ..\\ 
G263.66$-$22.53 & A3404  & no &0104.A-0801(B) &  970$\times$3 & 0.66 & .. & A.~Edge & .. \\
G266.84+25.07 & A3444  & yes &  0104.A$-$0801(B) &  970$\times$3  & 1.86 & Filaments &A.~Edge & (7) \\
G269.51+26.42 &  A1060  & yes & 094.B$-$0711(A) &  730$\times$2 & 0.91 & Compact disk &  M. Arnaboldi & (7, 4, 9)\\ 
G303.75+33.65 & A3528  & yes &  095.A$-$0159(A) & 600$\times$3 & 0.51 & Compact & T.~Green &..\\
G304.89+45.45 & A1644  & yes &  094.A$-$0859(A) &  900$\times$3 & 0.72 & Filaments & S.~Hamer & ..\\
G313.36+61.11 & A1689  & no & 094.A-0141(C) & 1500$\times$4 & 0.93 & .. & A.~Swinbank & (5)\\ 
G316.34+28.54 & A3571  & no & 097.B-0776(A)  & 360$\times$4 & 0.56 & ..& E.~Emsellem& ..\\
G318.13$-$29.57 & RXCJ1947.3-7623     & yes &  109.23L4.001  & 900$\times$3 & 1.67 & Compact & A.~Edge &..\\
G340.95+35.11 & AS0780 & yes &  0104.A$-$0801(A)  & 970$\times$3 & 0.80 & Filaments & A.~Edge & (0) \\
G349.46$-$59.94 & AS1063 & no & 60.A-9345(A) & 1420$\times$8 & 1.33  &..  & MUSE Team
& (6)\\
\enddata
\tablecomments{(1) Planck cluster. (2) Alternative name. (3) Optical line-emitting cluster (``yes'') or (``no'') non-optical line-emitting cluster. (4) Proposal ID. (5) Total exposure time in seconds. (6) Average DIMM seeing in arcseconds. (7) Distribution of the optical emitting gas. (8) PI. (9) Previous IFU works. References: (1) MUSE \citep{olivares19}, (2) MUSE \citep{richard21}, (3) MUSE \citep{mehrgan19}, (4) MUSE \citep{barbosa18}, (5) MUSE \citep{bina16}, (6) MUSE \citep{karman14}, (7) VIMOS  \citep{hamer16}, (8) IFU observations \citep{wilman06}, (9) MUSE \citep{richtler20}.}
\end{deluxetable*}

\section{Clusters with X-ray cavities}\label{sec:cavities}

In \citet{olivares22b}, we searched for X-ray cavities in Planck selected clusters from \textit{Chandra} observations using three images -- 0.5-2.0~keV original image, unsharp masked, and double $\beta$-model subtracted image.

{The first two co-authors independently looked for X-ray cavities and  classified them based on their significance.} Each cavity was classified as ``Certain'' (C) or ``Potential'' (P). ``Certain" (C) cavities were classified as visible on the 0.5-2.0~keV image but also in the unsharp or $\beta$-model subtracted images. On the other hand, ``potential'' cavities were identified having only a hint of X-ray depressions in the original image but visible in the unsharp-masked or double $\beta$-model subtracted image. Additionally, each cavity was classified based on the significance of the detection, calculated as the surface brightness ratio between the cavity and the surrounding ``background'', {measured within the same aperture size as the cavity (see Table~1 of \citealt{olivares22b} for more details). ``Certain'' (C) cavities have an average significance of 2.2, while ``Potential'' cavities have an average significance of 1.5.} Cavities in which the central photon count is too low to be certain were classified as ``potential'' (P) cavities. Clusters without depressions were classified as lacking cavities.
We found 12 cool-core clusters with ``Certain'' (C) cavities and 17 with ``Potential'' (P) cavities. Only one NCC cluster (A1060/G269.51+26.42, BCG NGC\,3311) shows a pair of ``potential'' cavities, however, previous studies suggest the presence of a {coronae} in this source \citep{sun09}, with a central cooling time of 2.2~Gyr \citep[e.g.,][]{edge92}. Therefore for the purpose of this study, this cluster is treated as a CC.

In Figures~\ref{fig:chandra_images_C} and~\ref{fig:chandra_images_P}, we present \textit{Chandra} X-ray images of the clusters with ``certain'' (C) and ``potential'' (P) cavities, respectively. From left to right, we show a slightly smoothed 0.5--2.0~keV \textit{Chandra} image, a double $\beta$-model subtracted image, an unsharp-masked image, and, lastly, an optical image of the central BCG.

\subsection{Central Radio Source}
Cavities are believed to be filled with radio emission as they are formed due to the interaction between the central radio source and the surrounding hot ICM. Consequently, verifying whether a central radio source is present in clusters that contain X-ray cavities is crucial. In Table~\ref{tab:sample_properties}, we highlight the clusters with radio emission associated with the central galaxy. To identify these sources we used the 1.4 GHz VLA FIRST Survey (with a spatial resolution of $\sim$5.4$\arcsec$--6.4$\arcsec$), 74 MHz VLA Low-Frequency Sky Survey Redux (VLSSr, with a resolution of 75$\arcsec$ \citealt{Lane14}), 1.4 GHz continuum NRAO VLA Sky Survey (NVSS, with a beam size of 45$\arcsec$ \citealt{condon91}), 843 MH Sydney University Molonglo Sky Survey (SUMSS, with a beam size of $\sim$40$\arcsec$ \citealt{Mauch03}), Giant Metrewave Radio Telescope (GMRT) data product archive\footnote{https://naps.ncra.tifr.res.in/}, 887.5 MHz Rapid ASKAP Continuum Survey (RACS\footnote{https://www.atnf.csiro.au/research/RACS/RACS\_I1/}, with a beam size of 13--15$\arcsec$ \citealt{McConnell20}), 120-168MHz LOFAR Two-meter Sky Survey (LoTSS, with a resolution of 25$\arcsec$ \citealt{Shimwell22}) Deep fields DR1, the GMRT 150 MHz all-sky radio survey (TGSS ADR\footnote{https://tgssadr.strw.leidenuniv.nl/doku.php}, with a resolution of 25$\arcsec$ \citealt{Intema17}), and the literature.

\begin{figure}[htp]
\centering
    \includegraphics[width=0.45\textwidth]{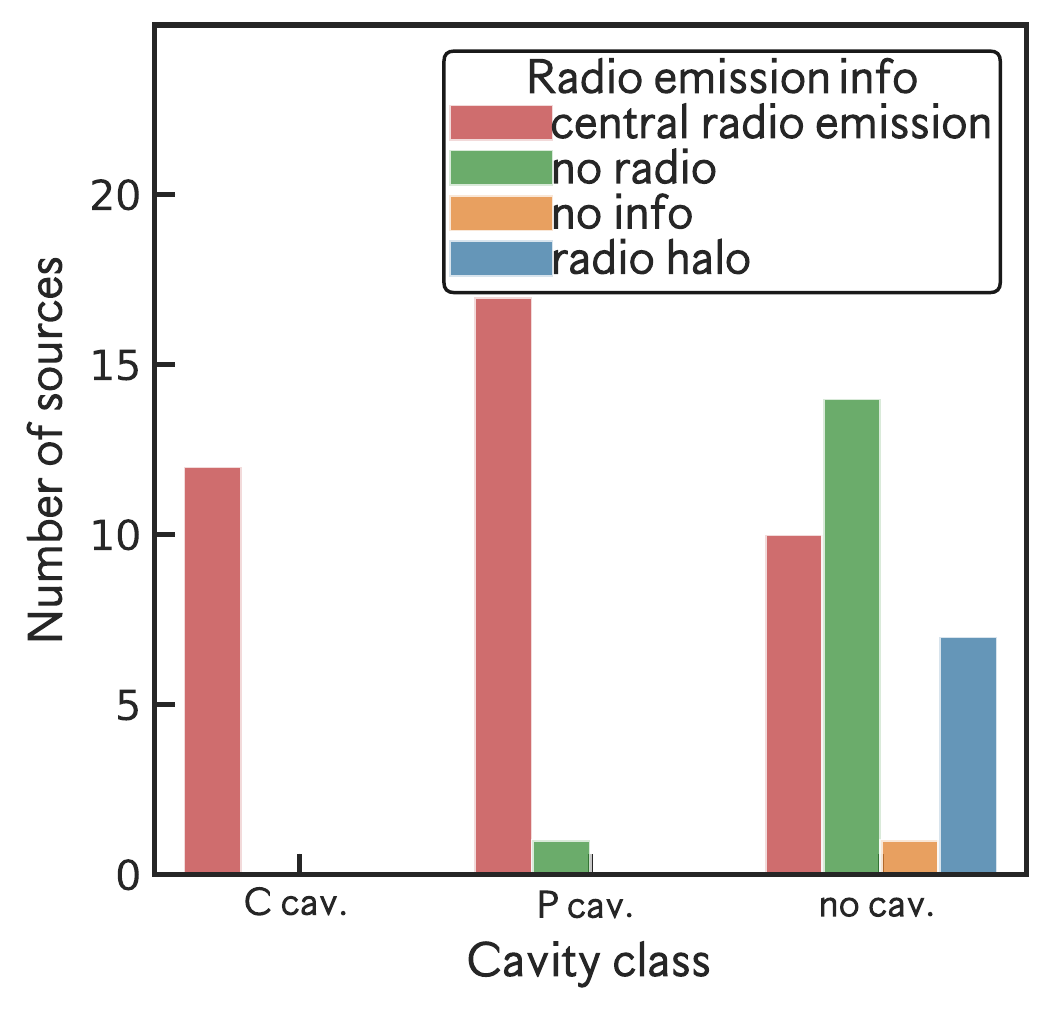}
    \caption{Distribution of clusters by cavity class (``certain'' (C), ``potential'' (P) cavities, and no cavities), and color-coded by the presence (red), and absence (green) of central radio emission. Clusters with radio halos are shown in blue, and those without radio data with orange. \label{fig:radio}}
\end{figure}

We found that 62\% of the CC Planck clusters display central radio emission, while the X-ray selected sample studied in \citet{hogan15}, which includes 246 clusters, revealed that all CC clusters display radio emission. As shown in Figure~\ref{fig:radio}, most clusters with cavities have radio emission associated with the central source. Since most radio data have large beam sizes, we cannot always resolve radio jet structures. Consequently, we use radio observations to verify whether clusters presenting X-ray cavities have radio-loud BCGs. With clusters with high-resolution radio observations, the radio emission appears to fill the detected X-ray cavities (see Figures.~\ref{fig:chandra_images_C} and \ref{fig:chandra_images_P}).


\startlongtable
\movetableright=-3cm
\tabletypesize{\footnotesize}
\setlength{\tabcolsep}{3pt}
\begin{deluxetable*}{lclclc}
\centering
\tablecaption{Summary of properties of the clusters\label{tab:sample_properties}}
\tablehead{
{Cluster name} &
{Alt. name} &
{Emission} &
{Instruments} &
{Radio} &
{Cavities?}\\
{} & {} &
{lines}&
{for emission lines}&
{emission} &
{}
}
\decimalcolnumbers
\startdata
G006.47+50.54 &            A2029  &   no (e,g,j) & MUSE, MMTF, IRAM  &  yes (a,j,p) & -- \\
G008.44$-$56.35 &          A3854 &    no (f) & SOAR & no (a,b,p) & P \\
G021.09+33.25 &            A2204 &    yes (e,m) &  MUSE, VIMOS, IRAM & yes (a,j,p) & C \\
G033.46$-$48.43 &            A2384 &    -- &  -- &  yes (p) & -- \\ 
G033.78+77.16 &            A1795 &      yes (e,g,j,k) & MUSE, MMTF, IRAM, ALMA   &  yes (b,j,p) & C\\
G036.72+14.92 &              RXCJ1804.4+1002 &          -- &  -- & no (b,h) yes (p) & P \\
G042.82+56.61 &            A2065 &     no (a) &  FOS & yes (a,j,p) & -- \\
G044.22+48.68 &            A2142 &    no (a,g,j) & FOS, MMTF, IRAM & yes (j)$^{\star}$ yes (b) no (p) & --\\
G049.20+30.86 &   RXJ1720.1+2638 &    yes (a) & FOS & yes (a,p) & -- \\
G049.66$-$49.50 &            A2426 &    no (a) & FOS & no (p) & -- \\
G055.60+31.86 &            A2261 & no (a,j) & FOS, IRAM & yes (p) & -- \\
G056.81+36.31 &            A2244 &     no (a,f) &  FOS, SOAR & yes (p) & -- \\
G057.92+27.64 &           Cl1742 &          yes (a,j) &  FOS, IRAM & yes (a,p) & P \\ 
G062.42$-$46.41 &               A2440 &             -- & -- &   no (p) & -- \\
G062.92+43.70 &            A2199 &  yes (a) no (j) &  FOS, IRAM & yes (b,j,p) & C \\
G067.23+67.46 &            A1914 &         -- & -- &  yes (a,l)$^{\star}$ no (p) & --\\ 
G072.80$-$18.72 &               RXJ2120.2+2258 &          no (b) & FOS & yes (a,e,p) & P \\
G073.96$-$27.82 &             A2390 &     yes (e,c) no (j) & MUSE, VIMOS, IRAM  & yes (b,k,p) & C \\
G080.99$-$50.90 &               A2552 &      no (e) & MUSE & yes (l)$^{\star}$ no (p) & --\\
G086.45+15.29 &               -- &             -- & -- &    no (b) yes (m,o)$^{\star\star}$ & -- \\
G094.01+27.42 &               E1821+644 &            yes (k) & CARMA &     yes (b,d) & P\\  
G096.85+52.46 &               A1995 &             -- &  -- &  no (n,m,o) & -- \\ 
G098.95+24.86 &               A2312 &    no (a) & FOS & yes (m,o) & -- \\
G115.16$-$72.09 &              A85 &     yes (e,c,j) & MUSE, VIMOS, IRAM &  yes (b,m,a,n,o) & C\\
G115.71+17.52 &               -- &            -- &  -- &  no (o,b) & -- \\
G124.21$-$36.48 &             A115 &    yes (a) &  FOS &yes (a) & C \\
G125.70+53.85 &               A1576 &             -- & -- & yes (b,c)$^{\star}$ & P \\
G139.59+24.18 &               -- &             -- &  -- &  no (o) & -- \\
G146.33$-$15.59 &               AWM7 &           -- & -- &  yes (b) no (p) & -- \\
G164.61+46.38 &               -- &            -- &  --  & no (b,o) & --\\
G166.13+43.39 &             A773 &    no (a) &  FOS &yes (f)$^{\star}$ no (m,o) & --\\ 
G176.28$-$35.05 &       2A0335+096 & yes (e,c,j,k) &  MUSE, VIMOS, IRAM, ALMA &yes (b,p,o,m) & C \\
G180.62+76.65 &            A1423 &      no (i) & IRAM & no (o,b)${\star\star}$ & -- \\
G182.44$-$28.29 &             A478 &     yes (e,c,g,j) & MUSE, VIMOS, MMTF, IRAM &  yes (b,j,p) & C \\
G182.63+55.82 &               A963 &     no (a) & FOS &   yes (o,p) & -- \\
G195.62+44.05 &              A781 &      no (a) & FOS &  yes (b,p)$^{\star\star}$   & -- \\
G209.56$-$36.49 &             A496 &     yes (e,c,g,j) & MUSE, VIMOS, MMTF, IRAM &  yes (b,j,p) & C \\
G226.24+76.76 &            A1413 & no (a) & FOS & yes (l)$^{\star}$ yes (p)$^{\star\star}$ no (o) & --\\
G228.49+53.12 &         Cl1023.3+1257 &      yes (c) & VIMOS &  yes (b,p) & P\\ 
G241.74$-$30.88 &    RXCJ0532.9-3701 &     no (e) & MUSE & yes (m,o,p)$^{\star\star}$ & --\\
G241.77$-$24.00 &            A3378 &    yes (e,c) & MUSE, VIMOS & yes (b,m,o,p) & P\\
G244.34$-$32.13 &           RXCJ0528.9-3927 &          yes (e) & MUSE  &  yes (h,m,p) & -- \\
G252.96$-$56.05 &            A3112 &   yes (e,c) no (k) &  MUSE, VIMOS, ALMA & yes (j,o,p) & --\\
G253.47$-$33.72 &            A3343 &      -- &  -- & no (b,h) & -- \\ 
G256.45$-$65.71 &            A3017 &    yes (e,c) & MUSE, VIMOS& yes (a,o,p) & C \\
G257.34$-$22.18 &              A3399 &    -- &  -- & no (a,b,g) & --\\
G260.03$-$63.44 &  RXCJ0232.2-4420 &     yes (e) & MUSE & yes (a,o,p) & C\\
G263.16$-$23.41 &           AS0592 &    no (e) & MUSE & yes (p) &     P \\
-- &               -- &    yes (e) & MUSE & yes (d) & --\\
G263.66$-$22.53 &           A3404 &  no (e) &  MUSE &  no (o,p) & -- \\
G264.41+19.48 &               -- &      -- &  -- & yes (p) & P \\
G266.84+25.07 &  A3444 & yes (e,c) &  MUSE, VIMOS &yes (a,o,p) & P \\
G286.58$-$31.25 &             -- &     -- &     -- &     yes (p)$^{\star\star}$  &-- \\
G303.75+33.65 &            A3528 &   yes (e) & MUSE &  yes (a,b,m,o,p) & P\\
G304.89+45.45 &            A1644 &    yes (e,g,) & MUSE, MMTF, ALMA&  yes (b,j,o,p) & C\\
G306.68+61.06 &            A1650 &      no (g) &  MMTF & yes (j) no (o,p) & --\\
G306.80+58.60 &            A1651 &    no (e,b) & MUSE, FOS &  yes (a,b,j,p) & P \\
G313.36+61.11 &            A1689 &   yes (e) & MUSE &  yes (j) yes (a,o,p)$^{\star\star}$ & -- \\
G313.87$-$17.10 &             RXCJ1601.7-7544 &        -- &  --  &  no (p) & --\\
G316.34+28.54 &            A3571 &   no (e) &  MUSE & yes (g,j)$^{\star}$ yes (p) no (o) & P \\
G318.13$-$29.57 &                RXCJ1947.3-7623 &     yes (e) & MUSE & yes (h) & P\\
G324.49$-$44.97 &               S150-06 &      -- &   -- & yes (p) & P\\
G340.95+35.11 &            AS0780 &   yes (e,c,g) & MUSE, VIMOS, MMTF&  yes (i$^{\star}$,m,o,p) & P \\
G349.46$-$59.94 &               AS1063 &   no (e) & MUSE & no (o) yes (p) & -- \\
G269.51+26.42   & A1060 &  yes (e,c,j) & MUSE, VIMOS, IRAM& yes (p) & P\\
\enddata
\tablecomments{\scriptsize
(1) Cluster name (the prefix PLCKESZ is omitted for simplicity). (2) Alternative name.\\
(3) Presence (yes) or absence (no) of warm and/or cold gas either by optical (H$\alpha$, $\sim$10,000~K) or millimeter (CO, $\sim$100~K) emission, respectively.\\
(a) FOS (Faint Object Spectrograph) observations \citep{Crawford_1999}, \\
(b) FOS observations \citet{allen92}, \\
(c) VIMOS (VIsible MultiObject Spectrograph) observations \citep{hamer16}, \\
(d) VIMOS observations \citep{wilman06},\\
(e) MUSE observations (see Table~\ref{tab:MUSE}), \\
(f) SOAR (Southern Astrophysical Research Telescope) Goodman Spectrograph observations  \citep{donahue10},\\
(g) MMTF (Maryland Magellan Tunable Filter) images \citep{mcdonald10,mcdonald12},\\
(i) IRAM 30m observations \citep{castignani20} \\
(j) IRAM 30m observations \citep{salome03,edge01,pulido18}\\
(k) ALMA (Atacama Large Millimeter Array) observations \citep{olivares19,russell19,baek22,rose19} and references therein.\\
(l) CARMA (The Combined Array for Research in Millimeter-wave Astronomy) \citep{aravena11}. \\
(4) Radio emission at the position of the BCG. References:\\
(a) GMRT data archive,
(b) The VLA FIRST or VLSSr,  
(c) \citet{kale13}, (d) \citet{blundell01}, (e) \citet{cuciti21}, (f) \citet{barrena07},
(g) \citet{venturi02}, 
(h) SUMSS, (i) \citet{giacintucci19},
(j) \citet{birzan12} and references therein,
(k) \citep{birzan20},
(l) \citet{cuciti21},
(m) NVSS,
(n) LoTSS,
(o) TGSS ADR,
(p) RACS.\\
$^{\star}$ Diffuse radio emission likely associated with a radio halo or mini radio halo.\\
$^{\star\star}$ Radio emission offset from the BCG position.\\
(5) Presence or lack of X-ray cavities. ``P'' and ``C'' correspond to clusters with ``potential'' or ``certain'' cavities, respectively. Taken from \citep{olivares22b}.
}
\end{deluxetable*}


 \section{Results}\label{sec:results}

\newcommand{\mylengthncc}{0.45}
\begin{figure*}
\centering
    \includegraphics[width=0.46\textwidth]{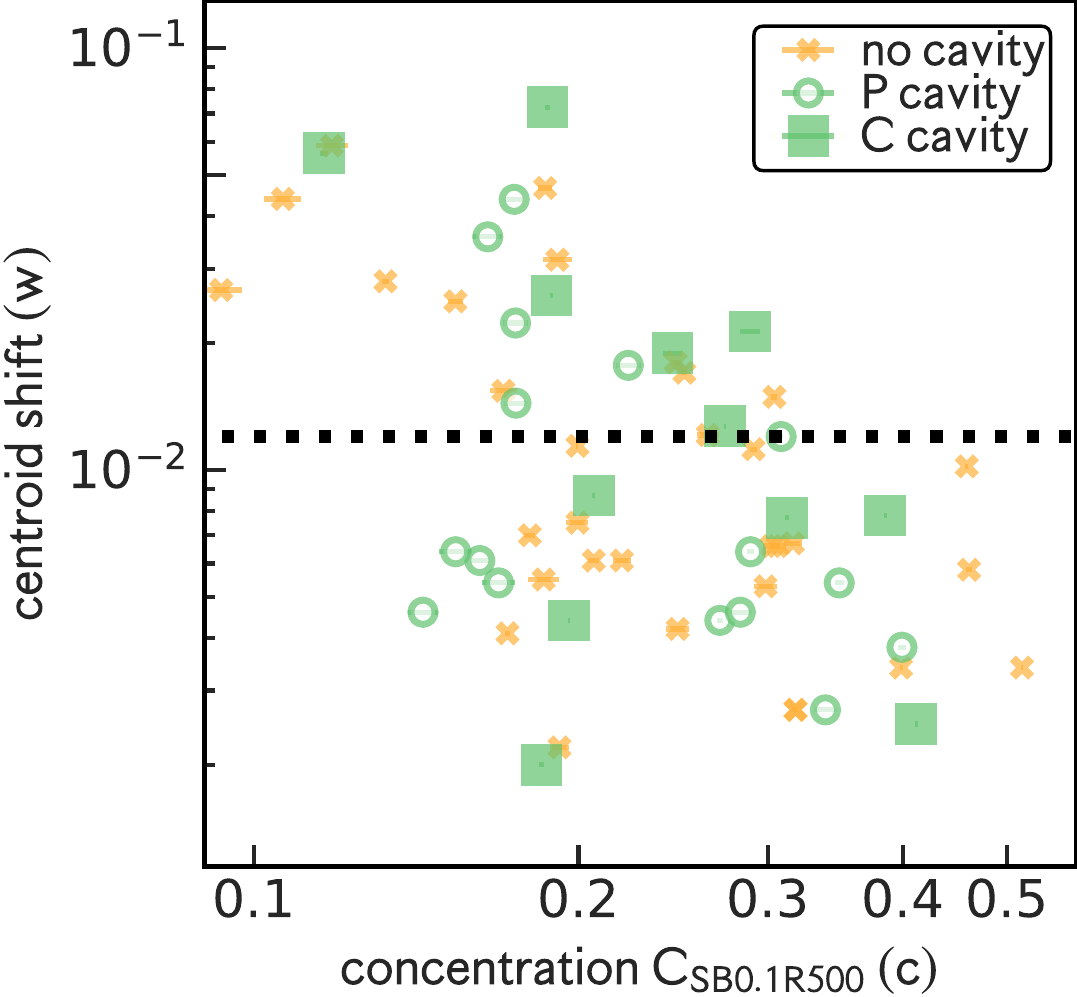}
    \includegraphics[width=0.52\textwidth]{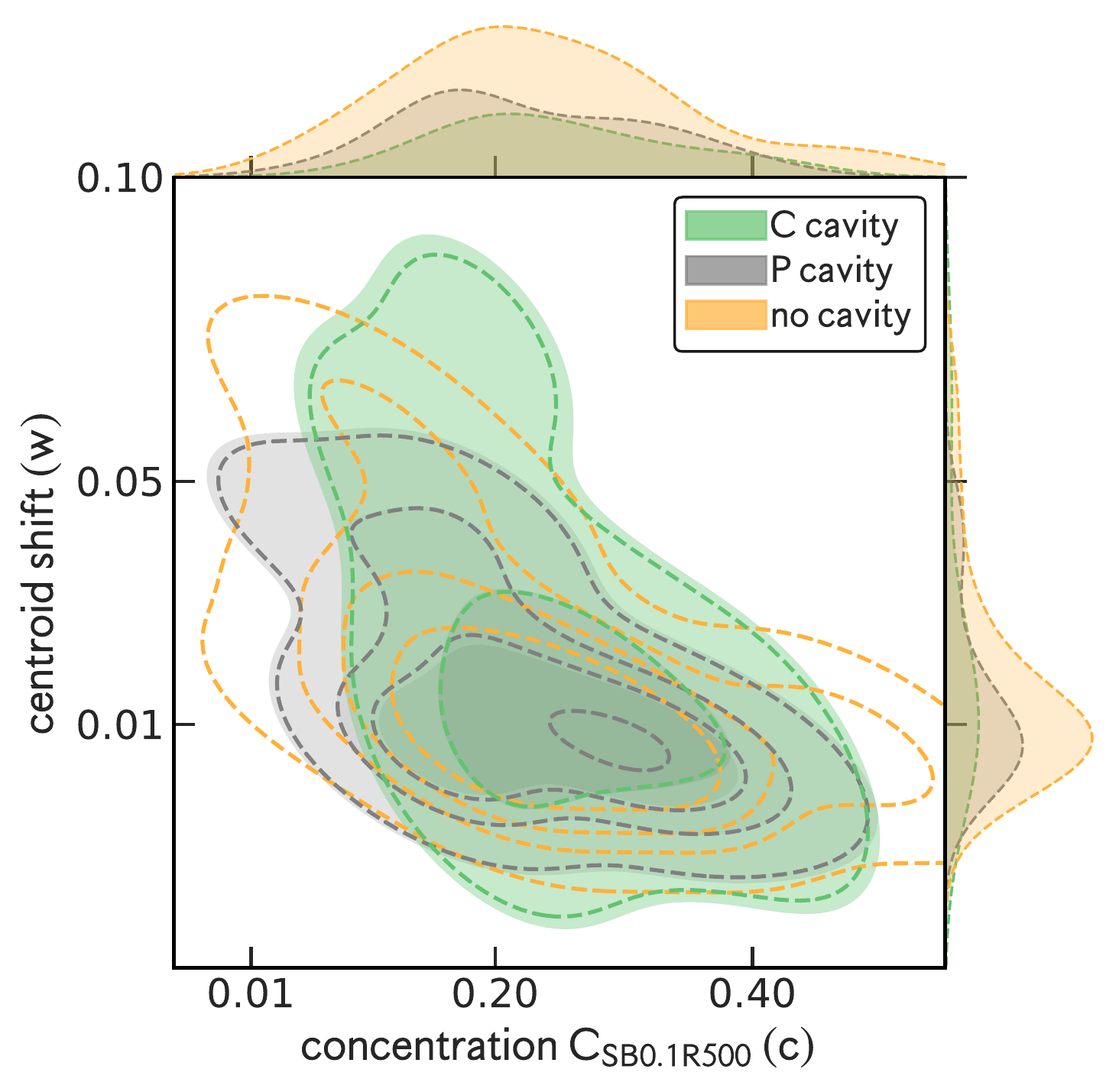}
    \caption{Dynamical state of the clusters as a function of cavity presence. Left panel: centroid shift versus concentration for the CC cluster sub-sample. Following \citet{lovisari17,cassano10}, we adopted the value $w>0.012$ (dotted black line) as an indication of disturbed clusters. Color-coded by cavity class, clusters with ``certain'' cavities are shown with green \textit{squares}, with ``potential'' cavities with empty green circles, and clusters without cavities are displayed with orange crosses. The right panel shows the kernel density estimate to visualize the distribution of the clusters with cavities (green and grey) and without cavities (orange). For visualization, we have added the 1D kernel distribution for each cluster classification. We recall that 27\% and 19\%  of the clusters show ``certain'' (C) and ``potential'' (P) cavities, while 54\% lack cavities.}\label{fig:disturbed_clusters}
\end{figure*}

\subsection{Dynamical state of clusters and their effects on cavities and AGN feedback.}\label{sec:dynamical_state}

\subsubsection{Presence of cavities and dynamical state}

{We explore whether the dynamical state of the cluster may influence the presence of cavities.} The dynamical state of clusters describes whether galaxy clusters are dynamically relaxed or are undergoing a merging (unrelaxed clusters). 
According to \citet{lovisari17}, the best parameters to separate relaxed and disturbed systems are by combining concentration ($c$) and centroid shift ($w$).

Figure~\ref{fig:disturbed_clusters} (left panel) shows the centroid shift ($w$) versus the concentration ($c$) values for the CC cluster sub-sample, color-coded by the presence (green {squares for ``Certain'' cavities and green empty-circles for ``Potential'' cavities} ) or absence (orange circles) of X-ray cavities. To visualize the distribution of the observations, we plot the kernel density estimate (KDE) {(see right panel of Figure~\ref{fig:disturbed_clusters})} that expresses the dependence of the presence (green curves) or absence (orange curves) of cavities on the centroid shift versus concentration. {For visualization, we have added the 1D kernel distribution for each cluster classification.}

The figure also shows that cavities are seen in clusters with various dynamical states -- relaxed, sloshing (minor merger), and merging. It also shows that more concentrated clusters are less perturbed with smaller centroid shifts. {The latter could be interpreted in two ways, either relaxed clusters are more difficult to shift, or the perturbation that causes a large centroid shift (w) in disturbed clusters can weaken the concentration (c).}

Overall, {the likelihood of a cluster having a detectable cavity is not related to its X-ray-based dynamical state. Our sample includes clusters with high centroid shift values ($w\geq0.012$), indicative of mergers or sloshing motions, of which a significant fraction, 35\% (16/45), display cavities.} Regardless, clusters with cavities appear to be {slightly shifted to the bottom-right of the diagram toward more CC systems, with concentration $\leq$ 0.2.} The latter are expected as they have better-resolved cores than clusters with ``potential'' cavities allowing a high detectability rate of X-ray cavities \citep[e.g.,][]{panagoulia14,olivares22b}.

\subsubsection{Effects of the dynamical state on the cooling luminosity versus cavity power relation}\label{sec:Lcool-Pcav}

\begin{figure*}
    \centering
    \includegraphics[width=0.49\textwidth]{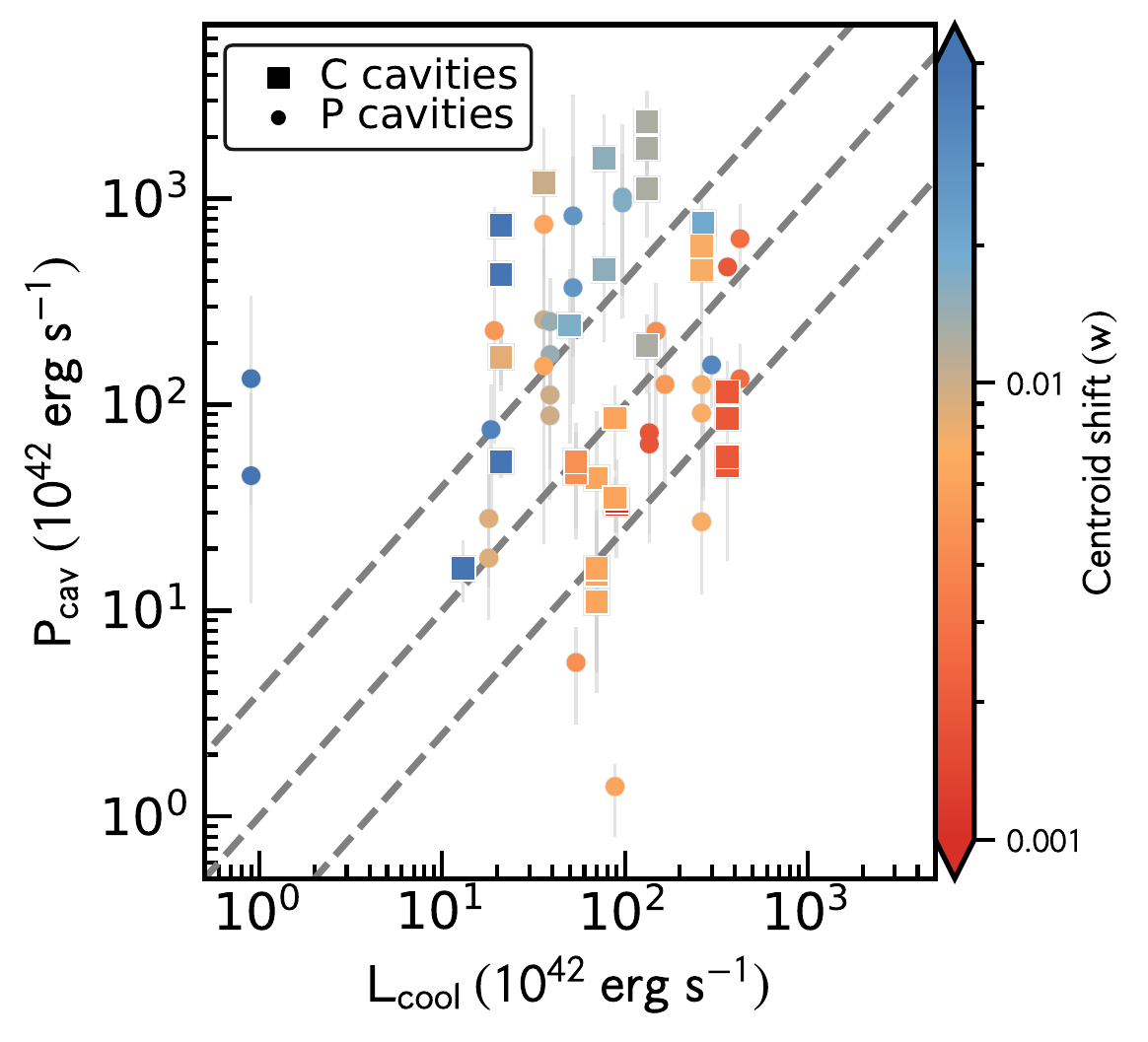}
    \includegraphics[width=0.47\textwidth]{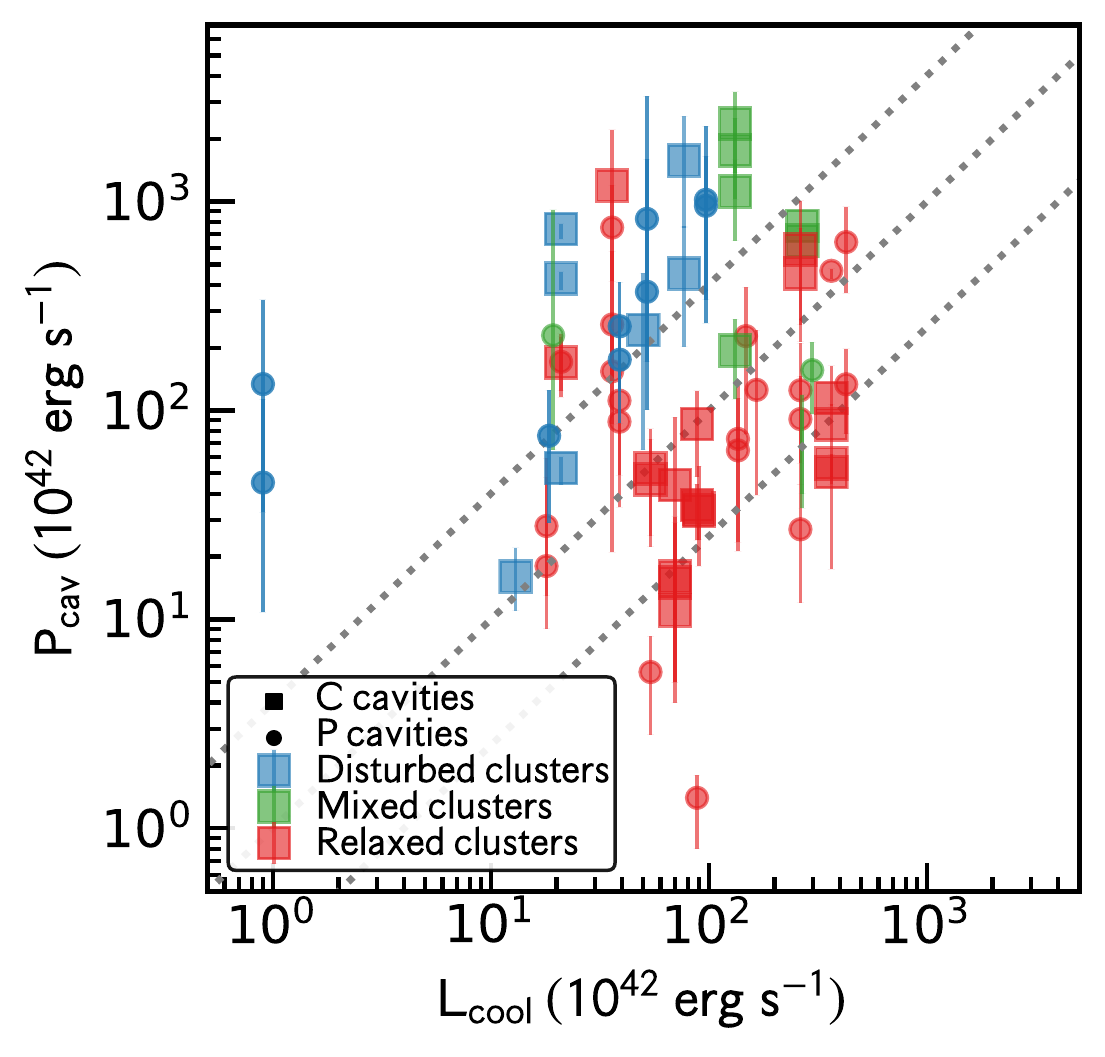}
    \caption{Left Panel: Cooling luminosity versus cavity power color-coded by centroid shift as a proxy for the dynamical state of each cluster. Disturbed clusters are marked with bluer colors, and relaxed clusters with redder colors. Right panel: Cooling luminosity versus cavity power color-coded by dynamical state based on the centroid shift, $w$, and concentration, $c$, classification of \citet{lovisari17}. Relaxed, disturbed and mixed clusters are shown with red, blue, and green colors, respectively. In both plots, ``certain'' (C) and ``potential'' (P) cavities are shown with {squares and circles} symbols, respectively. From top to bottom, the dotted black lines are pV, 4pV, and 16pV.}\label{fig:Lcool_Pcav_dyn}
\end{figure*}

{Previous studies \citep{rafferty06, birzan04, olivares22a, Hlavacek-Larrondo15} have shown that AGN can compensate for the cooling losses of the ICM by producing heating through radio jets, as demonstrated by the plot of cavity power, $P_{\rm cav}$, versus cooling luminosity, $L_{\rm cool}$.} In this section, we investigate how the dynamical state of the clusters affects the relationship between cavity power and cooling luminosity. Sloshing and mergers induce large bulk motions of the ICM, which may disrupt the cool core and reduce the cooling luminosity, impact the morphology (thus volume) of the X-ray cavities, or bring the X-ray cavities to regions with different ambient ICM pressures.

{Sloshing and mergers} may, thus, generate scatter in the $L_{\rm cool}$ -- $P_{\rm cav}$ relation. In Figure~\ref{fig:Lcool_Pcav_dyn} (Left panel), we plot this relation, color-coded by the centroid shift ($w$) as a proxy for the dynamical state of the hot atmospheres, for each cluster. Estimates of cavity powers ($P_{\rm cav}$) and cooling luminosities ($L_{\rm cool}$) for each detected cavity in the Planck SZ sample were taken from \citet{olivares22b}. The cooling luminosity, $L_{\rm cool}$, was calculated within a volume where the (isobaric) cooling time, $t_{\rm cool}$, is 7.7~Gyrs (see \citealt{olivares22a} for details). {The cavity power, $P_{\rm cav}$, which corresponds to the mechanical power released by the AGN, was estimated for each cavity of the clusters, by dividing the total enthalpy of each cavity ($E_{\rm cav} = 4 p V$) by its age. Here $p$ is the thermal pressure of the ICM at the projected location of the cavity, defined as the center of each ellipse, and $V$ is the cavity volume. For the purpose of this work, we used the buoyant rise time ($t_{\rm buoy}$) as the age of the cavity \citep{mcnamara00,birzan04,McNamara_2005}. }

As a matter of comparison, in Figure~\ref{fig:Lcool_Pcav_dyn} (Right panel), we plot the $P_{\rm cav}-L_{\rm cool}$ relationship with each cluster being classified as relaxed (red), disturbed (blue) or mixed (green), based on the centroid shift and concentration classification discussed in Section~4.2. Both figures show that disturbed and mixed clusters have systematically larger cavity powers than relaxed clusters, by about a factor of two, for a given cooling luminosity. Consequently, these clusters are located at the top of the $P_{\rm cav}$--$L_{\rm cool}$ relation. On the other hand, relaxed clusters tend to follow better the one-to-one relation. We have visually examined the dynamically disturbed clusters (e.g., A2390 (G073.96-27.82), G057.92+27.64, A85 (G115.16-72.09), A115 (G124.21-36.48), 2A0335+096 (G176.28-35.05), and A3017(G256.45-65.71)), and most of them display clear signatures of sloshing or recent interaction (see Figure~\ref{fig:chandra_images_C}).

\subsubsection{The effects of ICM ``weather'' on the cavities}\label{sec:dynamical_state_morphology}

\begin{figure}[htp]
    \includegraphics[width=0.49\textwidth]{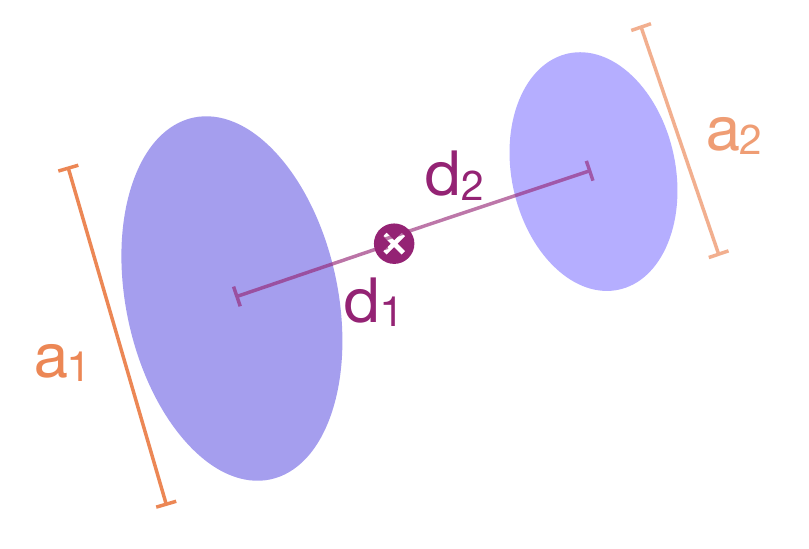}
    \caption{Diagram showing two pairs of cavities illustrating how we measure the major axis of the two cavities, $a_{1}$ and $a_{2}$, and the distances to the center, $d_{1}$ and $d_{2}$.} \label{fig:cartoon_cavities}
\end{figure}

\begin{figure*}
    \includegraphics[width=0.5\textwidth]{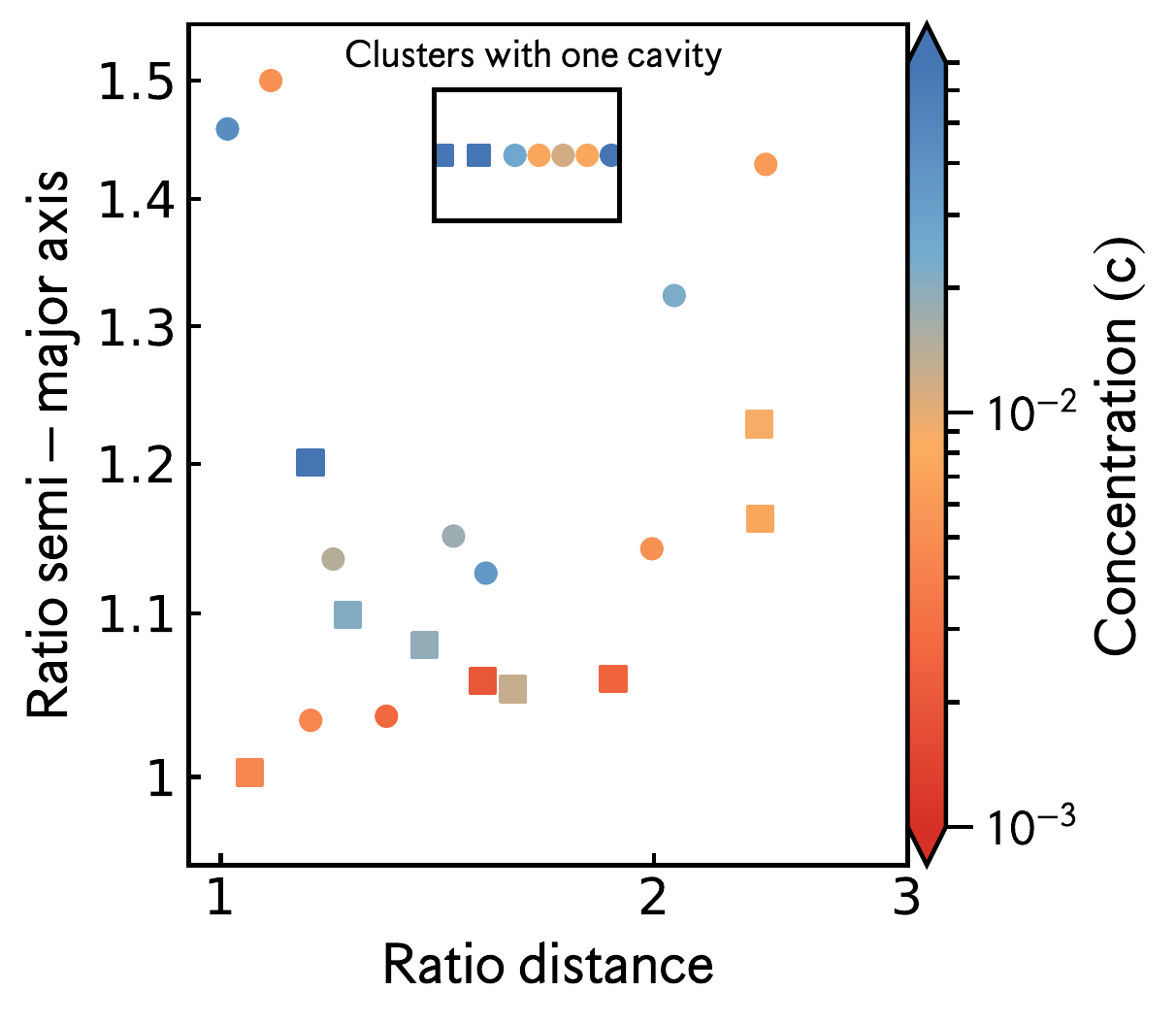}
    \includegraphics[width=0.45\textwidth]{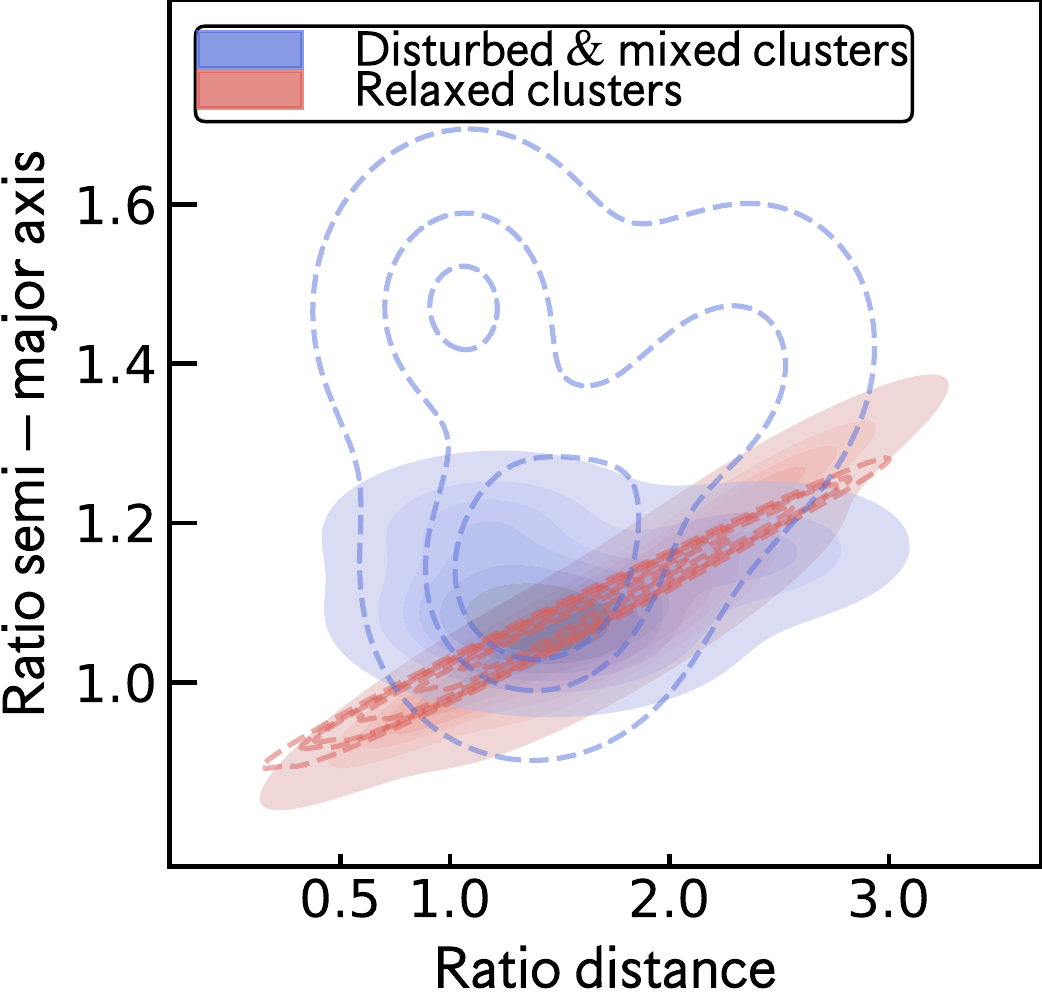}
    \caption{Left panel: The ratio of the semi major-axis (a) of a pair of cavities versus the ratio of the distance (D) of each cavity to the BCG center. Each cluster is color-coded by centroid shift. Clusters with only one cavity are shown in the inserted subplot at the top-left. ``Certain'' (C) and ``Potential'' (P) cavities are shown with {squares and circles}, respectively. Disturbed clusters display slightly more asymmetric cavities compared to relaxed clusters. Right panel: kernel density estimate to visualize the distribution of the asymmetry on the cavities in disturbed (blue) and relaxed (red) clusters. {``Certain'' (C) cavities are shown with a shaded region, while ``Potential'' (P) cavities with dashed lines.}
    }\label{fig:cav_asym}
\end{figure*}

High-resolution hydrodynamical simulations have shown that the cluster ``weather'', affected by sloshing and turbulent motions, drives more disturbed and asymmetric cavities (and, consequently, radio lobes) due to the interaction of the buoyant bubbles with the surrounding ICM \citep[e.g.,][]{mendygral12,gaspari12,wittor20}. To quantify how asymmetric the cavities are in a given cluster with respect to their dynamical state, in Figure~\ref{fig:cav_asym}, we compare the ratio of the cavity sizes (semi-major axis) with the ratio of the distances to the center (d) of pairs of cavities color-coded as a function of centroid shift ($w$). {In Figure~\ref{fig:cartoon_cavities}, we illustrate how we measure the  major axes and the distances to the center of each pair of cavities. This allows us to quantify the ratio of major axes, $a_{\rm 1}/a_{\rm 2}$, and the ratio of distances, $d_{\rm 1}/d_{\rm 2}$.} When a cluster displays more than two cavities, we plot the two largest cavities located on opposite sides of the BCG. For visualization propose, in Figure~\ref{fig:cav_asym} (right panel), we have included the kernel distribution categorized by their dynamical state in disturbed (including mixed) and relaxed clusters.

Based on the mean ratio of a pair of cavities, we find that unrelaxed clusters, including dynamically mixed clusters, have roughly 20\%, in fraction, more asymmetric cavities than relaxed clusters. No significant difference is found in the ratio of the distance between the two cavities in the two populations. We attribute that to the impact of ICM ``weather'' on the morphology of the cavities. In this fashion, the cavity shapes might be sensitive to the jet-ambient density contrast, and their asymmetry could be driven by differences in the underlying hot environment \citep{smith21,patrick21}. It is worth noting that the cavity morphology may also be related to the properties of the jets, such as velocity, density, duty cycle, and plasma micro-physics (\citealt{Gaspari_2011,guo15,yang19}). However, this is outside the scope of this paper, and we leave this for a future work.

3D simulations of relativistic high-power jets show that differences in the underlying environment can drive large-scale radio lobe asymmetry \citep{Yates-Jones21}. Specifically, more asymmetric environments have approximately 10-25\% more asymmetric radio lobes than less disturbed environments (see Fig.~11 of \citealt{Yates-Jones21}), which is consistent with our findings. The authors reported a length asymmetry ratio of 1.3 -- 1.1 for systems moving through environments with a high density asymmetry ratio ($\sim$3) (see Fig.~11 of \citealt{Yates-Jones21}). 

\subsubsection{Clusters with multiple generations of cavities}

\begin{figure}[htp]
    \includegraphics[width=0.49\textwidth]{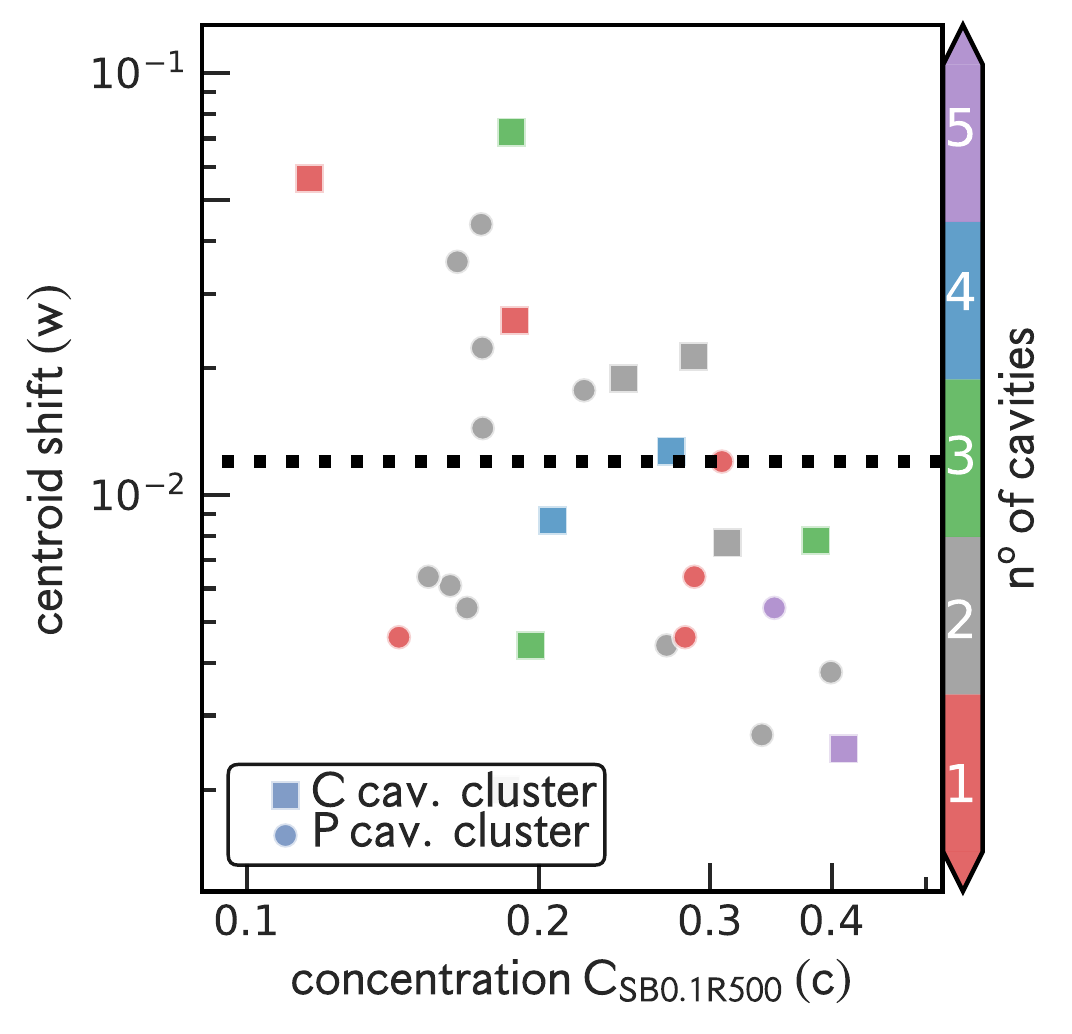}
    \caption{Centroid shift (w) versus concentration (c) for clusters with cavities. We color-coded the number of cavities found in each cluster as follows: red for one cavity, and blue, green, and purple for two, three, and four cavities, respectively. Systems with ``Certain'' (C) and ``Potential'' (P) cavities are shown with {squares and circles}, respectively. The dotted black line corresponds to $w=0.012$ as a separation between disturbed and relaxed clusters. } \label{fig:number_cav_centroid_shifts}
\end{figure}

We explore whether large motions of the ICM could have disrupted the AGN jets and bubbles, leading to the appearance of multiple disconnected X-ray bubbles, as predicted by numerical simulations presented in \citet{morsony10}. Figure~\ref{fig:number_cav_centroid_shifts} shows the dynamical state of each cluster versus concentration, color-coded by the number of cavities. {We found no correlation, a Pearson correlation coefficient of $-$0.18, between the number of cavities and the dynamical state of clusters, hinting that cavities are difficult to break apart.}

Instead, the high cavity numbers ($>2$) could be due to multiple AGN outbursts, reflecting the AGN duty cycle, with reoriented radio jets \citep[e.g.,][]{mckinley22}. This process can occur due to several agents, such as jet precession or instabilities in the accretion disc \citep[e.g.,][]{Dennett-Thorpe02, Gopal-Krishna12}. \citet{qiu20} argue that filaments in central cluster galaxies tend to deflect the outflowing radio plasma that inflates the cavities in different directions from the jet axis. \citet{gaspari17} show that the CCA condensation rain is able to induce fractal feeding events, with potentially multiple generations of AGN bubbles generated in a few tens of Myr. Moreover, \citet{beckmann19} show that the AGN spin evolves following the dynamics of the precipitation of the cold gas onto the cluster center, again producing bubbles in different directions. In summary, it seems more likely that multiple pairs of cavities are a by-product of several AGN outbursts, as clearly glimpsed in relaxed clusters, such as AS0780 (G340.95+35.11, {see Figure~\ref{fig:chandra_images_P}}), rather than a consequence of ICM disruption.

Of particular interest are the clusters with single cavities, {as most of them (4/6) tend to be more dynamically disturbed, with $w>0.01$} (see Figure~\ref{fig:number_cav_centroid_shifts} top-left inner panel). 
When interacting with sloshing cold fronts, one of the bubbles in a pair can be destroyed (and perturbed) as shown in numerical simulations \citep{fabian21}. It is worth mentioning that the sloshing simulations presented in \citet{fabian21} show remarkable similarities with some of the single-cavity clusters (see Figure~\ref{fig:chandra_images_C} of A1644 (G304.89+45.45) and A85 (G115.16-72.09) clusters). Consequently, sloshing cold fronts may have destroyed the undetected counter cavity in some of these clusters. Bulk motions from mergers could move the counter X-ray cavity far {from the cluster core} to the outskirts where it could be undetected \citep{giacintucci20}. {It may also have propagated to a region of less dense ICM on the opposite side, making the counter ICM bubble more challenging to detect.}

As discussed above, it is more probable that multiple cavities are a consequence of several AGN outbursts rather than the disintegration of large cavities, {and each pair of cavities can be used to quantify the AGN outburst age}. Our sample has four clusters with multiple generations of X-ray cavities, A2204 (G021.09+33.25, see \citealt{sanders09}), AS0780 (G340.95+35.11), 2A0335+096 (G176-35.05, see  \citealt{sanders09b,mazzotta03}), and A115 (G124.21-36.48). Cavities, from each pair, are located at roughly equal distances from the AGN in the BCG nucleus.

The pairs of cavities can be used to quantify the AGN outburst age. Using the buoyancy timescale, each cavity is between 9~Myr and 70~Myr old (see \citealt{olivares22b} for more details). We compute the time between each AGN outburst as the difference between the average cavity ages of the different pairs. The suggested interval times are 5--9~Myr, 10--22~Myr, 10--30~Myr, and 7--18~Myr for A2204 (G021.09+33.25), AS0780 (G340.95+35.11), 2A0335+096 (G176.28-35.05), and A115 (G124.21-36.48), respectively. {The interval timing estimates} indicates that the central AGN inflates a new pair of bubbles every 5 to 30~Myr, varying from cluster to cluster. We note that no correlation was found between such timescales and the mass of each cluster. Our mean outburst interval period of $\sim$20~Myr is comparable to the timescale found in known clusters with multiple generations of cavities (e.g., $\sim$20~Myr for Perseus \citep{fabian00} and M87 \citep{babul13}).



\subsection{Cold gas precipitation and the role of cavities}\label{sec:cold_gas}

\begin{figure*}[htp]
\centering
    \includegraphics[width=0.32\textwidth]{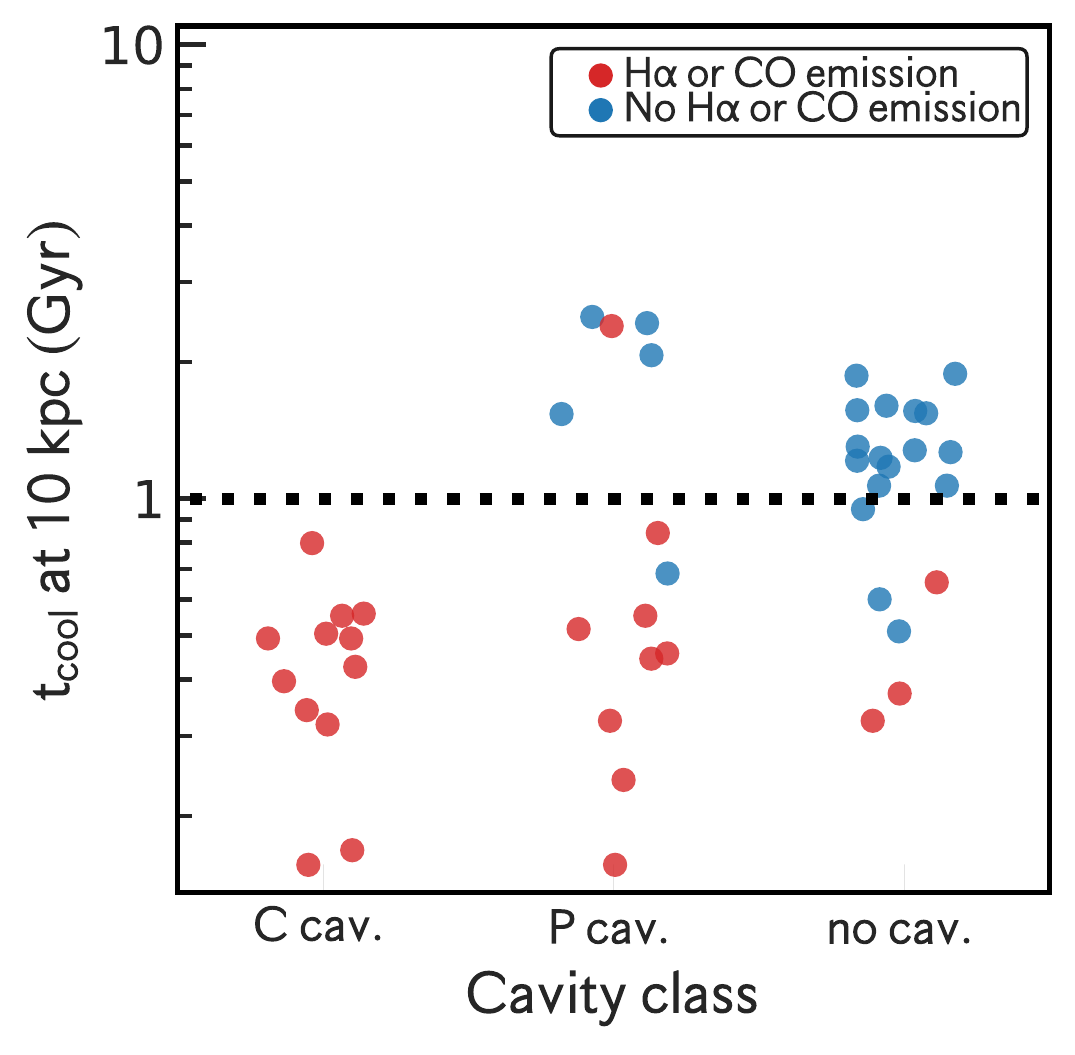}
    \includegraphics[width=0.33\textwidth]{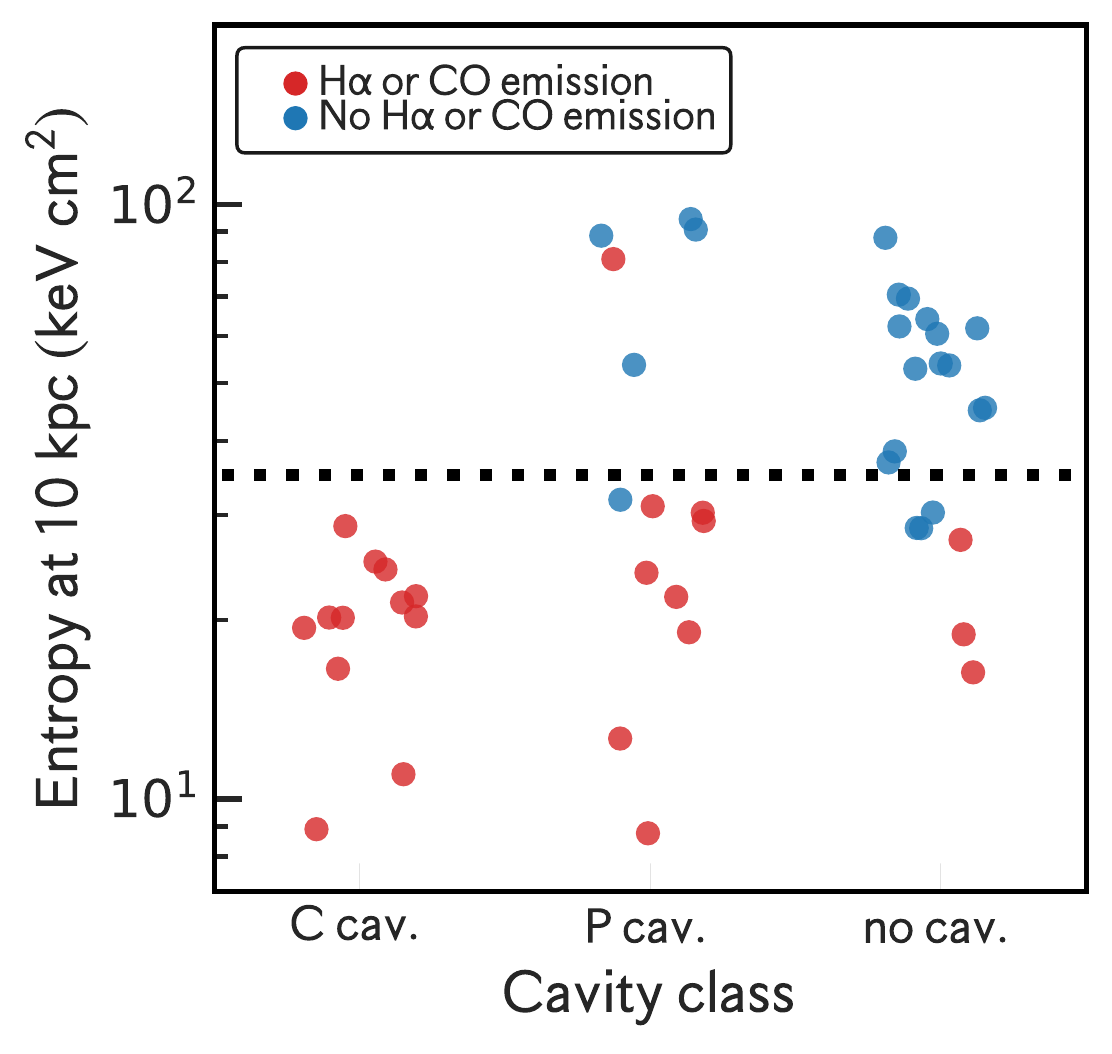}
    \includegraphics[width=0.32\textwidth]{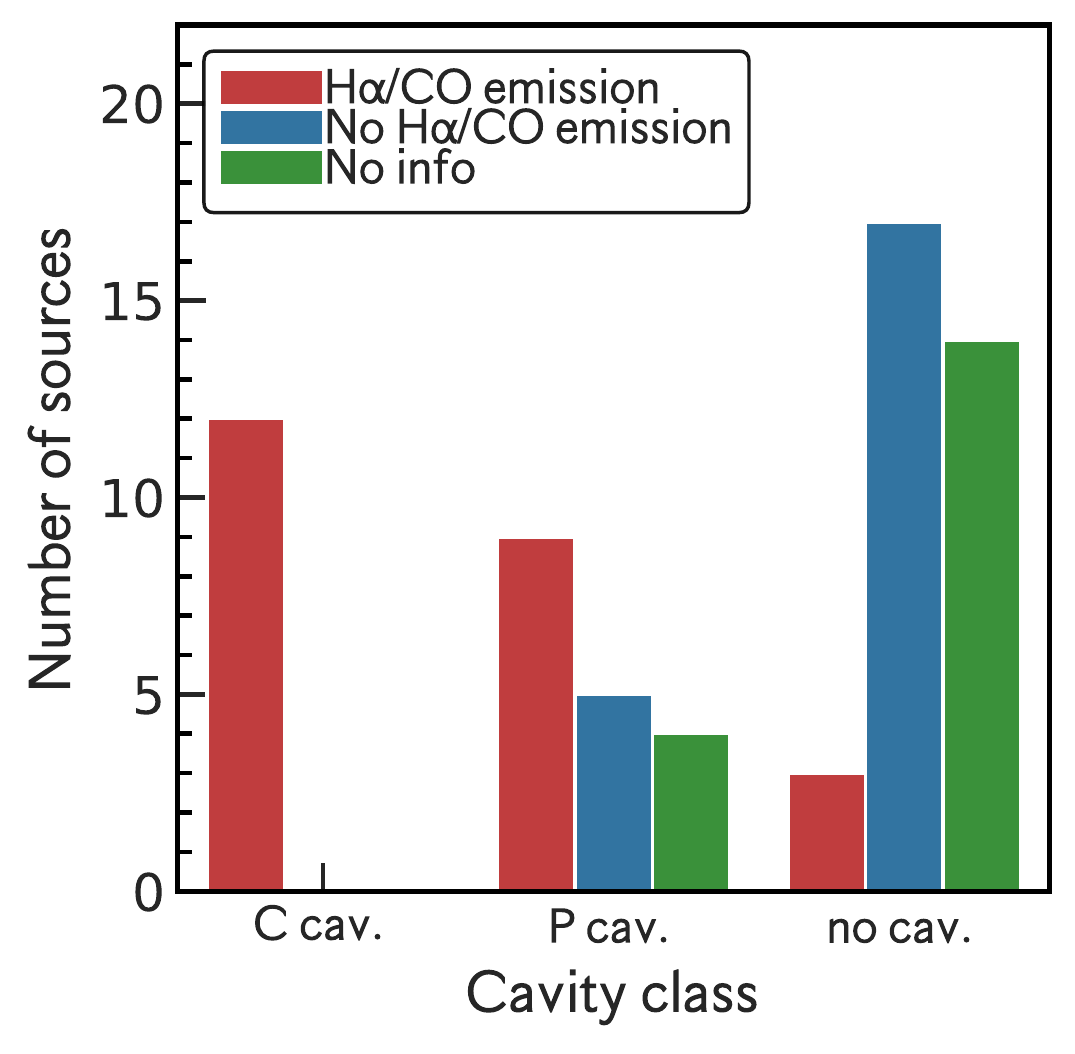}
    \caption{Left panel: Isochoric central cooling time measured at 10~kpc as a function of cavity class, color-coded by the presence or absence of H$\alpha$ emission. The dotted line marks the cooling time of 1~Gyr. Middle panel: Central entropy values measured at 10~kpc. The dotted line corresponds to the entropy threshold of 30~keV~cm$^{2}$ \citep[e.g.,][]{cavagnolo08}. The cluster with a higher central cooling time and H$\alpha$ emission corresponds to A1060 (G269.51+26.42).
    Right panel: Histogram of the clusters with ``certain'' (C), ``potential'' (P), and no X-ray cavities. The plots are color-coded by the presence (red), absence of H$\alpha$ emission (blue), and no information available (green). 
    }
    \label{fig:Cav_Halpha_relation}
\end{figure*}

Warm and cold gas in the form of filaments is common in clusters and groups with both short central cooling times, below 1~Gyr, and low central entropy values, $\lesssim$30~keV~cm$^{-2}$ \citep[e.g.,][]{cavagnolo08,rafferty08,hogan17b,main17,olivares22a}. This correlation has been interpreted as a natural outcome of the top-down condensation of hot atmospheres and subsequent rain/CCA \citep[e.g.,][]{gaspari12,prasad15,Voit_2017,beckmann19}.

We calculated the deprojected entropy and cooling time profiles to explore whether these findings hold for our sample. We use the deprojected density and temperature profiles, as follows:
\begin{equation}
   t_{\rm cool} = \frac{3}{2} \frac{nkT}{n_{\rm e} n_{\rm H} \Lambda(T,Z)},
\end{equation}
and
\begin{equation}
   K = kTn_{\rm e}^{-2/3}
\end{equation}
Here $n_{\rm e}$ and $n_{\rm H}$ are the electron and hydrogen number densities, $n$, is the total number of gas particles per unit volume, and $kT$ is the gas temperature. 
{We use the cooling functions, $\Lambda(Z,T)$, from \citet{OGnat07}, assuming metallicity of $Z=1~Z\odot$, since typically CC clusters have solar or nearly solar metallicity within their cores \citep[e.g.,][]{molendi01,mernier16}}. 

As shown in Figure~\ref{fig:Cav_Halpha_relation} (left and middle panel), we find that most Planck clusters with cold molecular gas or H$\alpha$ emitting gas have a short central cooling time, below 1~Gyr, and a low central entropy value, $<$30~keV~cm$^{2}$, measured at 10~kpc (see also \citet{pulido18} for similar results).

Nonetheless, {three} CC clusters with short central cooling times and low central entropy values lack cold molecular or H$\alpha$ emitting gas. One of these outliers is the well-known A2029 (G006.47+50.54, see \citealt{mcnamara16}), also AS0592 (G263.16-23.41) and A2261 (G055.60+31.86). It is worth mentioning that AS0592 (G263.16-23.41) is in the process of merging, and there is a compact (4~kpc, 1.2$\arcsec$) distribution of optical emitting gas, offset by 14~kpc (4$\arcsec$) from its BCG (J063845.17-535822.4). {The absence of a strong relationship between the central ICM properties and the presence of cold gas could be attributed to variations in metallicity. For instance, if the metallicity is 0.3~Z$_{\odot}$, then the cooling time is 1.5 times longer than with 1~Z$_{\odot}$ metallicity. The lack of CO/H$\alpha$ emitting gas may be due to insufficient sensitivity of the optical/millimeter observations \citep{Crawford_1999}. However, deep MUSE observations show no evidence of ionized gas in the case of A2029 or AS0592.}

{Other distinguishing features of A2029 and A2261 which could help us understand why these strong CC clusters lack cold/warm gas, are that they host some of the most massive black holes \citep{Dullo17,Gultekin21}, and have steep gravitational potentials (with large stellar velocity dispersion of 1044 km~s$^{-1}$ and 778 km~s$^{-1}$, respectively).}

\subsubsection{The distribution of multiphase gas and its connection to cavities}

\begin{figure*}[ht]
   \textit{\hspace{0.2cm}\large H$\alpha$ filaments trailing cavities}\\
       \vspace{0.2cm}
    \includegraphics[width=0.495\textwidth]{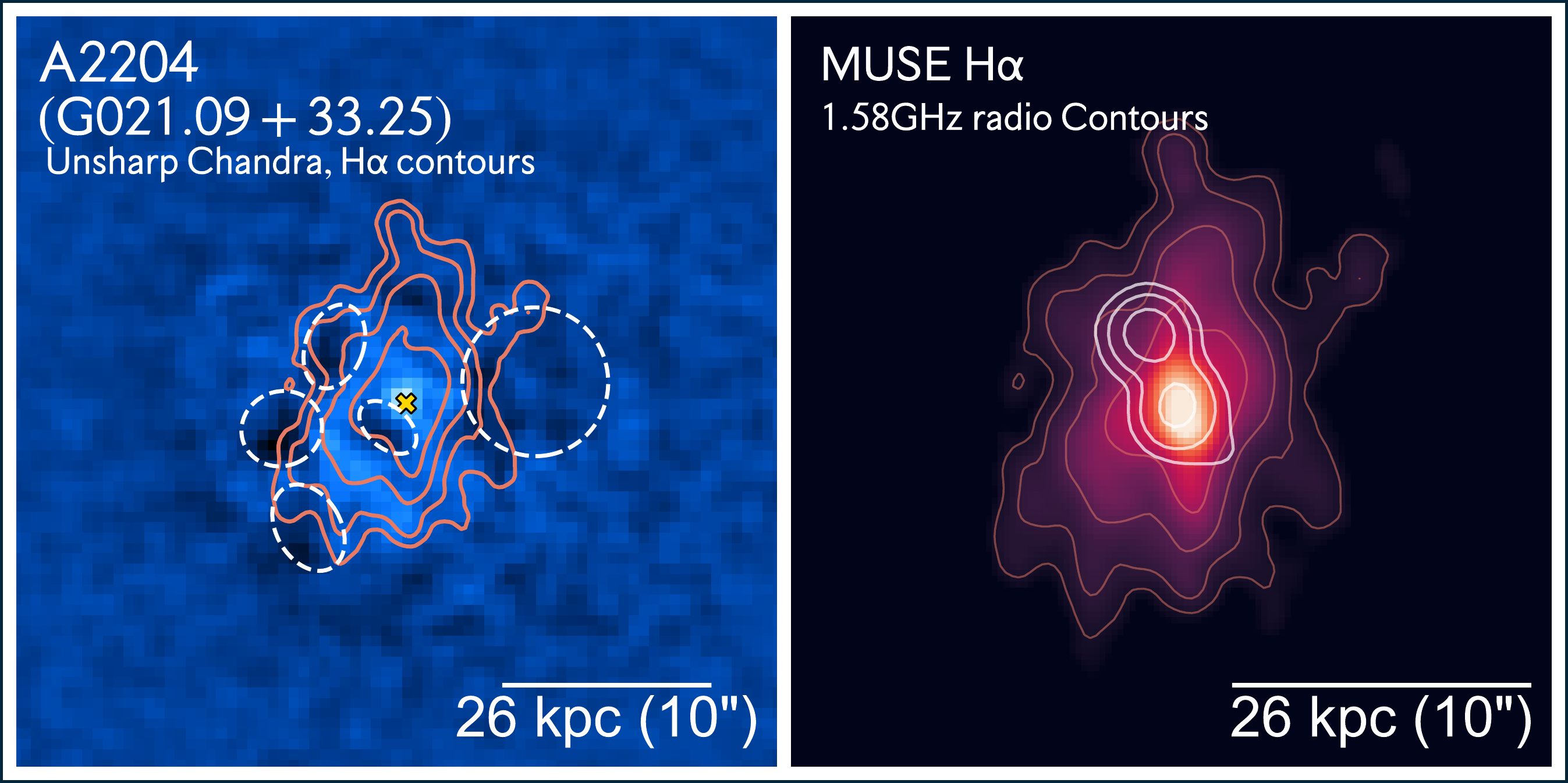}
    \includegraphics[width=0.495\textwidth]{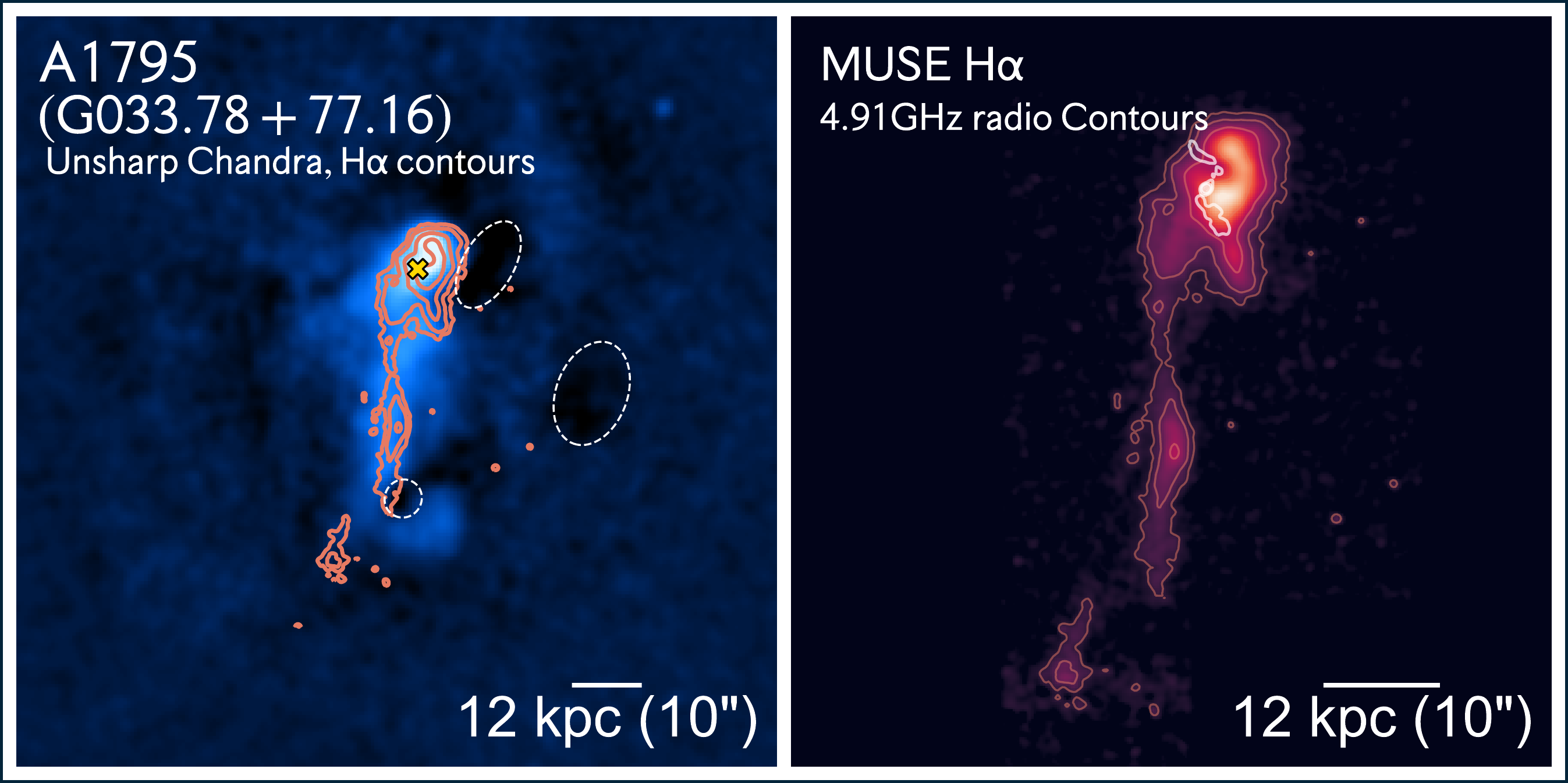}\\
    \vspace{0.2cm}
    \includegraphics[width=0.495\textwidth]{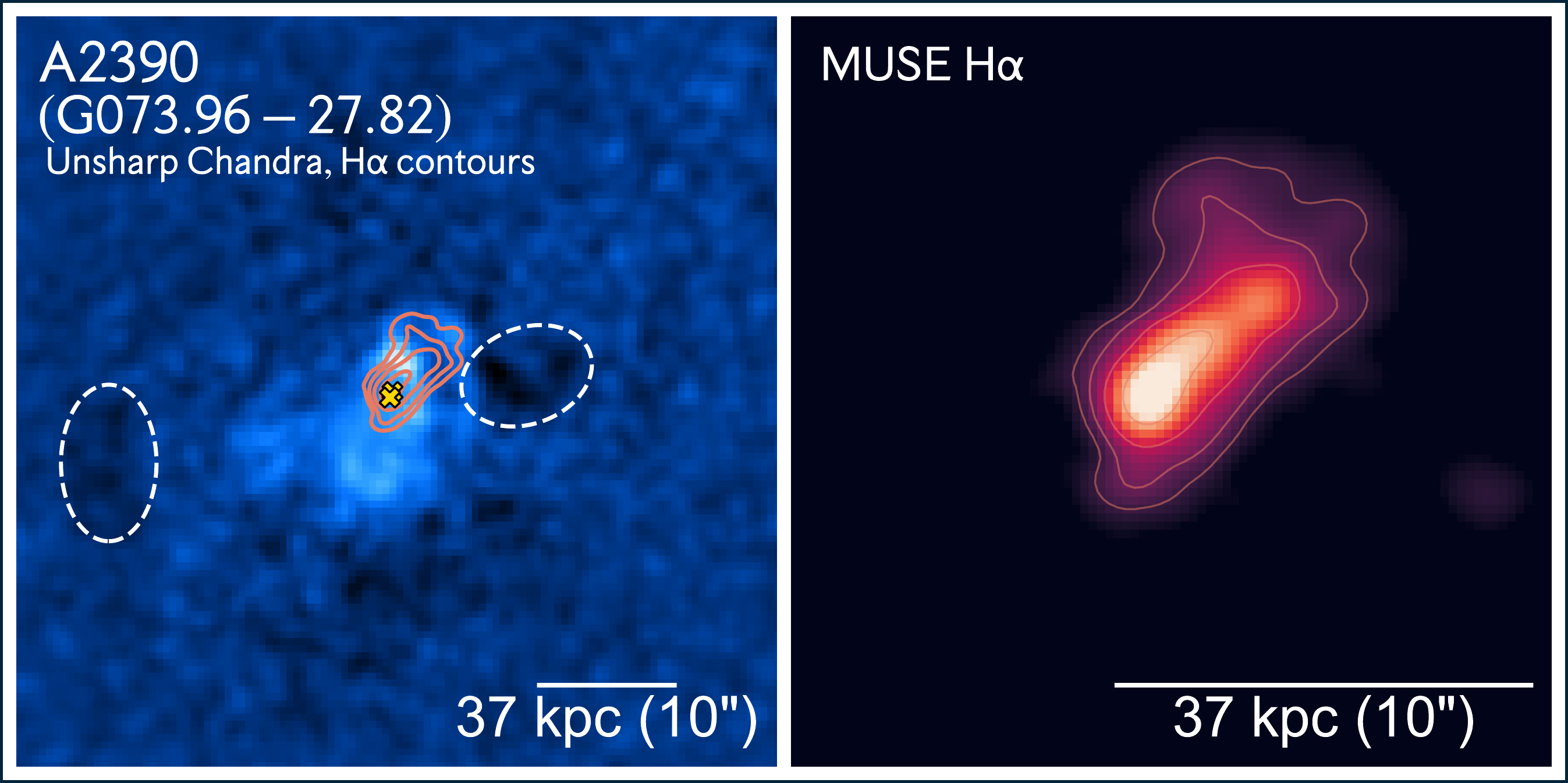}
    \includegraphics[width=0.495\textwidth]{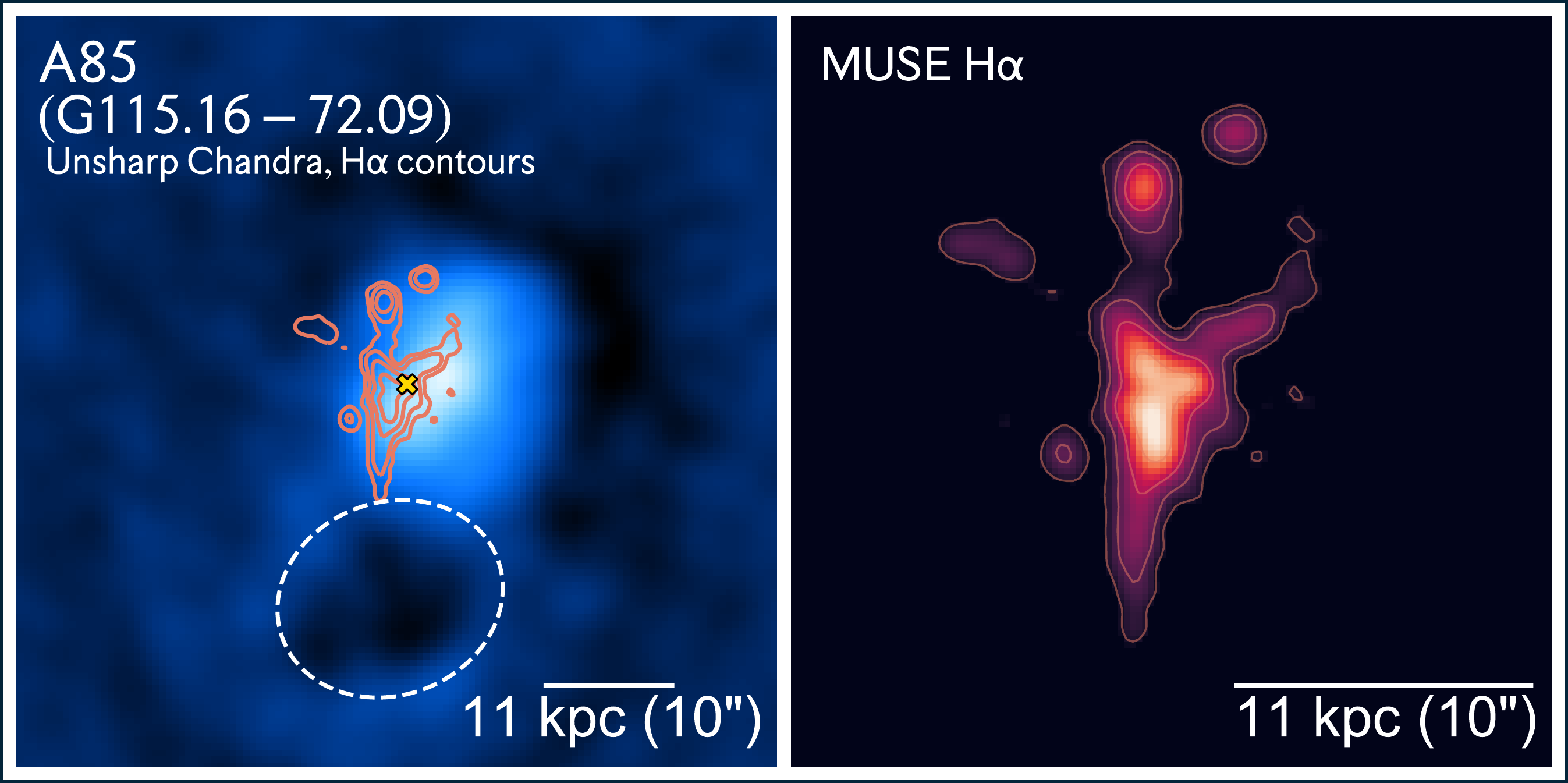}\\
    \vspace{0.2cm}
    \includegraphics[width=0.495\textwidth]{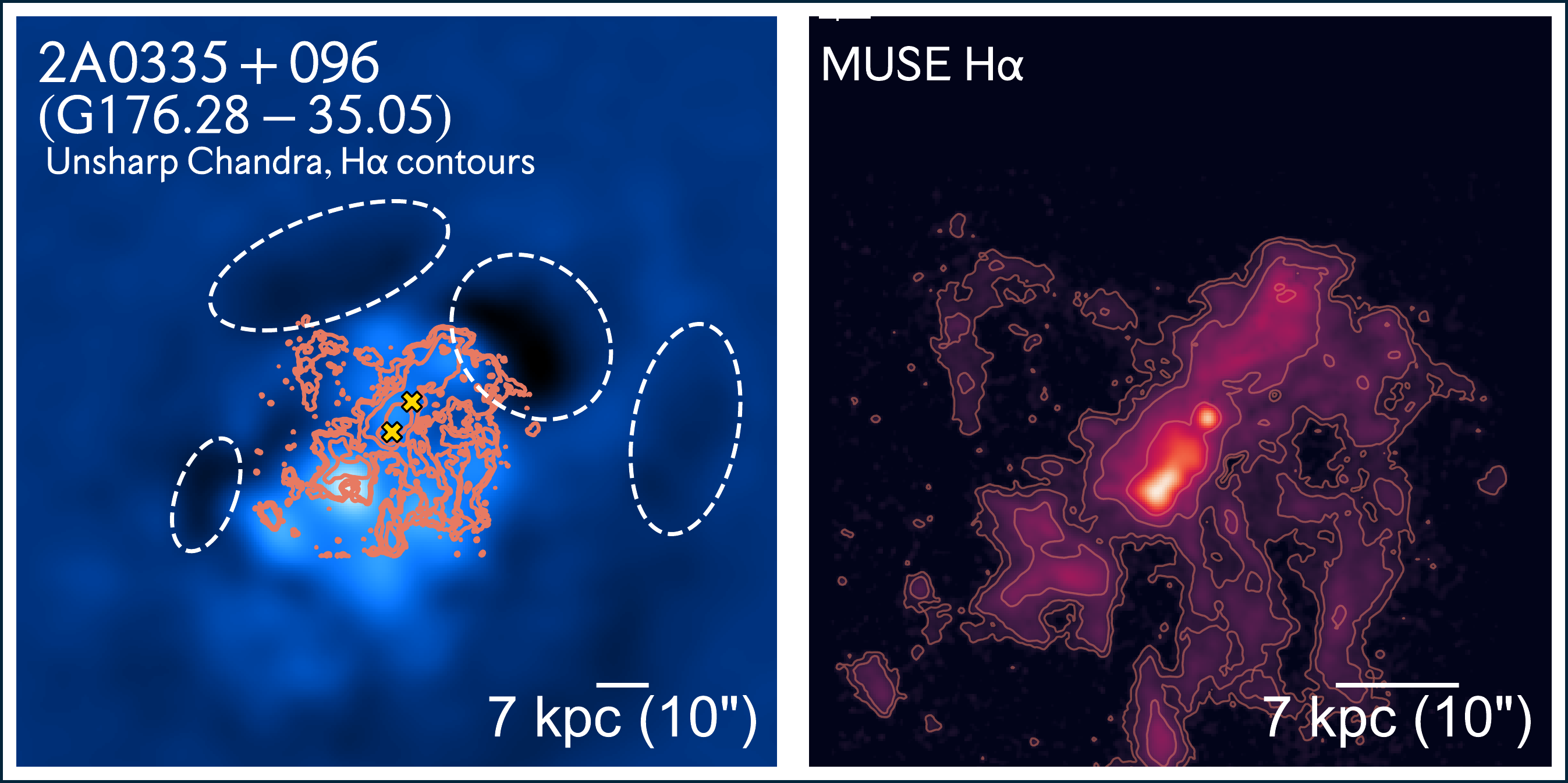}
    \includegraphics[width=0.495\textwidth]{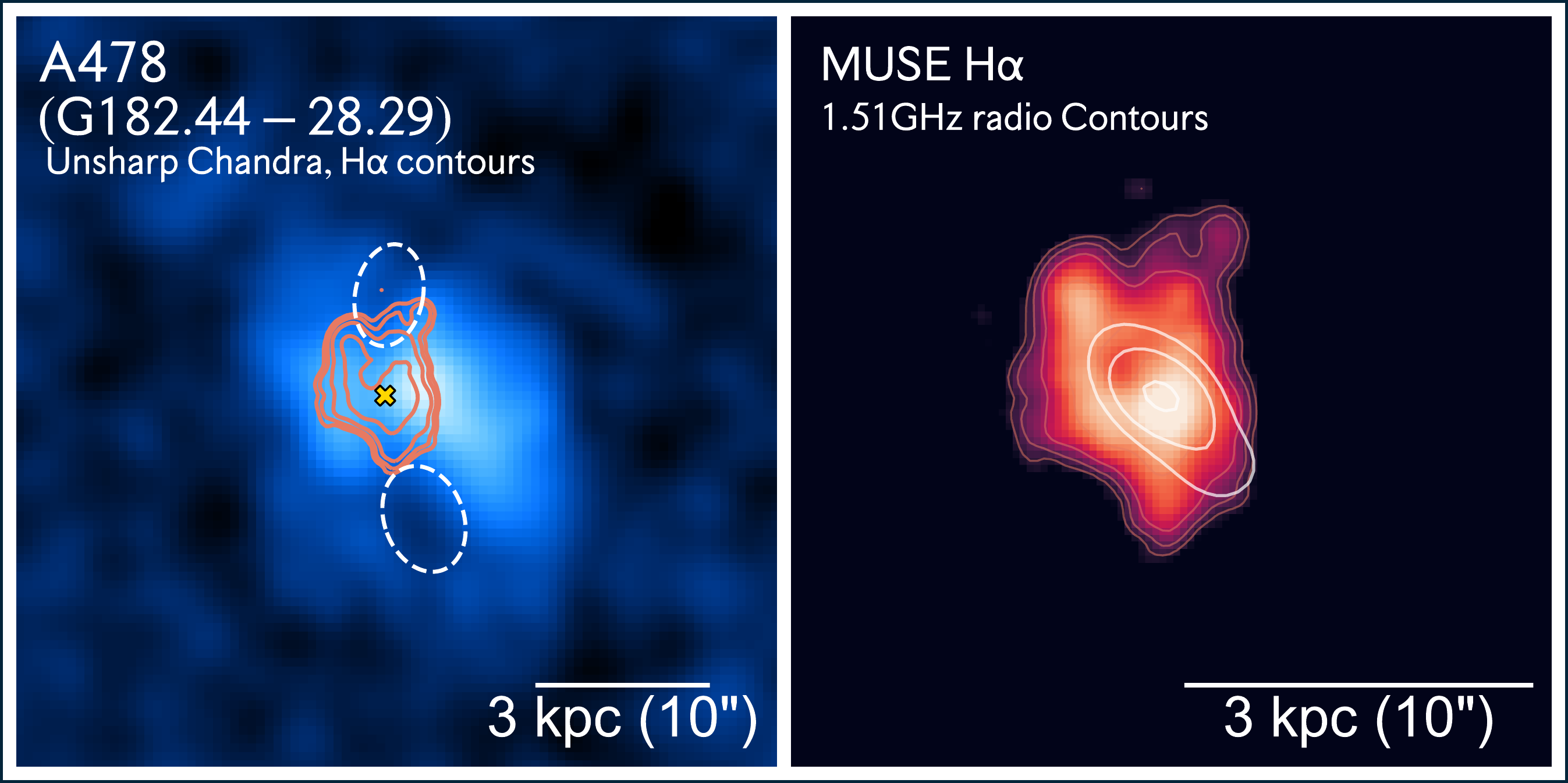}\\
    \vspace{0.2cm}
    \includegraphics[width=0.495\textwidth]{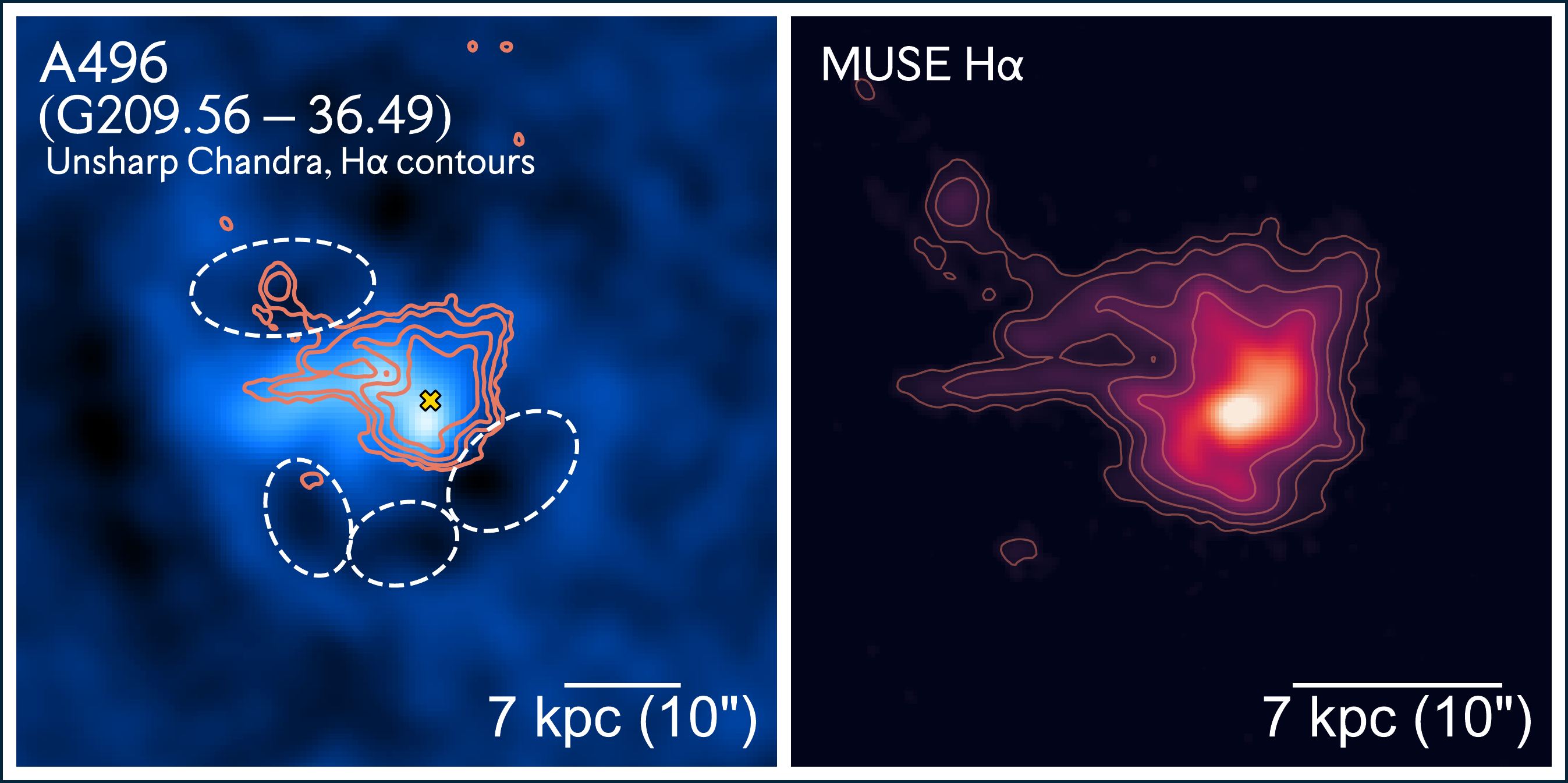}
    \includegraphics[width=0.495\textwidth]{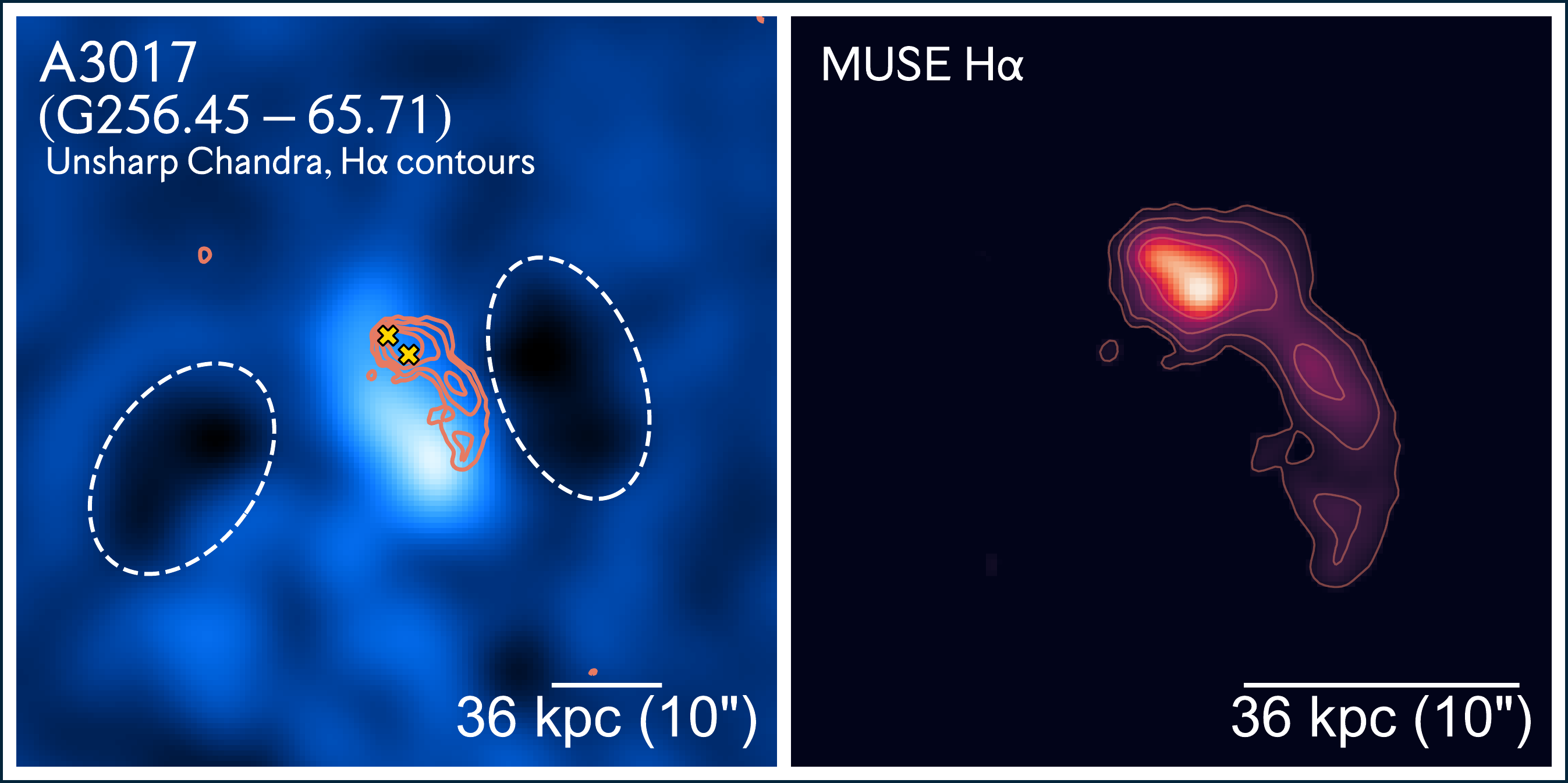}\\
    \caption{ \textit{Chandra} X-ray images of the sources that have optical emission lines detected by MUSE observations where the H$\alpha$ filaments are trailing cavities. Left panel: \textit{Chandra} unsharp masked image. X-ray cavities are shown with white dashed ellipses, while the H$\alpha$ distribution is shown with red contours. Right panel: A zoom-in image of the H$\alpha$ emission distribution from MUSE observations. The central BCG(s) is marked with a yellow cross. The scale bar corresponds to 10$\arcsec$. North is up, and east is left.
    \label{fig:Halpha_maps}}
\end{figure*}

\begin{figure*}[ht]
    \figurenum{8}
    \includegraphics[width=0.495\textwidth]{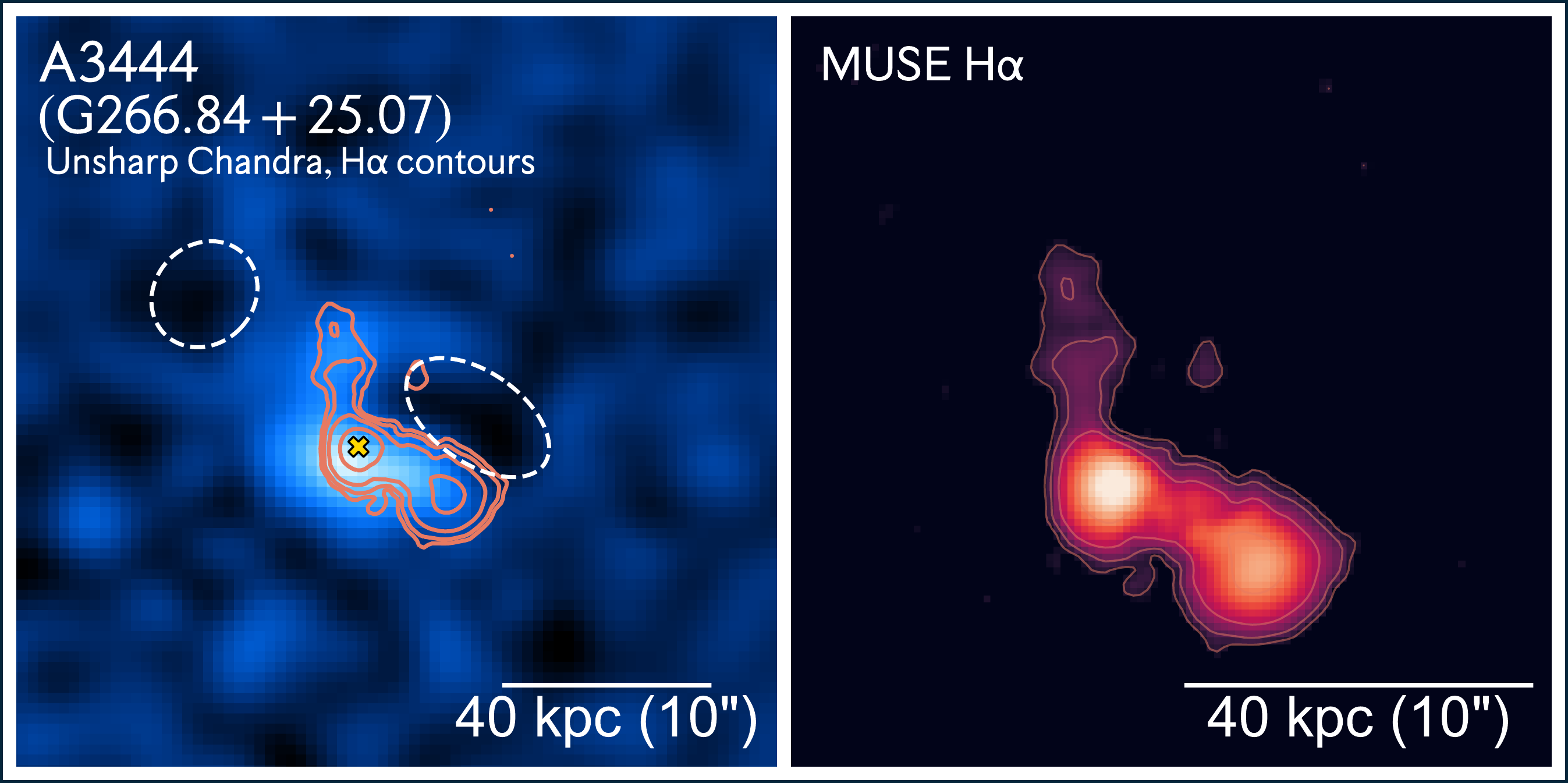}
    \includegraphics[width=0.495\textwidth]{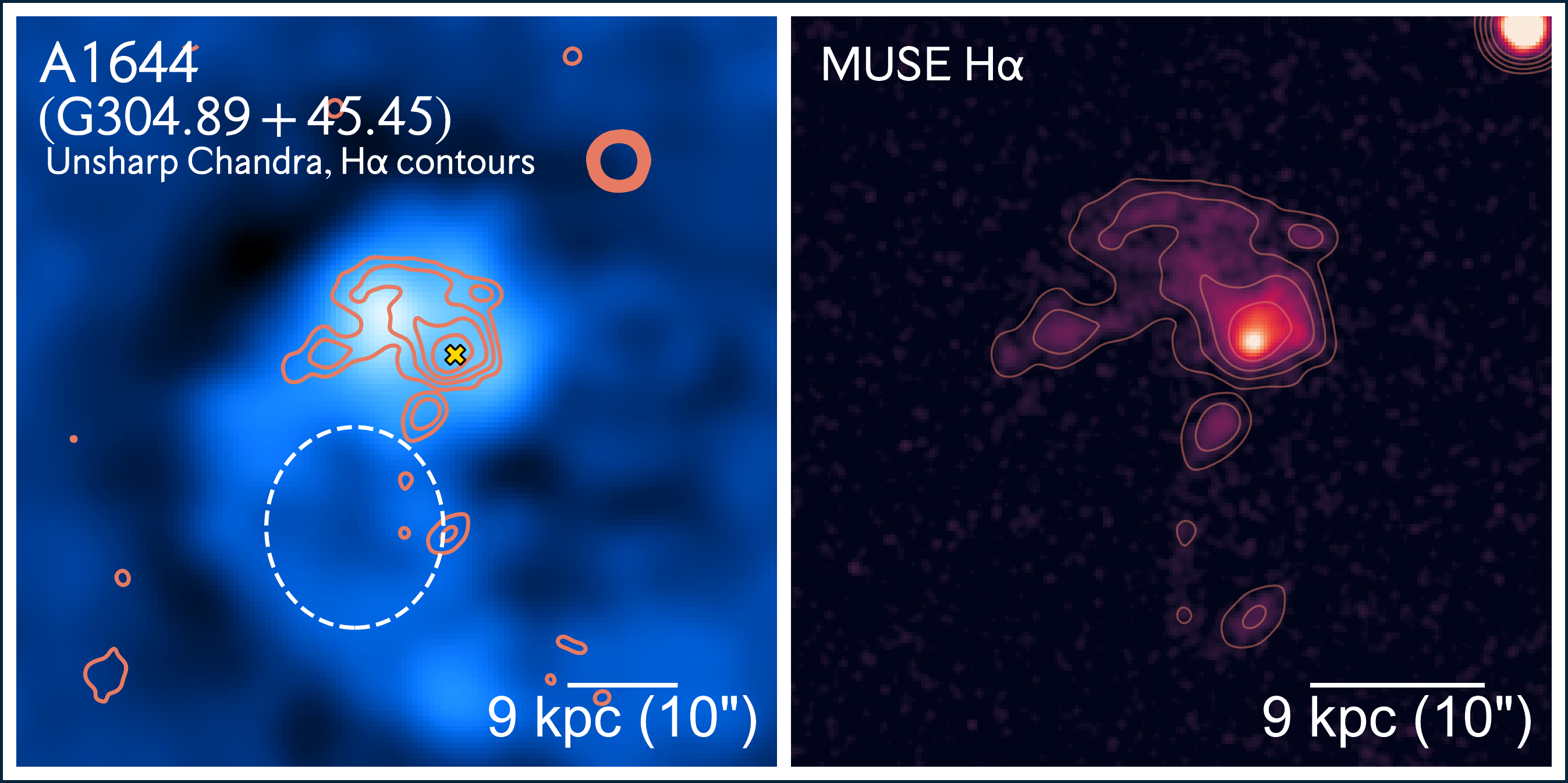}\\
    \vspace{0.2cm}
    \includegraphics[width=0.495\textwidth]{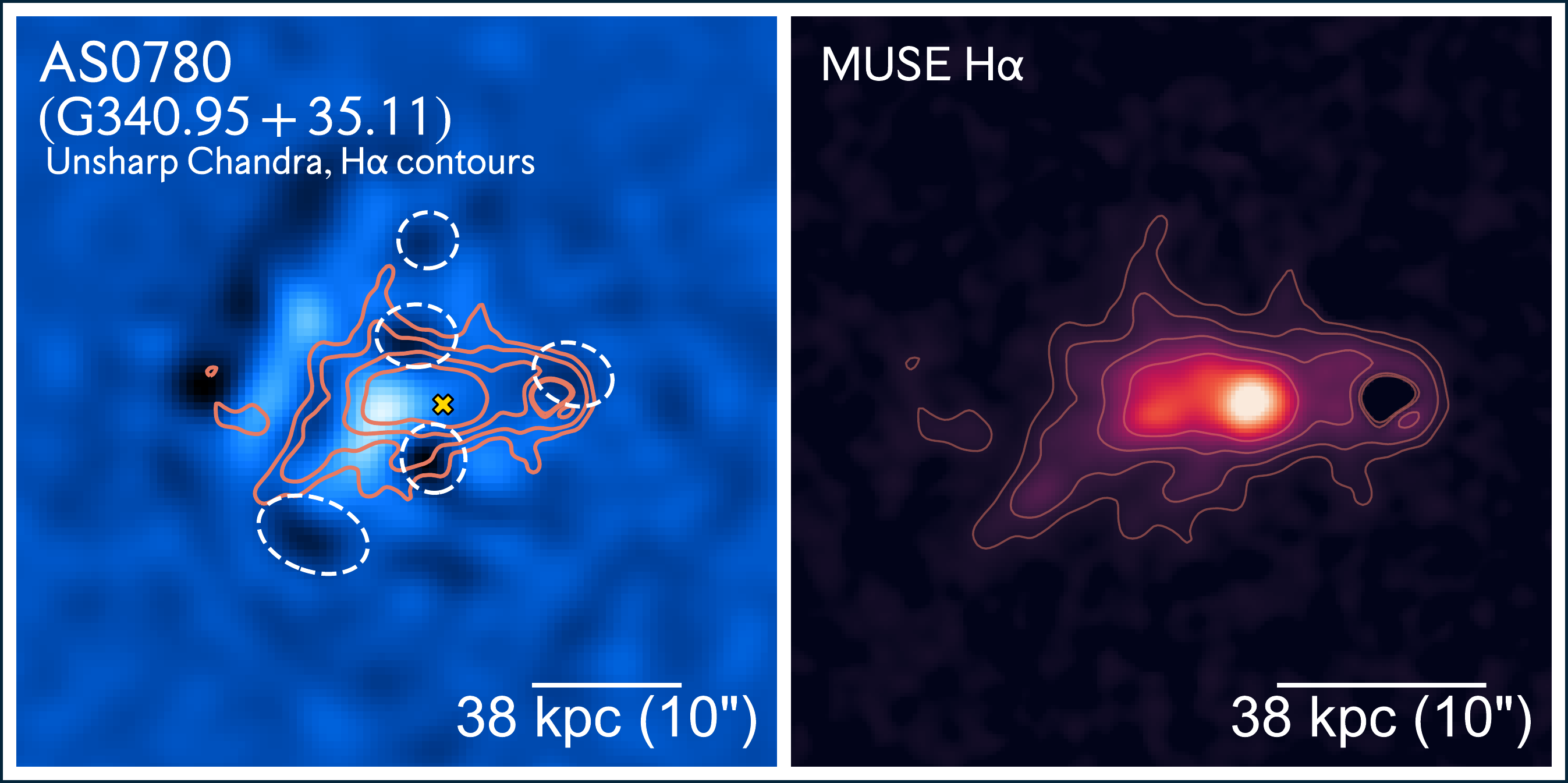}
    \caption{Continuation of Figure \ref{fig:Halpha_maps}.}
\end{figure*}

\begin{figure*}[ht]
\textit{\large Connection between the H$\alpha$ gas and the cavities is not obvious}\\
    \vspace{0.2cm}
    \includegraphics[width=0.495\textwidth]{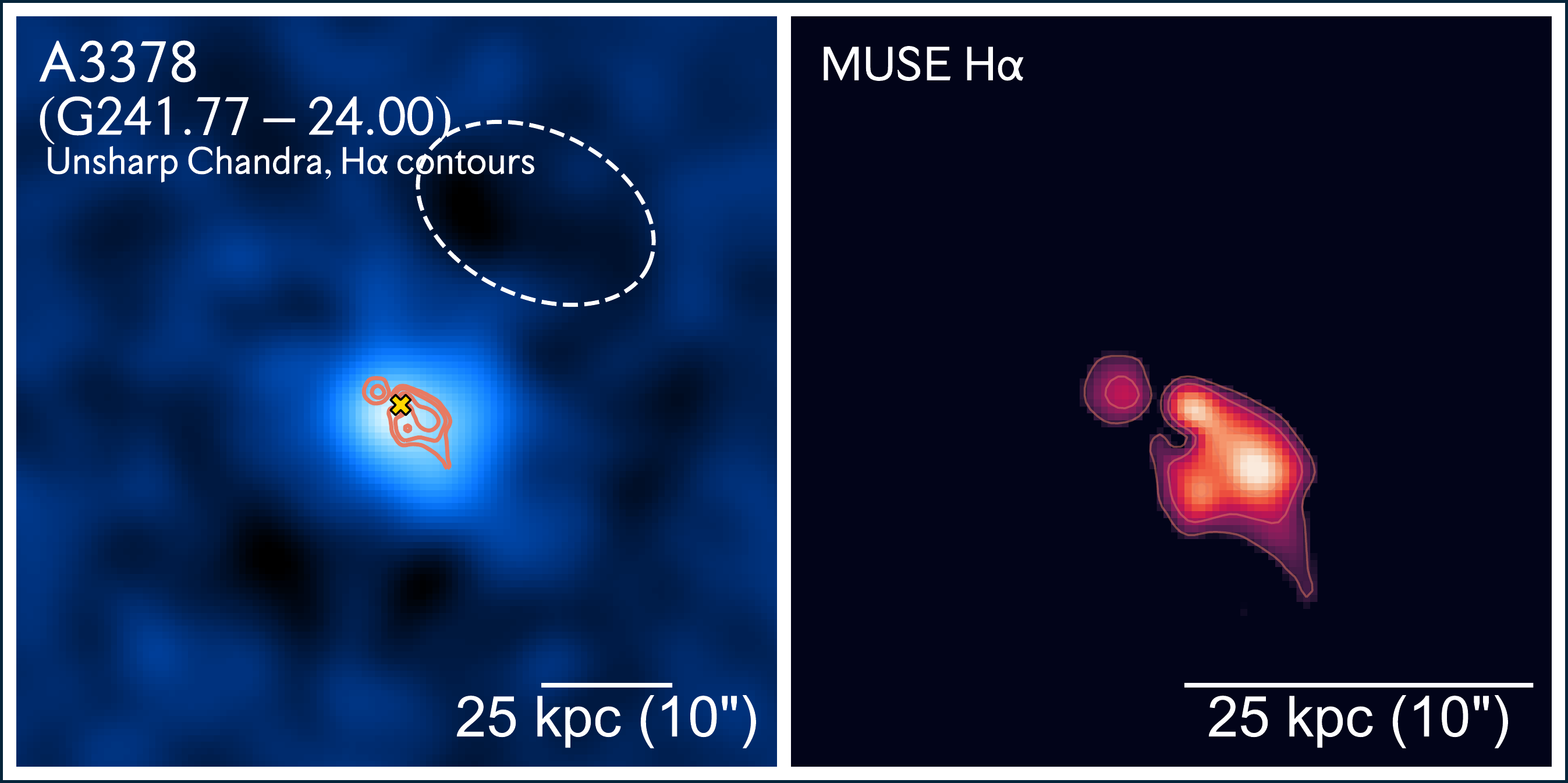}
    \includegraphics[width=0.495\textwidth]{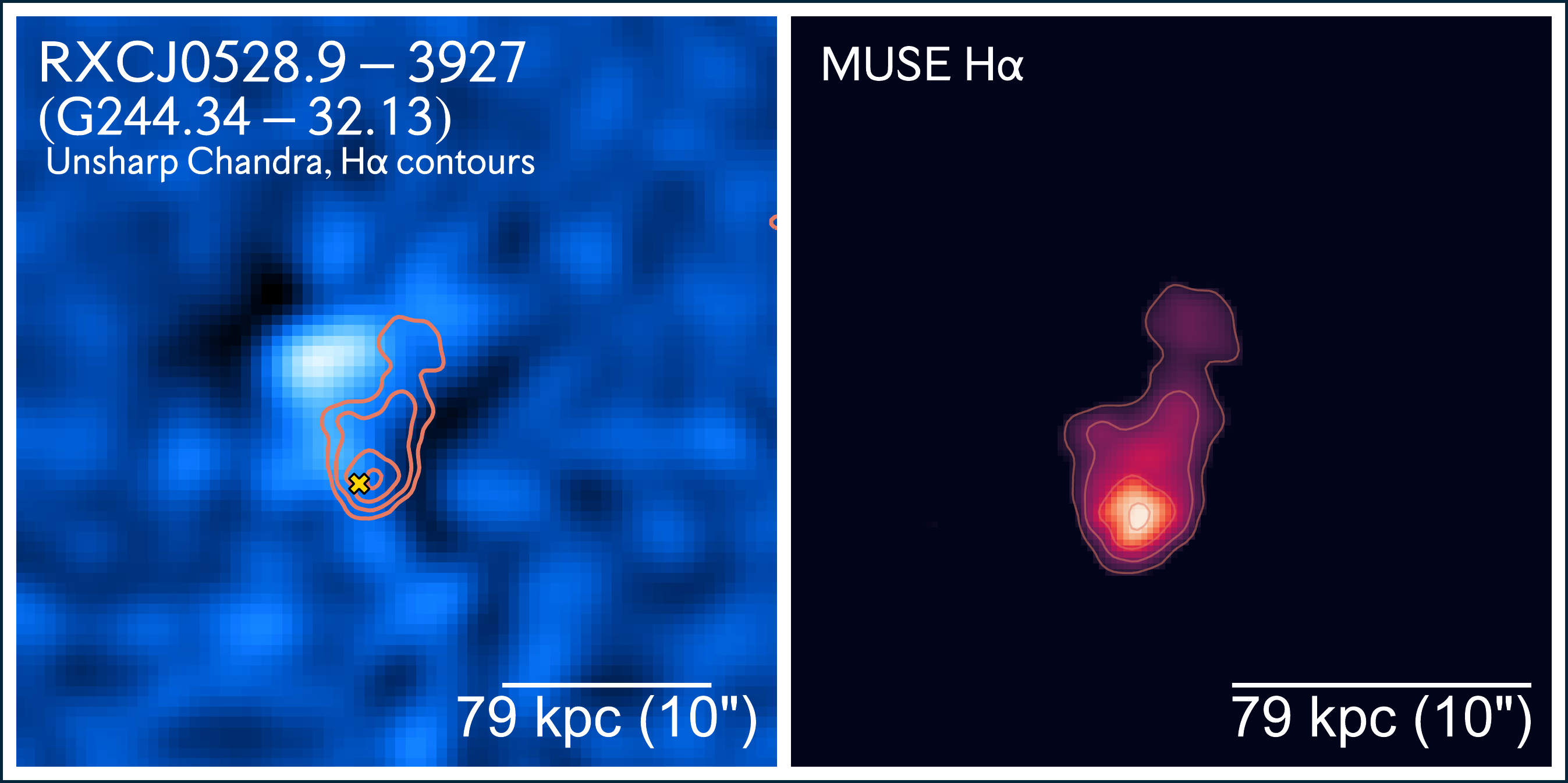}\\
    \vspace{0.2cm}
    \includegraphics[width=0.495\textwidth]{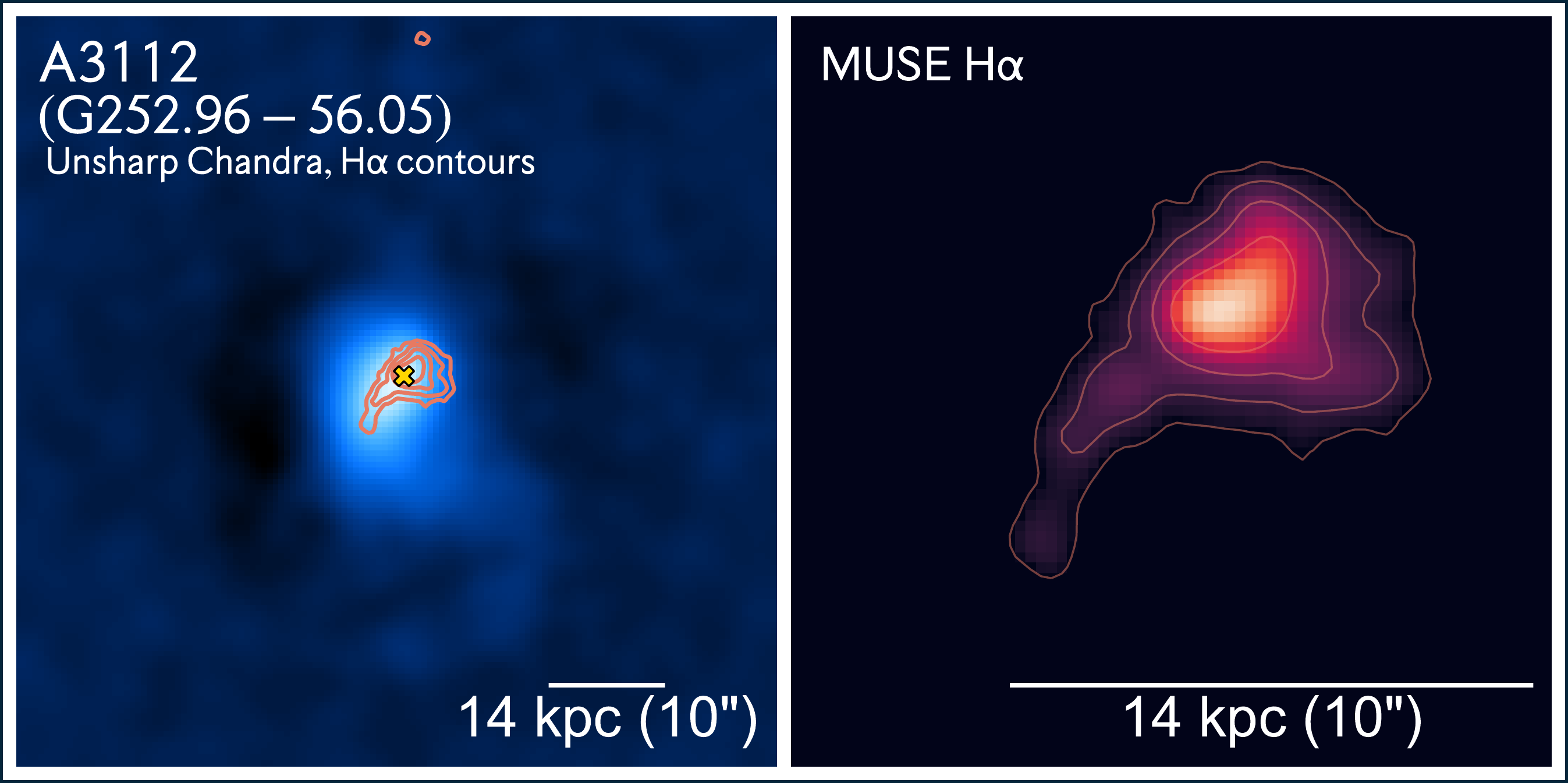}
    \includegraphics[width=0.495\textwidth]{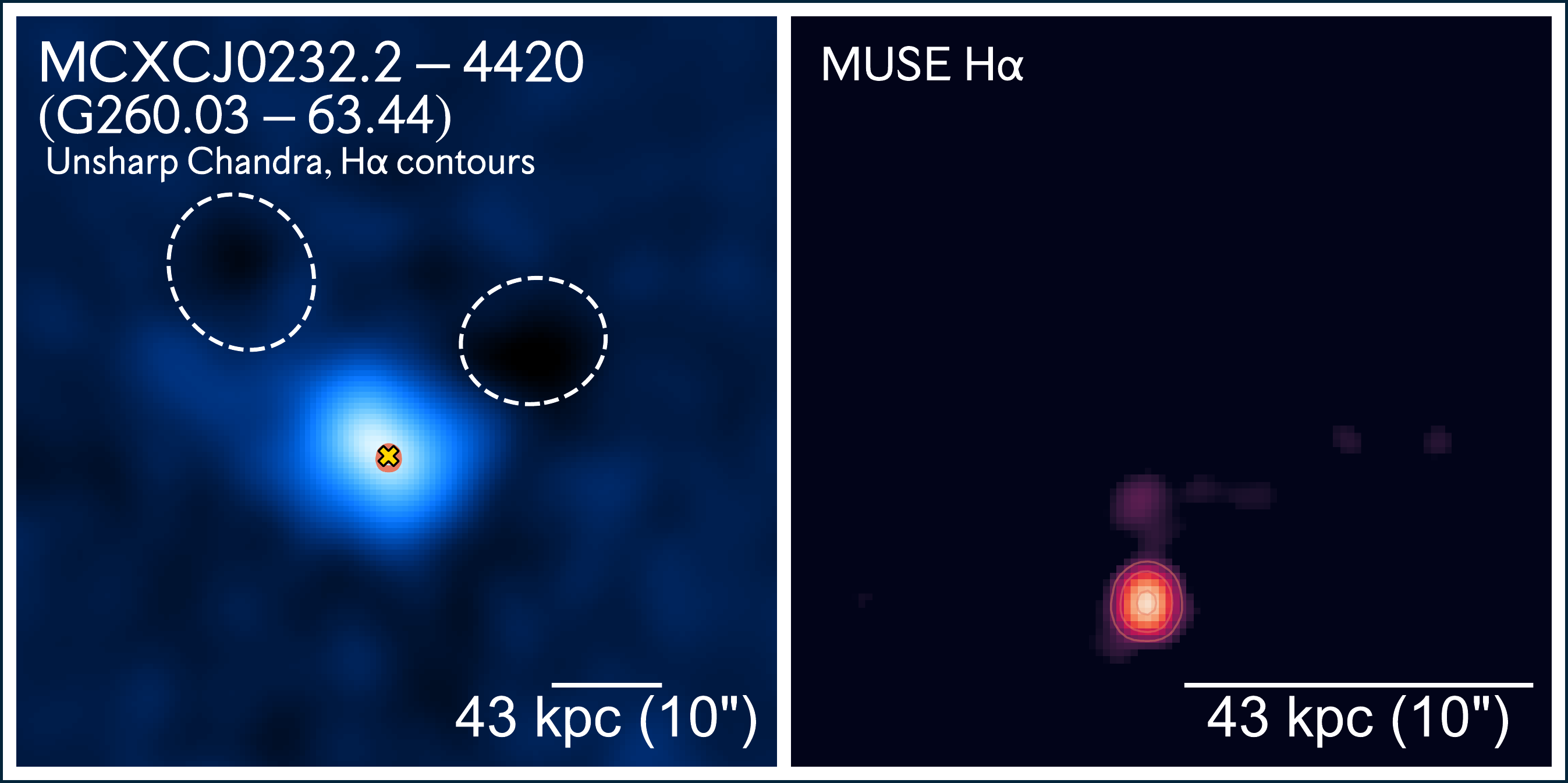}\\
    \vspace{0.2cm}
    \includegraphics[width=0.495\textwidth]{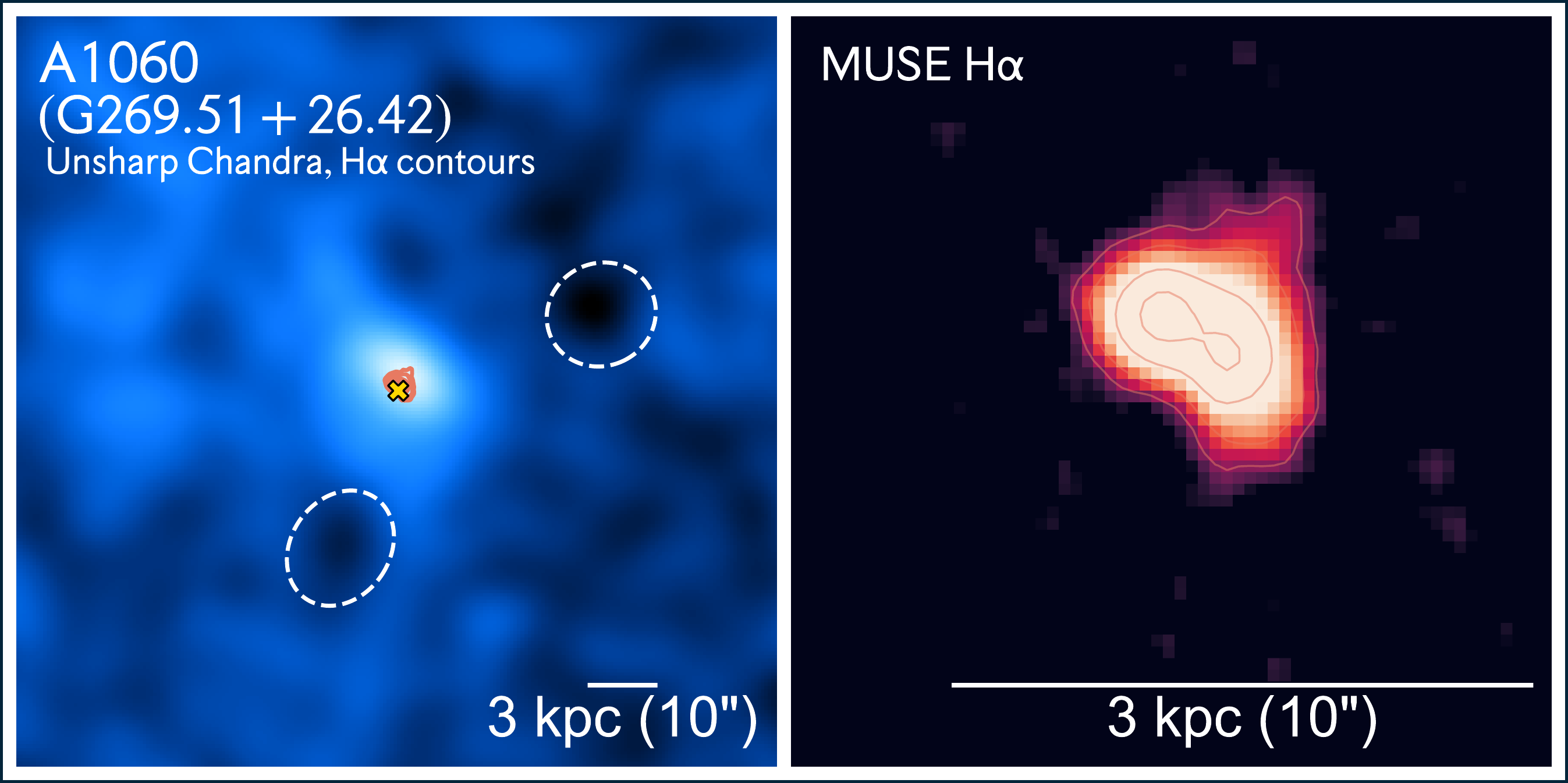}
    \includegraphics[width=0.495\textwidth]{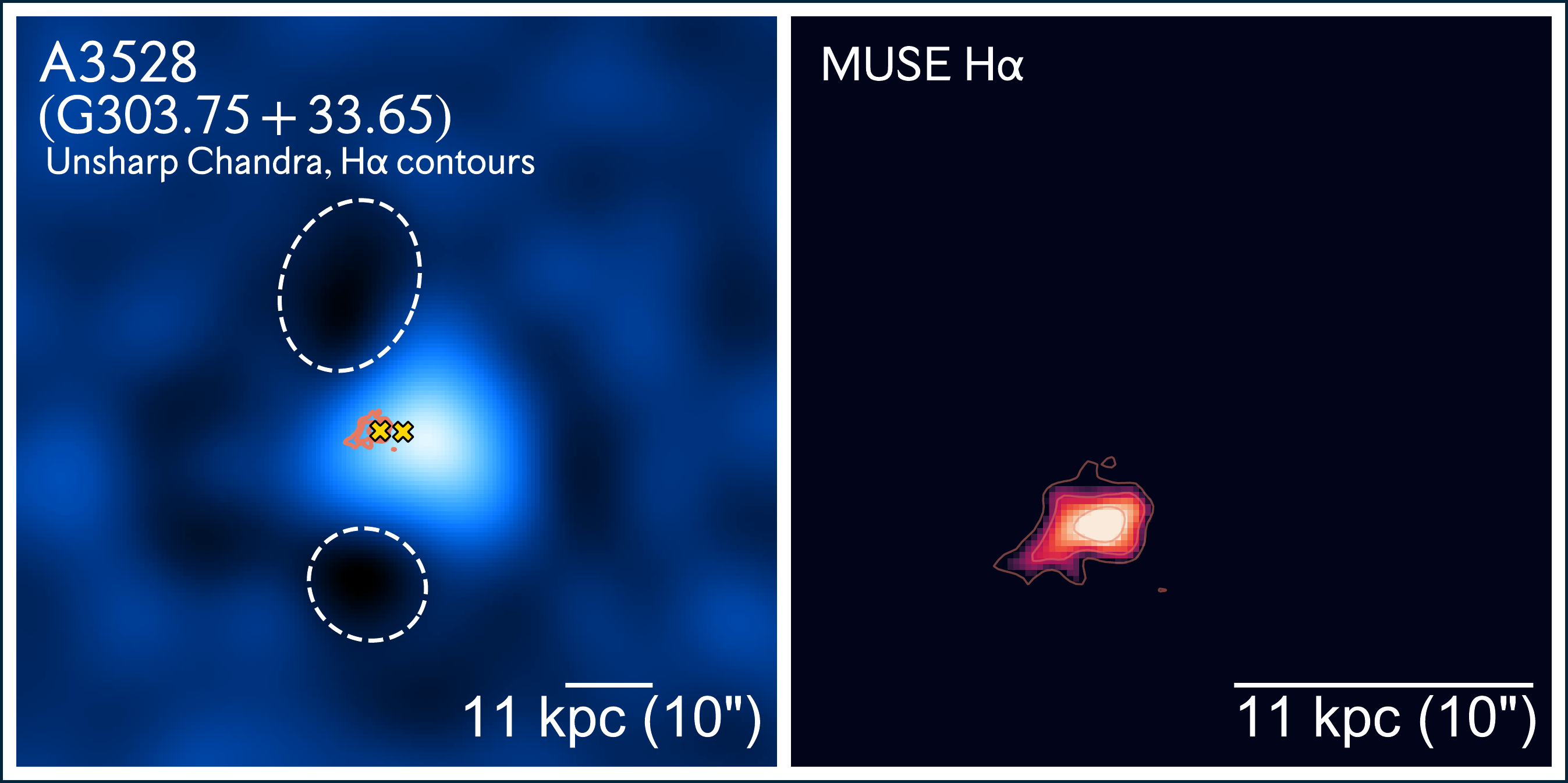}\\
    \vspace{0.2cm}
    \includegraphics[width=0.495\textwidth]{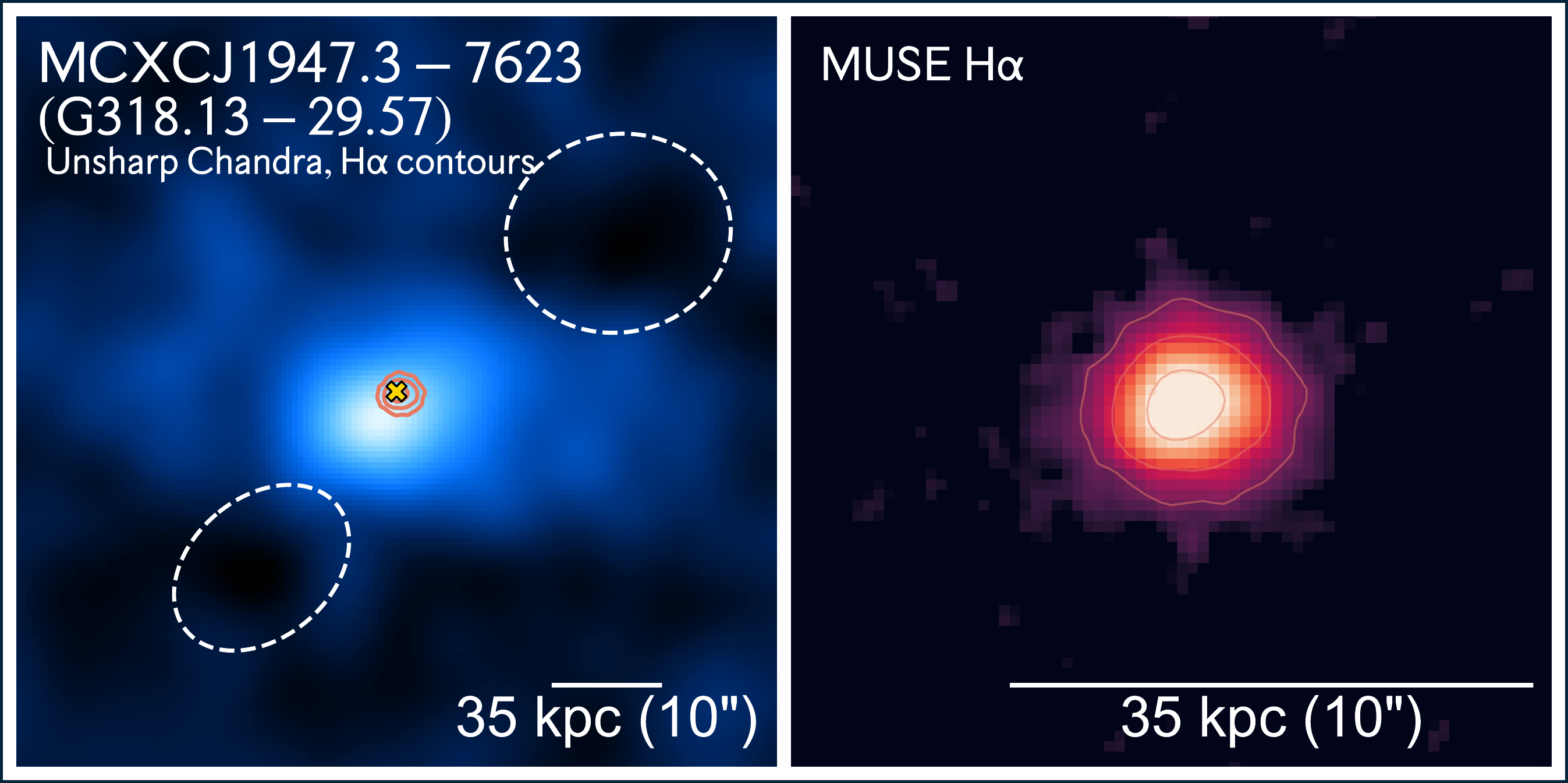}
    \caption{Sources where the association between the H$\alpha$ emitting gas and the cavities is not obvious. Left panel: \textit{Chandra} unsharp masked image, with ``potential'' cavities displayed with white dashed ellipses. The H$\alpha$ distribution is shown with red contours. Right panel: A zoom-in image of the H$\alpha$ emission  distribution from MUSE observations. The central BCG(s) is marked with a yellow arrow. The scale bar corresponds to 10$\arcsec$.}
    \label{fig:Halpha_maps2}
\end{figure*}

It is still an open question whether precipitation models necessarily require the presence of AGN bubbles for thermally unstable cooling to occur in the ICM, even if there has been a growing number of BCGs displaying a spatial association between cold gas and AGN bubbles \citep[e.g.,][]{salome04,salome06,salome08,olivares19,russell19,pasini21,Jimenez-Gallardo22}. The turbulence induced by the rise of AGN bubbles may also increase the local compression and facilitate cooling of the surrounding hot gas \citep[e.g.,][]{Gaspari_2011,gaspari12,prasad15,Prasad_2017,li15,li20,zhang22}.

To explore the uplift hypothesis, we first quantify how many H$\alpha$ emitting clusters have detected X-ray cavities in their ICM. According to Figure~\ref{fig:Cav_Halpha_relation} (right panel), all clusters (100\%, 12/12) with ``certain'' (C) cavities also display H$\alpha$ emission-line gas (red bars). In the same vein, 85\% (17/20) clusters that lack cavities (with available information about their nebular emission) also lack H$\alpha$ emission-line gas (blue bars). The correlation appears to be less strong for the clusters with ``potential'' (P) cavities, of which nine (9/14, 64\%) display H$\alpha$ emission-line gas (red bar), while the rest (5/14, 36\%) lack the presence of warm gas.

{Based on our limited sample of clusters that have information on the presence or absence of warm or cold gas, we found that 88\% (or 21 of 24) of clusters emitting H$\alpha$ and/or CO display cavities, which may suggest a potential relationship between AGN feedback and condensation of gas.}

There are three clusters that lack X-ray cavities, but harbor H$\alpha$ emission-line gas or cold molecular gas (see Figure \ref{fig:cluster_nocav_halpha}) --  RXJ1720.1+2638 (G049.20+30.86), A3112 (G252.96-56.05), RXCJ0528.9-3927 (G244.34-32.13). {We note that these clusters show central radio emission associated with the BCG. Moreover, as shown by \citet{hogan15}, the X-ray cavity power correlates with both extended and core radio emission. The latter suggests that those sources might experience AGN feedback (see Figure~\ref{fig:cluster_nocav_halpha} and Appendix~\ref{sec:app_Ha_filaments} for details for each source).}  

{To explore this point further, we look at the spatial association between the H$\alpha$ and X-ray cavities for systems that have archival MUSE observations.} The MUSE observations of the 18 Planck galaxy clusters reveal a variety of optical-emitting structures with different sizes and morphologies. Most of the sources (14) reveal an extended set of filaments -- A2204 (G021.09+33.25), A1795 (G033.78+77.16), A2390 (G073.96-27.82), A85 (G115.16-72.09), 2A0335+096 (G176.28-35.05), A478 (G182.44-28.29), A496 (G209.56-36.49), A3378 (G241.77-24.00), RXCJ0528.9-3927 (G244.34-32.13), A3112 (G252.96-56.05), A3017 (G256.45-65.61), A3444 (G266.84+25.07), A1644 (G304.89+45.45), and AS0780 (G340.95+35.11), while four systems display compact ionized gas distributions -- RXCJ0232.2-4420 (G260.03-63.44), A1060 (G269.51+26.42), A3528 (G303.75+33.65), and RXCJ1947.3-7623 (G318.13-29.57, unresolved). For a thorough description of the H$\alpha$ emission-line gas distribution and its association with the X-ray cavity arrangement, we refer the reader to the Appendix~\ref{sec:app_Ha_filaments}. From a visual inspection, we found that in 60\% (11/18) of the clusters, the optical filaments tend to be located beneath or around the cavity rims (see Figure \ref{fig:Halpha_maps}).

{We found that in 60\% (11/18) of the clusters, the optical filaments tend to be located behind or around the cavity rims (see Figure \ref{fig:Halpha_maps}). Nevertheless, in (7/18) of the clusters, the optical filaments appear unrelated to any detected cavity. }

\section{Limitations}\label{sec:limitations}

\subsection{Bias in Cavity Identification}
Cavities are found primarily in clusters with short central cooling times. Clusters with denser ICM may host clear cavities, since their cores have higher surface brightness which facilitates the detection of cavities (see also Figure~1 of \citealt{olivares22b} and \citealt{panagoulia14b}). In other words, clusters with denser ICM may host clear cavities since a denser ICM creates more contrast between the surrounding hot gas and the surface brightness depressions, making it easier to identify X-ray cavities. Intriguingly, a few clusters with high central surface brightness and short central cooling times lack X-ray cavities. Among those sources are A2029 (G006.47+50.54), AWM7 (G146.33-15.59), A3112 (G252.96-56.05), and A2142 (G044.22+48.68). These clusters reveal radio jets or a central radio source. A possibility is that jets might be pumping out their energy far from the central regions, and thus creates X-ray cavities far from the center, where the contrast is too small to detect X-ray depressions (see \citealt{enblin02,mcnamara07}). Another possibility could be that the timescale of the AGN and the lifetime of cavities are not synchronized, as a consequence of the intermittent nature of the AGN jets. Lastly, the location, orientation, and angular size of the cavities may interfere with our capability to detect cavities (see \citealt{enblin02, diehl08, bruggen09,olivares22b}). {Additionally, smaller cavities are more difficult to detect, and thus deeper \textit{Chandra} observations are required for these clusters to detect any X-ray depression in their hot ICM.}

{Another limitation of this work is the {non-trivial} effect of projection. As pointed out by \citet{bruggen09}, most of the detected cavities come from systems where the AGN axis is between $90\degr$ and $45\degr$ with respect to the line of sight and depending on the angle they can look like horseshoes (at $45\degr$) or more rounded. Similarly, as noted by \citet{enblin02}, the detectability of cavities decreases with distance from the center, and for a cavity moving in the plane of the sky, the contrast decreases slowly. Thus, cavities that lie off the plane of the sky have reduced contrast and therefore are harder to detect. In \citet{olivares22a}, we quantify this projection effect, by assuming a random distribution of the angle of the cavity relative to the plane of the sky, a typical cavity size of 10~kpc, and a typical $\beta$ profile for the ICM distribution (r$_{c}$=20, $\beta$=3/4). At an average projected distance of 30~kpc, 20\%--30\% of the cavities would have a contrast below our detection limit and would have been missed in our study.}

As shown by \citet{olivares22b} and \citet{diehl08}, the physical quantities that involve the cavity power, $P_{\rm cav}$, such as density, temperature, and pressure are affected by projection effects. These quantities are measured at the projected distance from the cavity to the center, rather than the true distance. As shown in Figure~\ref{fig:cav_asym}, we do not see a difference in the ratio of cavity distances between disturbed and relaxed clusters. Therefore, the calculation of gas pressure may be affected by projection equally for disturbed and relaxed clusters. The volume of the cavity may be less reliable for disturbed clusters than for relaxed clusters, as suggested by the difference in the ratio of the semi-major axis (see Figure~\ref{fig:cav_asym}).

\subsection{Effect of Data Quality}

{Another bias in identifying cavities is the inhomogeneity of the X-ray data quality across the sample. This is evidenced by the large variation in the number of counts in \textit{Chandra} observations for each source. Most cavities are detected within a radius of 40 kpc from the cluster center because they are fainter and more difficult to detect at larger distances. In addition, numerical simulations suggest that cavities fragment and disrupt after a few 100 Myr \citep{brighenti15}, which roughly corresponds to a distance of 40 -- 50 kpc, considering an expansion velocity for the cavity of 400 -- 500 km\,s$^{-1}$ \citep{Jimenez-Gallardo22}.}

{The central number counts, which corresponds to the total number of source X-ray photons detected inside an aperture of 40 kpc, greatly vary for these clusters, ranging from a few $\times$10$^{5}$ to 500, and decreasing with redshift (see also Figure~2 right panel of \citealt{olivares22b}). Clusters with ``certain'' and ``potential'' cavities have central number counts greater than 10$^{4}$ and between 10$^{4}$ and 10$^{3}$, respectively. Clusters without cavities do not show any dependence on central number counts.}

\subsection{Incompleteness of MUSE Observations}
{Another caveat of this work is the absence of complete MUSE IFU observations of the optical-emitting gas, specifically for the Planck-selected sample. The available archival MUSE observations are from either X-ray-selected samples or individual observations of BCGs. Therefore, future optical IFU observations for the missing optically-emitting clusters are necessary for a comprehensive investigation of the emission-line properties of SZ-selected cluster centers.}

{Also in the category of observational biases,} filaments could be more extended, as evidenced in the {Centaurus cluster}, where a weak optical [NII] {emission line region} is found outside the filamentary nebula \citep{hamer19}.

\section{Discussion}\label{sec:discussion}

\subsection{AGN Feedback in Diverse Environments}

{AGN feedback in relaxed clusters is barely keeping up with the replacement of radiation losses, as the feedbacks appear to be below the $L_{\rm cool}$ to $P_{\rm cav}$ one-to-one relation. On the other hand, the AGN feedback in disturbed and mixed clusters more than compensates for radiative losses, with much more energy than needed to compensate for the radiative losses of the ICM. That excess of energy could be going into lifting the ICM at large radii. That trend is similar to what is found in AGN feedback in groups and individual galaxies \citep{eckert21}. {Disturbed clusters have a higher cavity power, $P_{\rm cav}$, for a given cooling luminosity value, $L_{\rm cool}$, compared to relaxed clusters, which could be due to many factors. First, for a given $L_{\rm cool}$, disturbed clusters tend to have larger volumes than relaxed clusters, and yet the ambient pressures are comparable. Disturbed atmospheres could perturb bubble shapes, causing volumes to be calculated larger than they actually are. Second, central AGNs in disturbed clusters may produce stronger outbursts \citep[e.g.,][]{MSolAlonso07}. In addition, the buoyancy times to determine cavity lifetimes may be more uncertain for disturbed clusters, since bubble velocities would have components not related to pure buoyancy (and jet lengths could be shorter/longer in relaxed vs. disturbed systems). The true explanation for the higher outburst power in disturbed systems, compared to relaxed systems, remains unknown however, and future simulations and observations of larger samples may help us to clarify this puzzle.}



\begin{figure}
    \centering
    \includegraphics[width=0.45\textwidth]{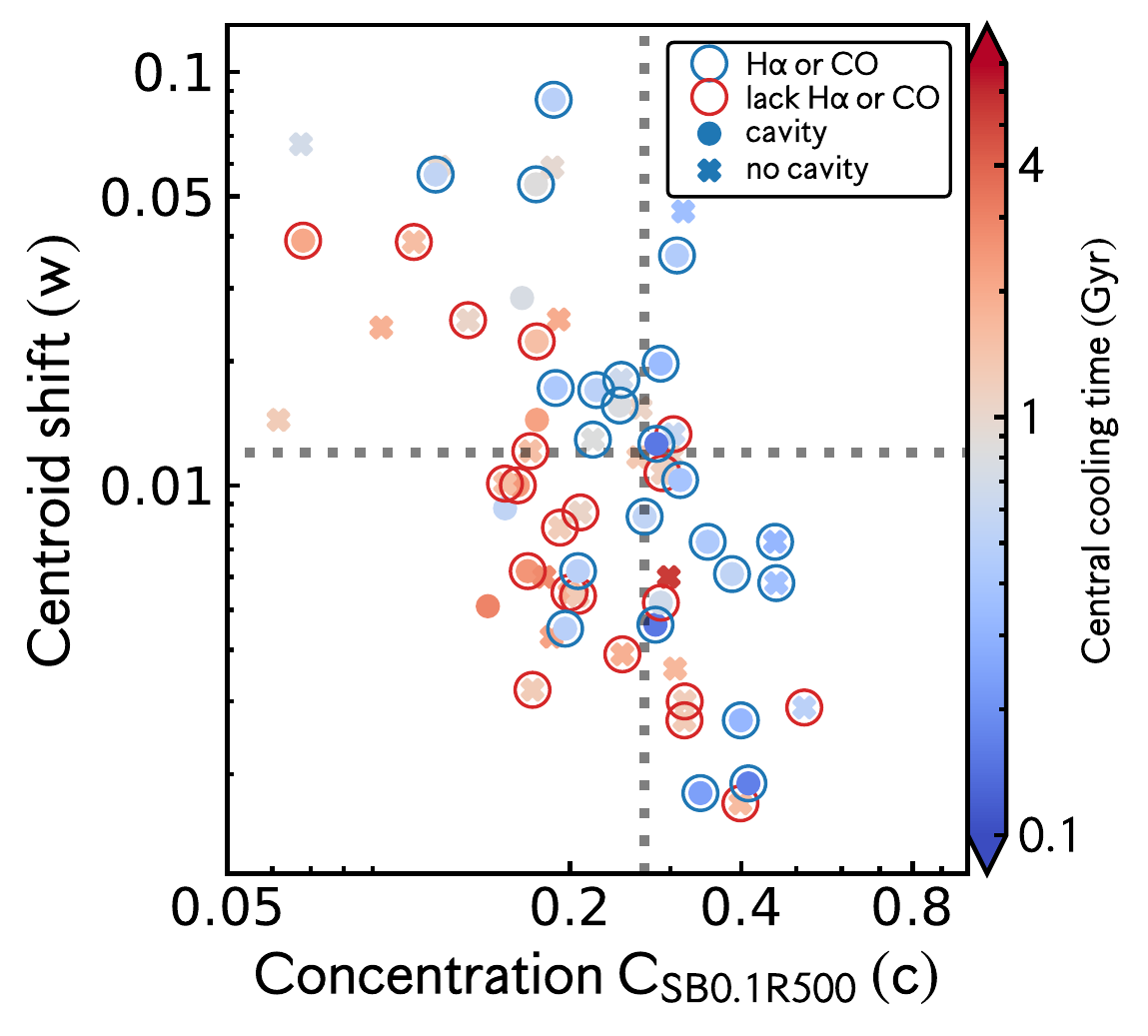}
    \caption{Centroid shift ($w$) versus concentration ($c$) in the 0.1--1.0~$R_{\rm 500}$ range (left panel) and 40--400 kpc range (right panel), color-coded by central cooling time at 10~kpc. Clusters with and without cavities are shown with circles and crosses, respectively. Clusters with warm ionized or cold molecular gas are marked with outer blue circles, while systems lacking colder gas are marked with red circles.}
    \label{fig:Lcool_Pcav_cold_gas}
\end{figure}

{The above statement implies that the feedback efficiency, represented by the ratio $P_{\rm cav}/L_{\rm cool}$, is higher ($P_{\rm cav}/L_{\rm cool}>$1) in disturbed clusters than in relaxed clusters ($P_{\rm cav}/L_{\rm cool}<$1). The cavities inject enough energy to overheat the cores, potentially depleting the central regions of their gas content (see also \citealt{eckert21} for a similar behavior in groups). Then we would expect a larger fraction of relaxed clusters with warm and cold molecular gas, compared to disturbed clusters.}

{To visualize the above trend along the entire sample}, including clusters without cavities, in Figure~\ref{fig:Lcool_Pcav_cold_gas}, we plot the centroid shift (w) as a function of concentration (c) color-coded by the central cooling time. Additionally, each cluster is marked with blue or red circle to show the presence or absence of warm ionized gas or cold molecular gas, respectively. As shown in Figure~\ref{fig:Lcool_Pcav_cold_gas}, we see a {slightly larger fraction} of relaxed clusters, {55\% compared to 45\%}, with H$\alpha$ or CO emitting gas compared to disturbed clusters. Short central cooling time, $<$ 1Gyr, is the main condition that dictates the presence of cold and warm gas in clusters as found by previous works \citep[e.g.,][]{pulido18,mcnamara16} and as discussed in Sec~\ref{sec:cold_gas}.


{Finally, we found that the central AGN inflates a new cavity pair every 5 to 30~Myr, depending on the clusters. Such a feedback frequency is also supported by CCA simulations, showing a flicker-noise power spectrum in the AGN accretion/power rates (\citealt{gaspari17}). Contrary to what was found for the central group galaxy, NGC~5813 \citep{randall15}, multiple pairs of cavities in this cluster sample are placed in different directions around the central BCG, due to either reorienting jets or ICM weather. This allows AGN jets to spread the energy nearly isotropically \citep{babul13}. However, symmetric bubbles are also capable of spreading the energy, as they rise in the atmosphere \citep{Zhang18}.}

\subsection{Formation of the multiphase filaments}
\subsubsection{AGN feedback and uplift}
\newcommand{\mylengthh}{0.49}
\begin{figure}[ht]
\centering
    \includegraphics[width=\mylengthh\textwidth]{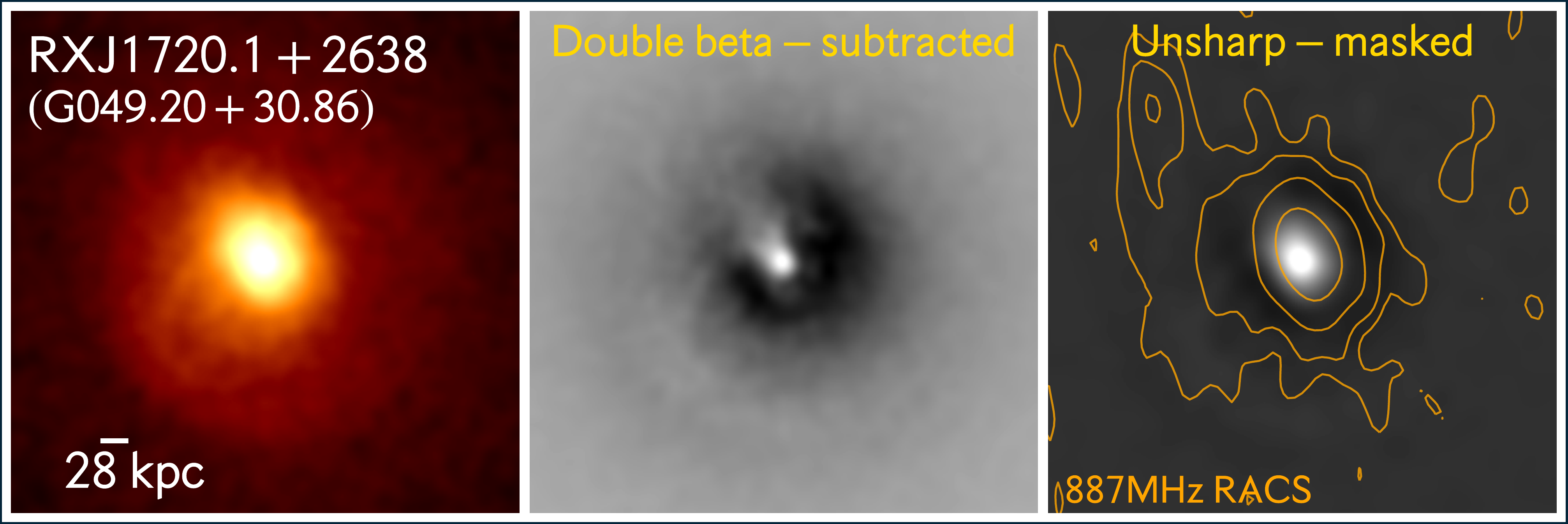}\\
    \vspace{0.1cm}
    \includegraphics[width=\mylengthh\textwidth]{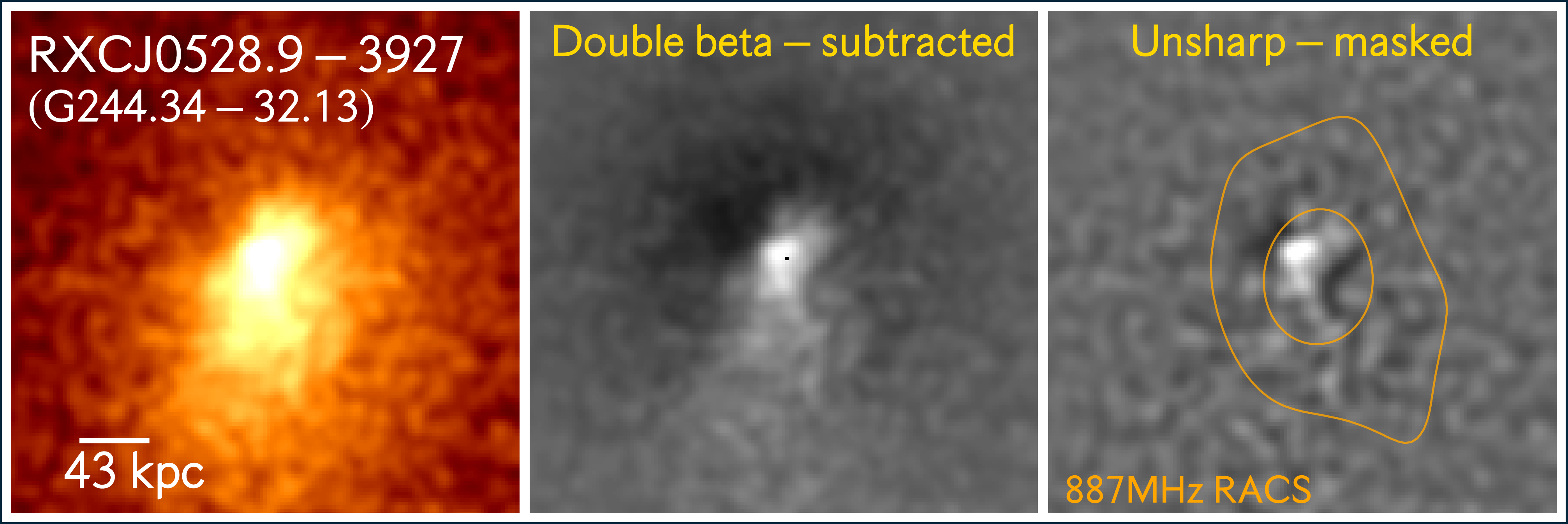}\\
     \vspace{0.1cm}
    \includegraphics[width=\mylengthh\textwidth]{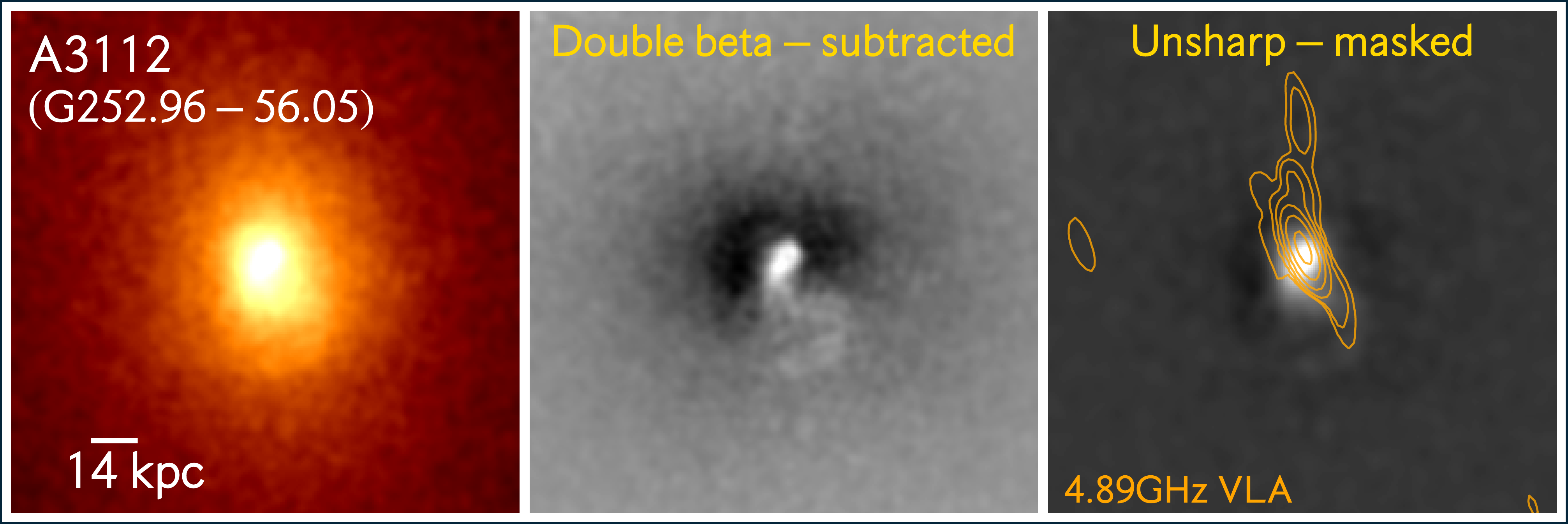}
    \caption{\textit{Chandra} X-ray images of the H$\alpha$ or CO line emitting clusters without X-ray cavities, from top to bottom: RXJ1720.1+2638 (G049.20+30.86), RXCJ0528.9-3927 (G244.34-32.13), and A3112 (G252.96-56.05). Left: smoothed 0.5-2.0~keV \textit{Chandra} image. Middle: double-$\beta$ subtracted X-ray image. Right: unsharp-masked X-ray image. Radio emission is shown in the unsharp-masked image with yellow contours when available.
     {RXJ1720.1+2638 (G049.20+30.86) also displays radio emission, and as recently shown by \citet{Perrott23} is consistent with sub-kpc radio jets from high-resolution e-MERLIN observations (see also \citealt{giacintucci14}).}
     }\label{fig:cluster_nocav_halpha}
\end{figure}

{AGN may induce cooling through ``stimulated feedback'', in which low-entropy warm gas is uplifted by the buoyantly rising radio bubbles \citep[e.g.,][]{mcnamara06,li14,brighenti15,qiu20}, promoting in-situ cooling to form cold molecular gas. Recent observations have shown a morphological association of the cavities with cold molecular filaments in a few clusters \citep{salome06,olivares19,russell19,qiu20}. 
However, it is still debated whether the gas condensation needs the presence of bubbles (cavities) for the cooling to happen, as AGN feedback in dynamically disturbed clusters has remained understudied. Alternative mechanisms, such as sloshing and mergers, might be able to move the hot gas to altitudes where it becomes thermally unstable simulating the uplifting mechanism. In particular, \citet{Gaspari_2018} proposed that the ICM condenses through turbulent non-linear instabilities (triggered mainly by AGN feedback but also mergers); this is expected when the ratio of the cooling time to the eddy-turnover timescale approaches $t_{\rm cool}/t_{\rm eddy}\sim 1$ \citep{Gaspari_2018}.}

{Feedback from the AGN may significantly contribute to the formation of gas condensations, as out of the 24 clusters emitting H$\alpha$ and/or CO, 88\% (21/24) display X-ray cavities, and the other 3/24 show only radio jets (without X-ray cavities). Among the H$\alpha$ and/or CO emitting clusters, 50\% are in relaxed clusters, while the other 50\% are in mixed or disturbed clusters.}

{By looking at the spatial correlation between the X-ray cavities and H$\alpha$, we found that in 60\% 11/18 of the clusters, the optical filaments tend to be located beneath the bubbles or around the cavity rims (see Figure \ref{fig:Halpha_maps}, as found in several other cooling flow clusters with joint ALMA and deep \textit{Chandra} observations. Nevertheless, in 7/18 of the clusters, the optical filaments appear unrelated to any detected cavity, or the correlation is less obvious (see Figure \ref{fig:Halpha_maps2}).}

{The poor correlation between cavities and H$\alpha$ filaments in the remaining clusters is expected from multiple causes. Among them are the different timescales for the presence of H$\alpha$-emitting gas and the dissipation of the AGN bubbles. Radio jets inflate cavities, triggering a turbulent cascade that induces gas condensation. The cavities will eventually expand and even dissipate. After the multiphase gas has condensed, the dense clouds will rain and accumulate toward the meso region {(inner kiloparsec region)}. Some of these clouds will then experience recurrent inelastic collisions, thus canceling angular momentum and boosting the CCA feeding  \citep{gaspari17}}. {A second possibility comes from observational biases related to our ability to detect X-ray cavities (see Section~\ref{sec:limitations}).}

{Based on the available observations, particularly the information presented in Figures~\ref{fig:Halpha_maps} and ~\ref{fig:Halpha_maps2}, we can infer that H$\alpha$ filaments tend to be located inside the radius at which cavities are observed (see also~\citealt{Calzadilla22}). Therefore, identifying and positioning X-ray cavities provides insight into the importance of AGN feedback on the formation of filaments, whether they form in-situ or through a turbulent cascade.}

\subsubsection{Sloshing, mergers, and ram pressure stripping}


In addition to the ``stimulated feedback'' or precipitation model described above, alternatively, the condensation of the cold gas can be triggered by sloshing motions or ongoing mergers through turbulence-driven precipitation. It has been pointed out that sloshing could produce enough heating to balance the radiative cooling losses within the central region \citep{zuhone10} and generate moderate levels of turbulence ($\sim$50 -- 200 km~s$^{-1}$) in the cool-core region and inside the sloshing front \citep{zuhone13}. Such a subsonic level of turbulence is still sufficient to induce top-down multiphase condensation and a CCA rain \citep{gaspari17,Gaspari_2018}, with filaments condensing out of the turbulent eddies, even without major AGN bubbles. Sloshing may also trigger condensation by emulating uplift, as galaxy interactions and mergers can move the peak of the ICM from the center of the potential well for several Gyrs, providing enough time for this gas to cool \citep{Osullivan21}. {Additionally, for subsonic sloshing to play a role in the creation of multiphase filaments, it would need to be ``re-energized'' within times shorter than the cooling time, $\sim$1~Gyr, and have dissipation times also shorter than the cooling time. As shown in \citet{Su2017b}, the sloshing timescales could be 10 times shorter than the cooling time over the entire cluster center, making it a promising mechanism for the formation of filaments.}

Evidence of turbulent-driven precipitation has been found at the core of the dynamically disturbed CHiPS1911+4455 and SpARCS1049+56 clusters, which both appear to have recently undergone a major merger, showing cooling gas, but lacking AGN feedback (\citealt{somboonpanyakul21} and \citealt{HlavacekLarrondo20}, respectively). For the SpARCS1049+56 cluster, a ram pressure origin of the cold gas cannot be discarded \citep{Castignani20_sparc}.

Sloshing motions may yield either a specific spiral pattern, often observed in X-ray images (see \citealt{markevitch07} for a review), or an offset between the peak of the ICM and the BCG position \citep[e.g.,][]{hamer16}. The latter indicates relative motions between the cooling ICM and the BCG. Evidence of sloshing is found in a few H$\alpha$ emitting clusters: A3444 (G266.84+25.07), RXJ1720.1+2638 (G049.20+30.86), 2A0335+096 (G176.28-35.05), A1644 (G304.89+45.45), A85 (G115.16-72.09), A2390 (G073.96-27.82), A1795 (G033.78+77.16), A3112 (G252.96-56.05) (see appendix~\ref{sec:app_Ha_filaments} for details on each source). Lastly, A1795 (G033.78+77.16), a sloshing source, shows a very long $\sim$50~kpc set of filaments extending from the BCG, that were possibly produced via gravitational focusing as the BCG passed through the ICM (see \citealt{fabian01} and \citealt{markevitch01}), not only shifting the filaments, but potentially also the smaller cavity.


A less likely scenario is the formation of filaments through galaxy mergers or interactions, via ram pressure stripping. {Ram pressure stripping might account for} some filaments, however, most of the line emitting gas would be stripped from an infalling galaxy before it gets to the center of clusters. 

At least four H$\alpha$ emitting clusters in this sample contain a companion galaxy within a projected radius of 10~kpc, suggesting that a galaxy merger or interaction is taking place. {One possible explanation for the presence of off-center cold/ionized gas in these merger systems is the existence of an undisrupted cool-core (see \citealt{Schellenberger23}).} The A3017 (G256.45-65.71) BCG is interacting with a neighbor galaxy located 7~kpc (in projection), which moves with a velocity of 330~km~s$^{-1}$ with respect to the main BCG. Both systems have optical line emitting gas. In the case of A3528 (G303.75+33.65), the companion is located at 2.7~kpc west of the BCG and is moving with a systemic velocity of $>$500~km~s$^{-1}$. 2A0335+096 (G176.28-35.05) also has two main galaxies. The secondary galaxy is situated 5~kpc in projection to the BCG, with a velocity offset of 210~km~s$^{-1}$. Again both galaxies have cooling gas.

{As found in sloshing and interacting cooling clusters, it is difficult to disentangle the role of cavities and sloshing/mergers in forming filaments, as both mechanisms may be taking place.} In sloshing systems, we expect a spiral-like morphology for the filaments. However, in some systems, the filaments also appear to be connected to the AGN bubbles. Therefore, it is also possible that filaments were formed through either uplifting or thermal instabilities (TI) cooling, triggered by AGN feedback, and subsequently, both bubbles and filaments evolved following the bulk motions generated by sloshing. However, when there are no cavities present but are radio jets, the warm ionized gas orientation appears to be related to the radio jets, showing that this might be the essential mechanism for the formation of filaments. In summary, sloshing may have contributed to the formation of filaments in a few specific clusters, although in some cases, the dominant mechanism is not clear. Therefore we need further evidence of the sloshing/merger mechanism to produce filaments. Future numerical simulations of sloshing cooling-flow clusters also are required to unveil the contribution of the sloshing mechanism on filament formation.

\section{Conclusions}\label{sec:conclusions}
In this paper, we have investigated the mechanical AGN feedback mechanism in central cluster galaxies using \textit{Chandra} X-ray observations of Planck selected clusters to study the effect of the ICM ``weather'' on the X-ray cavities, as well as explore the connection with the ionized gas traced with MUSE observations. Our results are summarized as follows:

    
\begin{itemize}
   \item We find that clusters with disturbed hot atmospheres have more asymmetric cavities than relaxed clusters, likely due to the influence of ICM weather, which may perturb the overall morphology and distribution of the cavities. The absence of a correlation between cavity numbers and cluster dynamical state indicates that the bubbles are not readily disrupted by ICM weather, and, therefore, multiple cavities are more likely to be a consequence of different generations of AGN outbursts.
    
   \item We find that sloshing and mergers induce a scatter on the $L_{\rm cool} - P_{\rm cav}$ relation, as dynamically disturbed clusters tend to have systematically larger cavity power, $P_{\rm cav}$, for a given cooling luminosity.
        
   \item Based on the clusters with multiple cavities, we find that AGN likely inflate a new pair of bubbles every 5 to 30~Myr, consistent with previous observational findings and a CCA-driven flickering variability.
    
   \item Our analysis suggests that AGN feedback plays an important role in the precipitation of the ionized warm or cold molecular gas by increasing the ICM turbulence or by uplifting gas, as all clusters with clearly (``certain'') detected cavities also contain either warm ionized and/or cold molecular gas. Along the same lines, 85\% (17 of 20) of the clusters that lack cavities also lack warm or cold gas.
    
    \item We analyzed the archival MUSE observations of 18 galaxy clusters. The optical line emitting gas is preferentially located beneath or around the bubbles for most sources (11/18), indicating that AGN feedback plays an important role in forming the warm filaments by uplifting gas which may adiabatically cool, but may also by enhancing turbulence and mixing as revealed by CCA simulations. We discuss several possible reasons for the lack of association of bubbles with the distribution of optical-emitting gas in a few cases, including a different timescale for the dissipation of cavities and optical filaments, or alternative turbulence-driven mechanisms, such as sloshing/mergers.


\end{itemize}

    Future observations of the optical filaments of the remaining disturbed Planck clusters will help resolve the dominant mechanism for the origin of the cold gas.
    
\begin{acknowledgments}
{The authors would like to thank the  referee who provided useful and detailed comments on the manuscript.} This research has made use of software provided by the \textit{Chandra} X-ray Center (CXC) in the application packages CIAO. V.O. and Y.S. were supported by NSF grant 2107711, Chandra X-ray Observatory grant GO1-22126X, and NASA grant 80NSSC21K0714.
M.G. acknowledges partial support by HST GO-15890.020/023-A and the \textit{BlackHoleWeather} program. W.F., C.J., and P.N. acknowledge support from the Smithsonian Institution and the Chandra High Resolution Camera Project through NASA contract NAS8-03060. WF also acknowledges support from NASA Grants 80NSSC19K0116, GO1-22132X, and GO9-20109X. P.S. acknowledges support by the Agence Nationale De La Recherche (ANR) grant LYRICS (ANR-16-CE31-0011).

Based on observations collected at the European Organisation for Astronomical Research in the Southern Hemisphere under ESO programme(s): 60.A-9345, 60.A-9100, 094.A-0859, 094.A-0115, 094.A-0141, 094.B-0711, 095.B-0127, 095.A-0159, 097.B-0776, 098.A-0502, 0101.B-0808, 0103.A-0777, and 0104.A-0801.
\end{acknowledgments}

\appendix
\restartappendixnumbering
\section{Chandra images}
\newcommand{\mylength}{0.9}
\begin{figure*}
\centering
    \includegraphics[width=\mylength\textwidth]{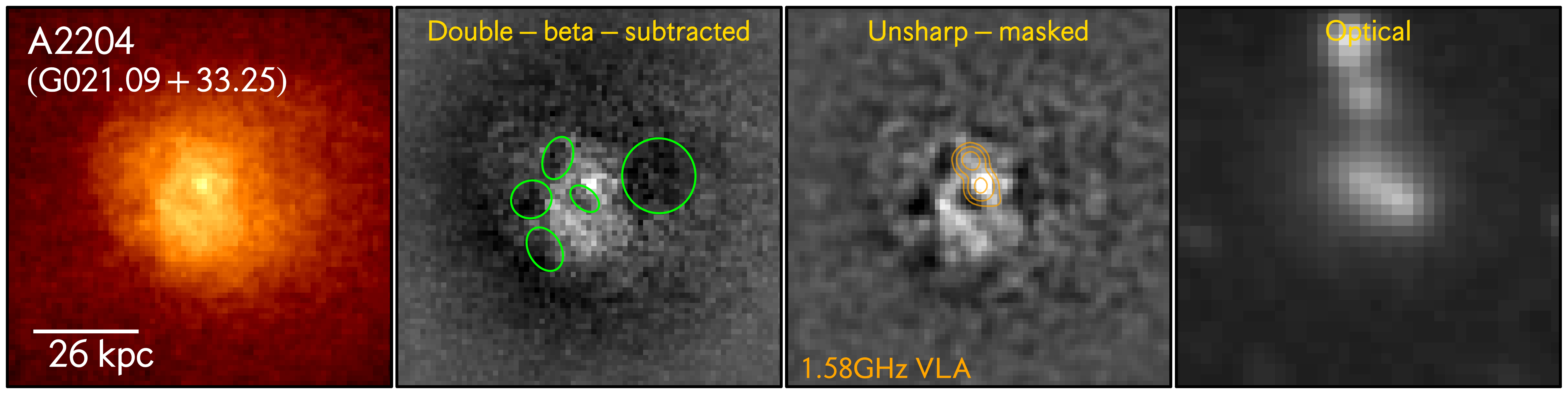}
        \includegraphics[width=\mylength\textwidth]{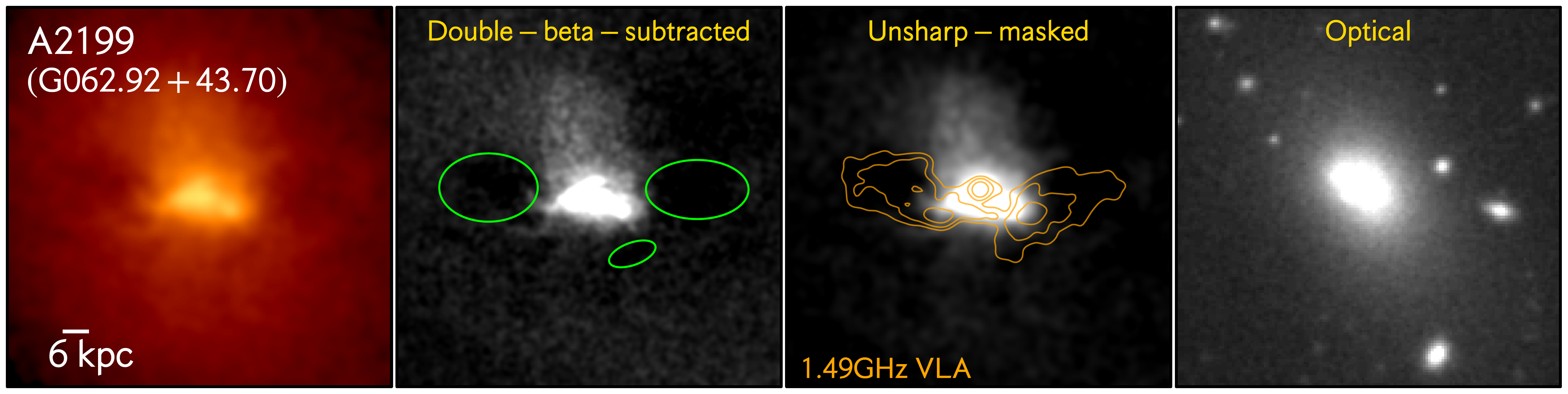}
        \includegraphics[width=\mylength\textwidth]{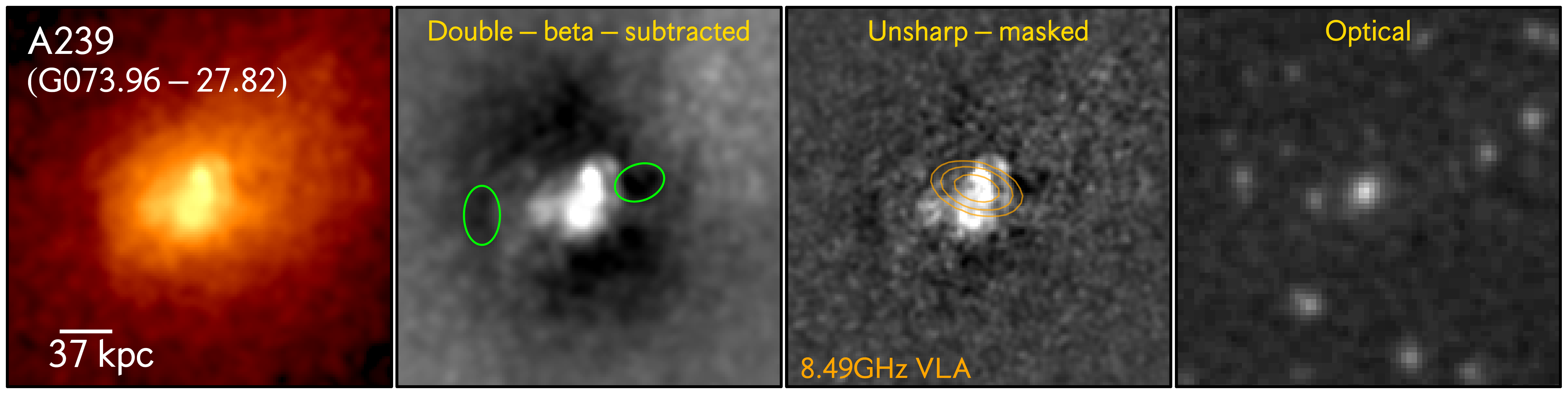}
        \includegraphics[width=\mylength\textwidth]{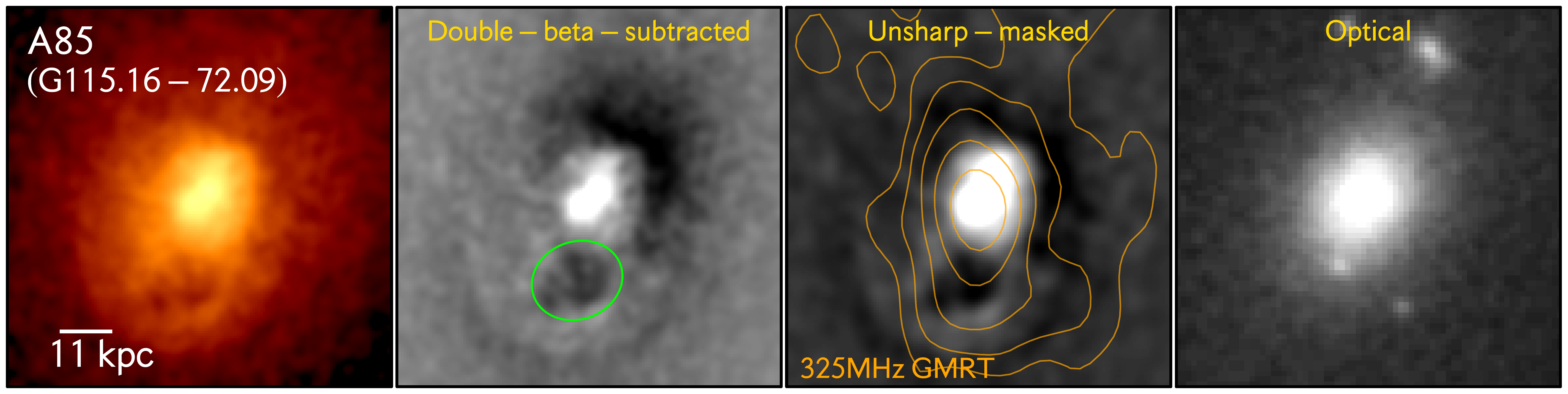}
    \caption{Images of the CC Planck clusters with ``certain'' (C) X-ray cavities. Left panels: 0.5-2.0~keV smoothed X-ray image. Left middle panels: 0.5-2.0~keV double $\beta$ model subtracted image. Right middle panels: 0.5-2.0~keV unsharp-masked image. Right panels: optical image. The cavities are highlighted with green ellipses. All images show the same region, and the scale bar corresponds to 10\arcsec. \label{fig:chandra_images_C}}
\end{figure*}

\begin{figure*}
    \figurenum{A.1}
    \centering
        \includegraphics[width=\mylength\textwidth]{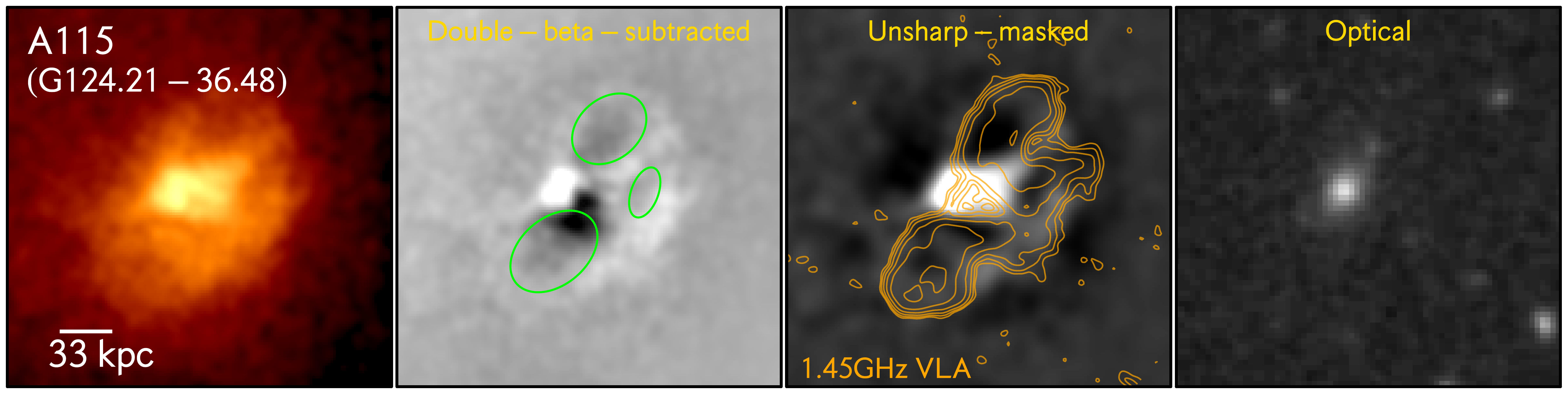}
        \includegraphics[width=\mylength\textwidth]{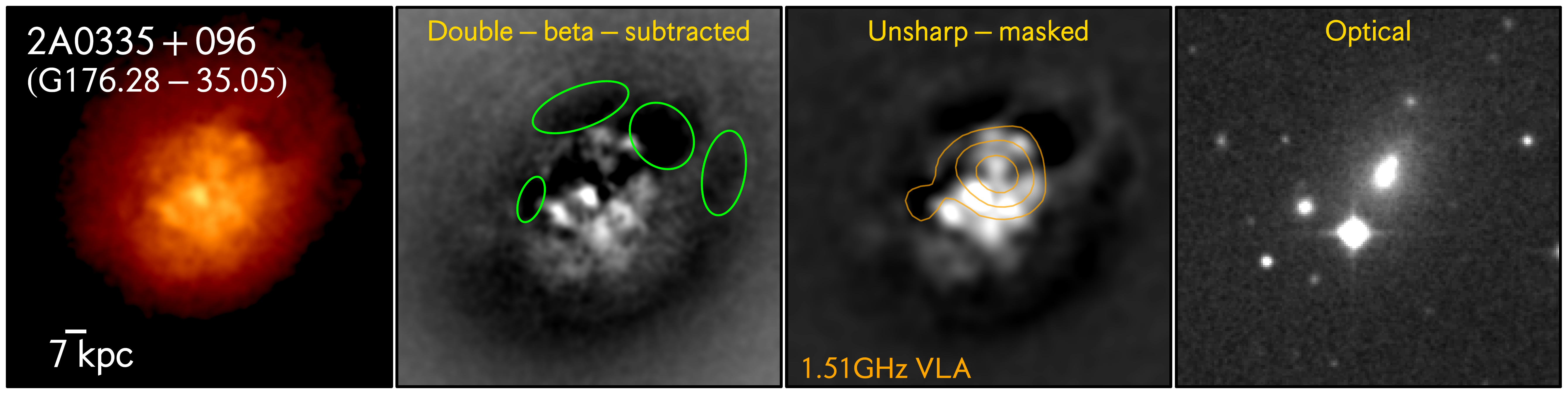}
        \includegraphics[width=\mylength\textwidth]{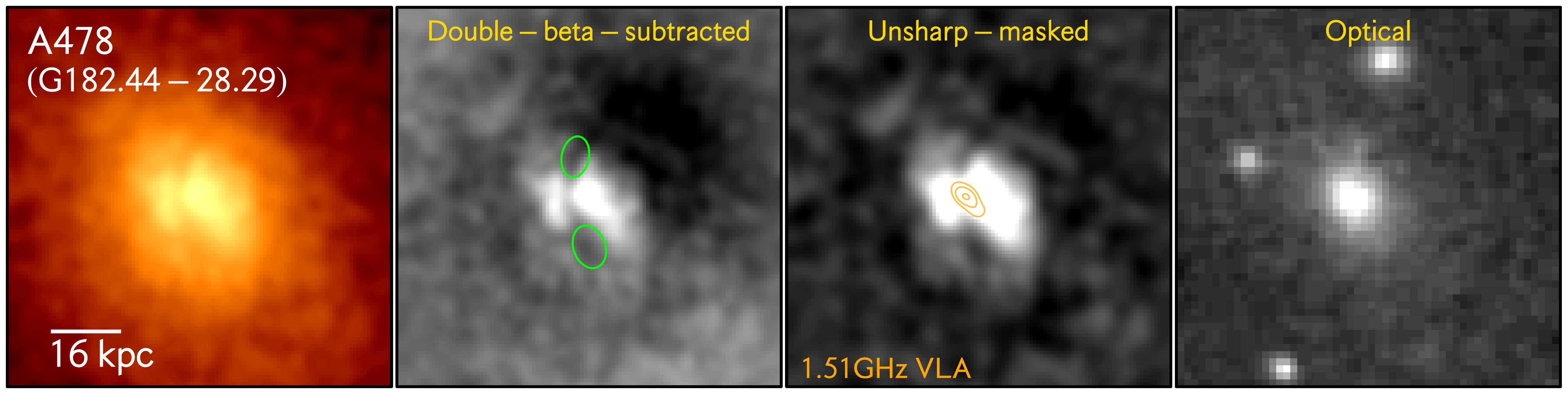}
        \includegraphics[width=\mylength\textwidth]{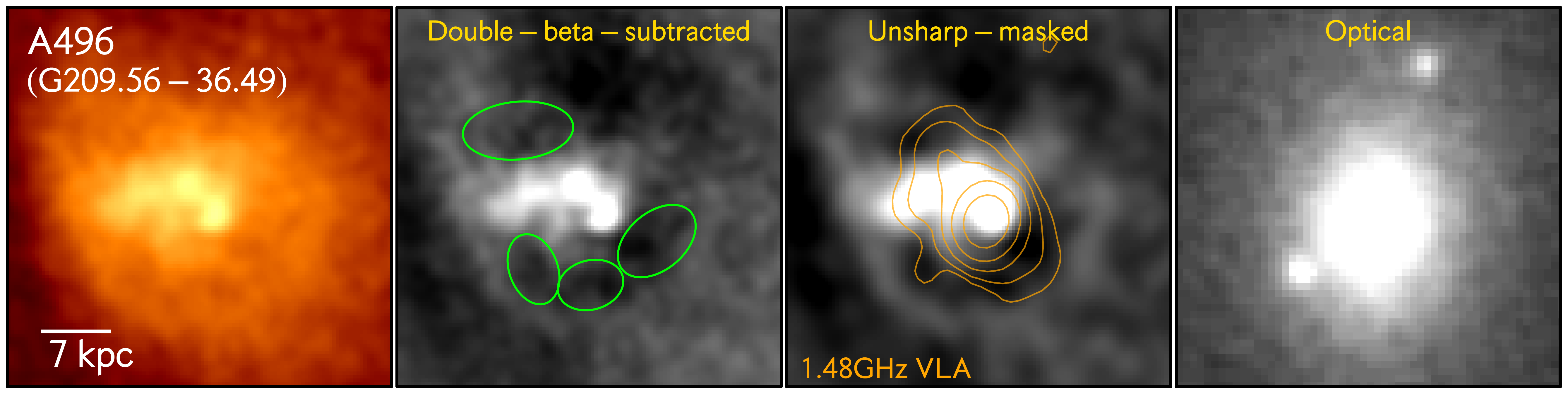}
        \includegraphics[width=\mylength\textwidth]{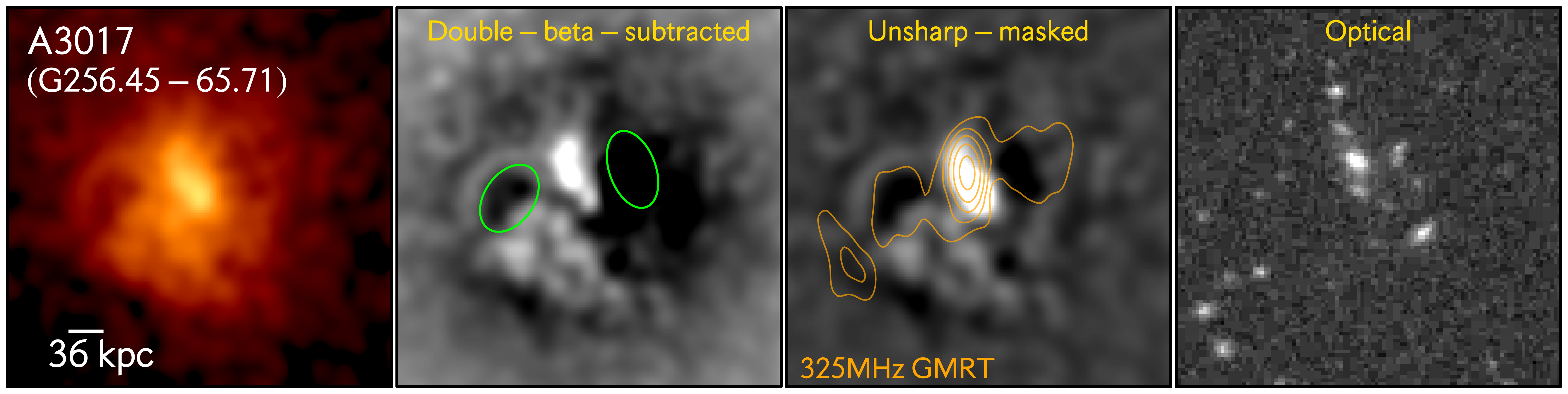}
    \caption{Continuation of Figure \ref{fig:chandra_images_C} (see text for details).}
\end{figure*}

\begin{figure*}
    \figurenum{A.1}
    \centering
        \includegraphics[width=\mylength\textwidth]{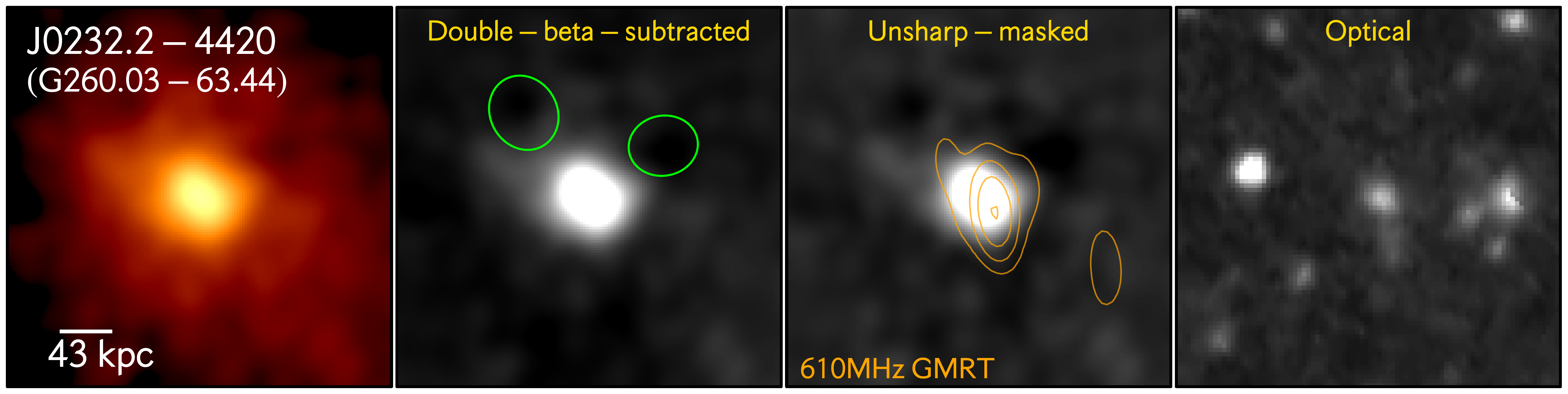}
        \includegraphics[width=\mylength\textwidth]{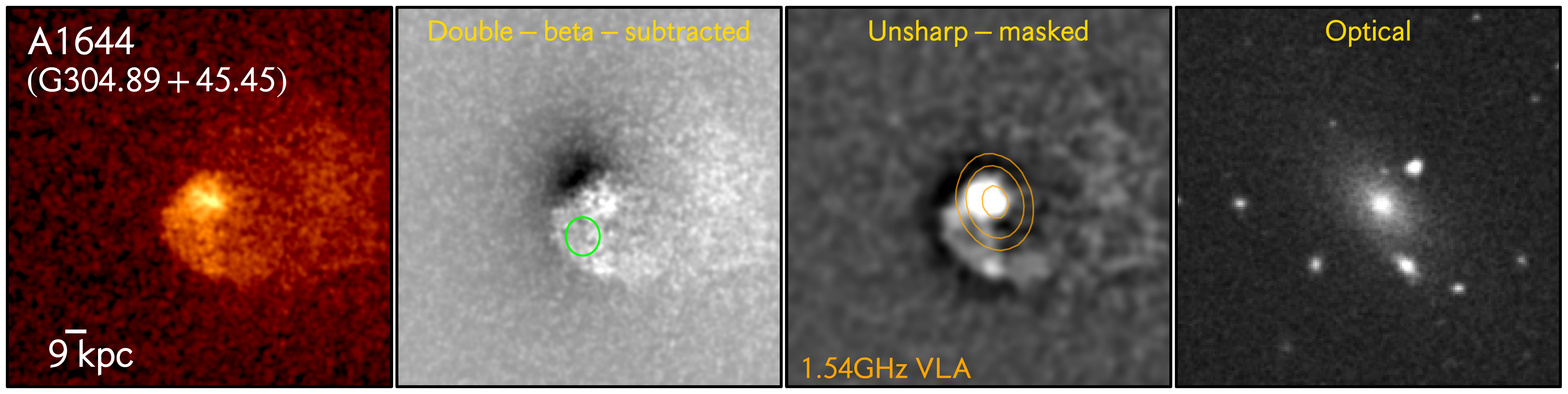}
    \caption{Continuation of Figure \ref{fig:chandra_images_C} (see text for details).}
\end{figure*}

\begin{figure*}
\centering
    \includegraphics[width=\mylength\textwidth]{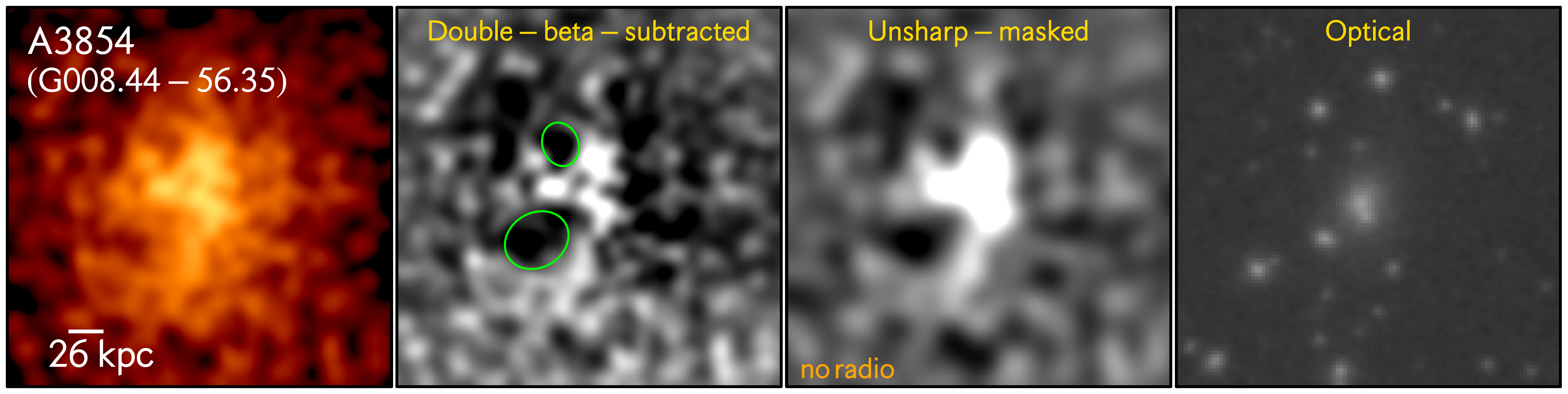}
    \includegraphics[width=\mylength\textwidth]{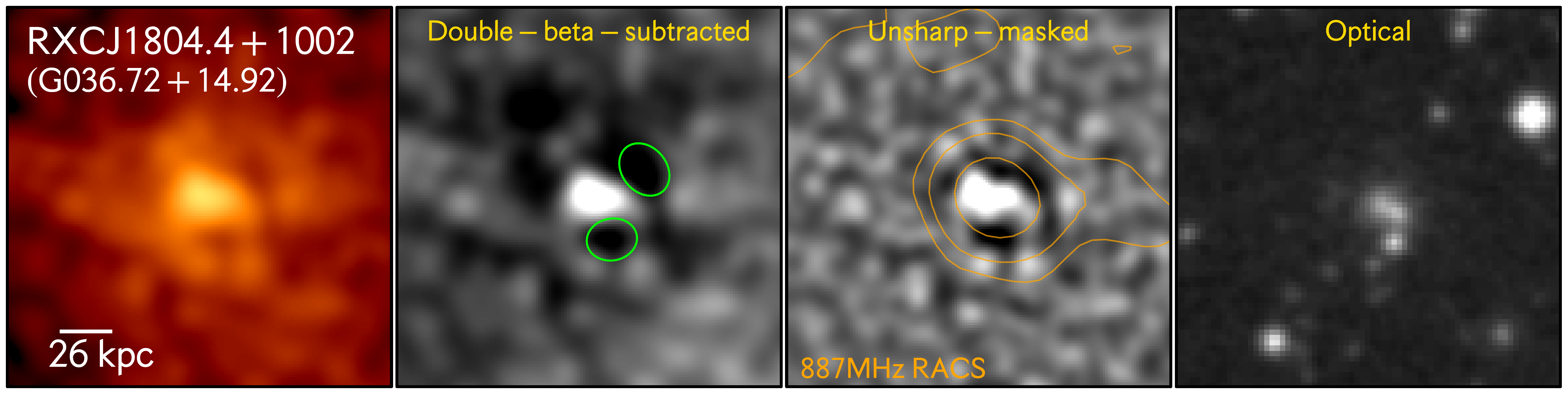}
    \includegraphics[width=\mylength\textwidth]{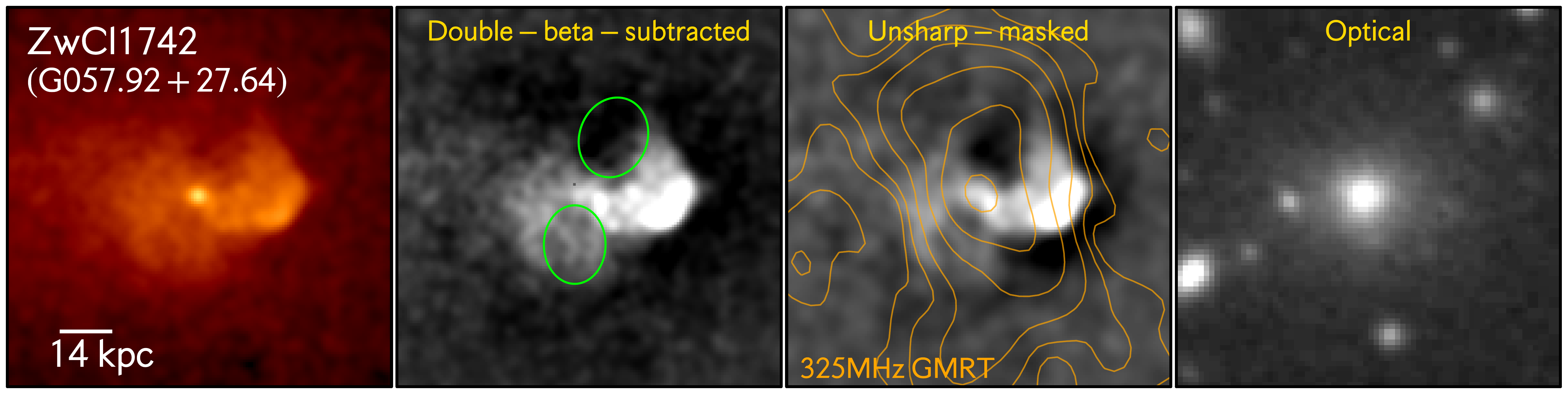}
    \includegraphics[width=\mylength\textwidth]{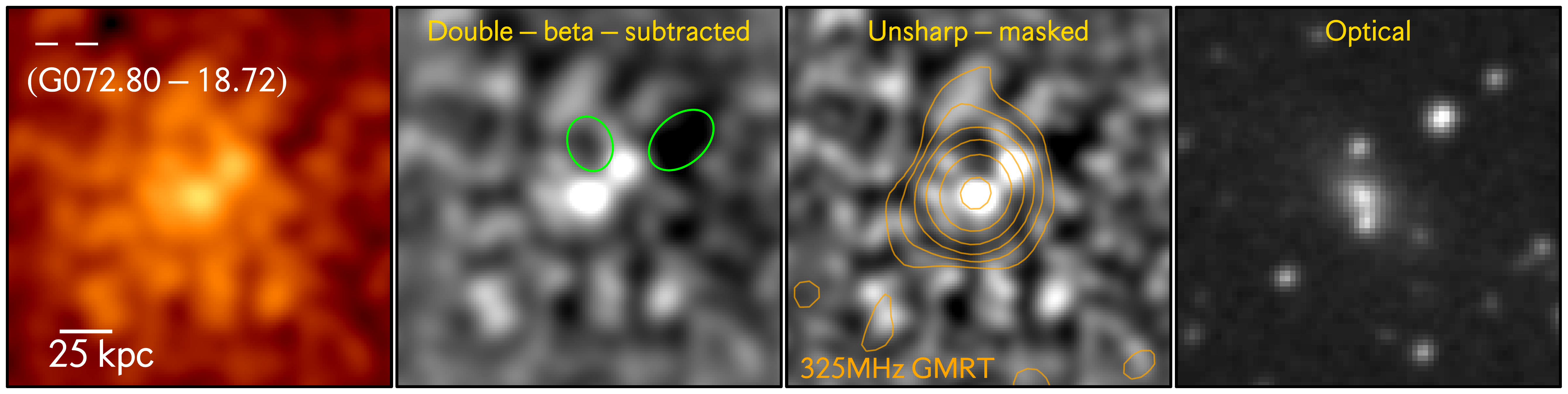}
    \includegraphics[width=\mylength\textwidth]{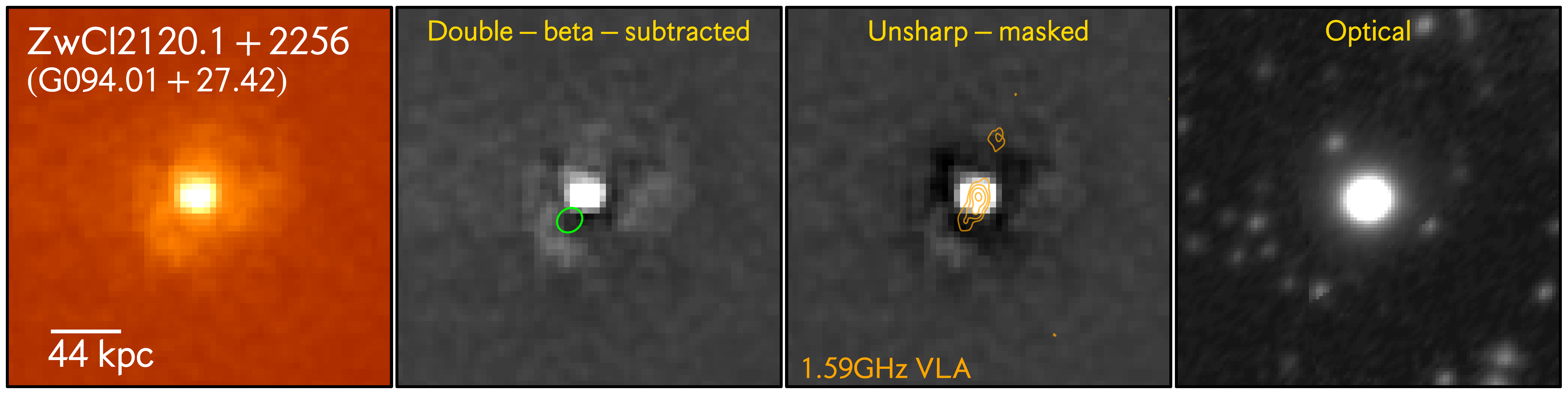}
    \caption{Images of the Planck clusters with potential (P) X-ray cavities. Left row: 0.5-2.0~keV smoothed X-ray image. Left-middle row: 0.5-2.0~keV double $\beta$ model subtracted image. Right middle row: 0.5-2.0~keV unsharp-masked image. Right row: optical image. The cavities are highlighted with green ellipses. All images show the same region, and the scale bar corresponds to 10\arcsec. \label{fig:chandra_images_P}}
\end{figure*}

\begin{figure*}
    \figurenum{A.2}
    \centering
    \includegraphics[width=\mylength\textwidth]{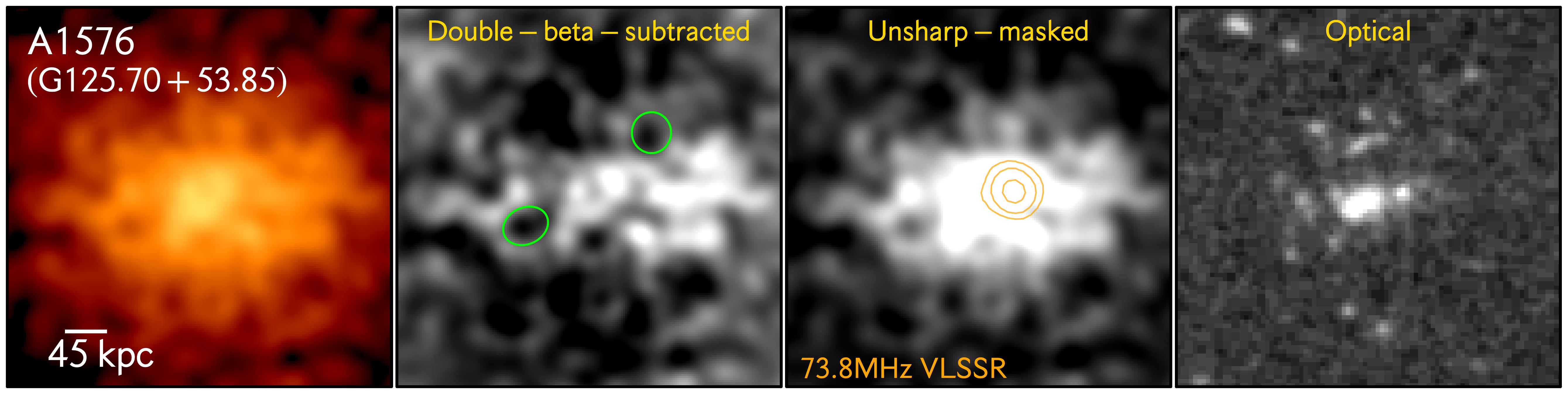}
        \includegraphics[width=\mylength\textwidth]{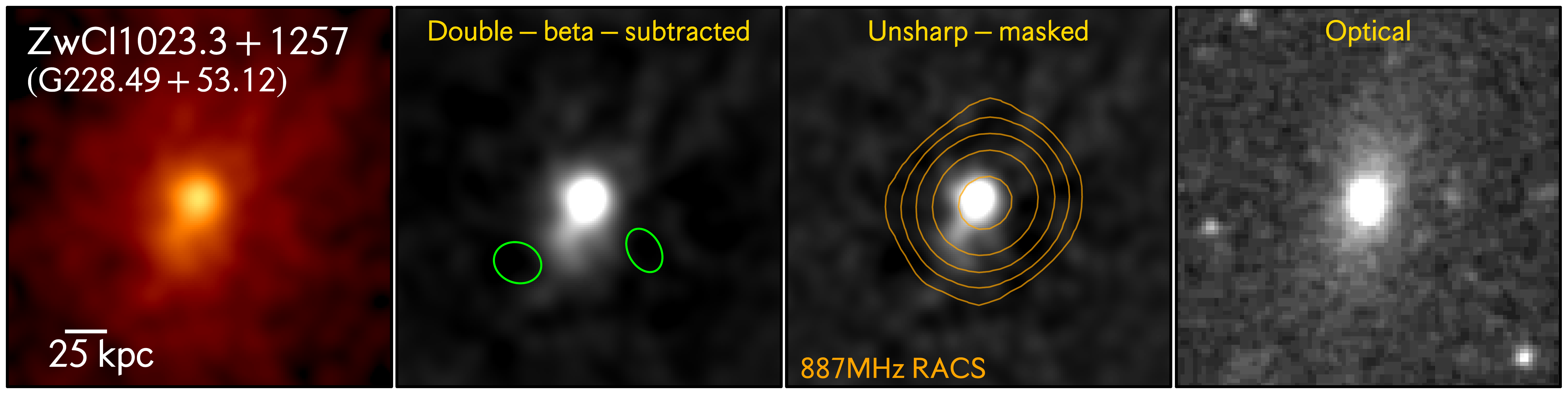}
        \includegraphics[width=\mylength\textwidth]{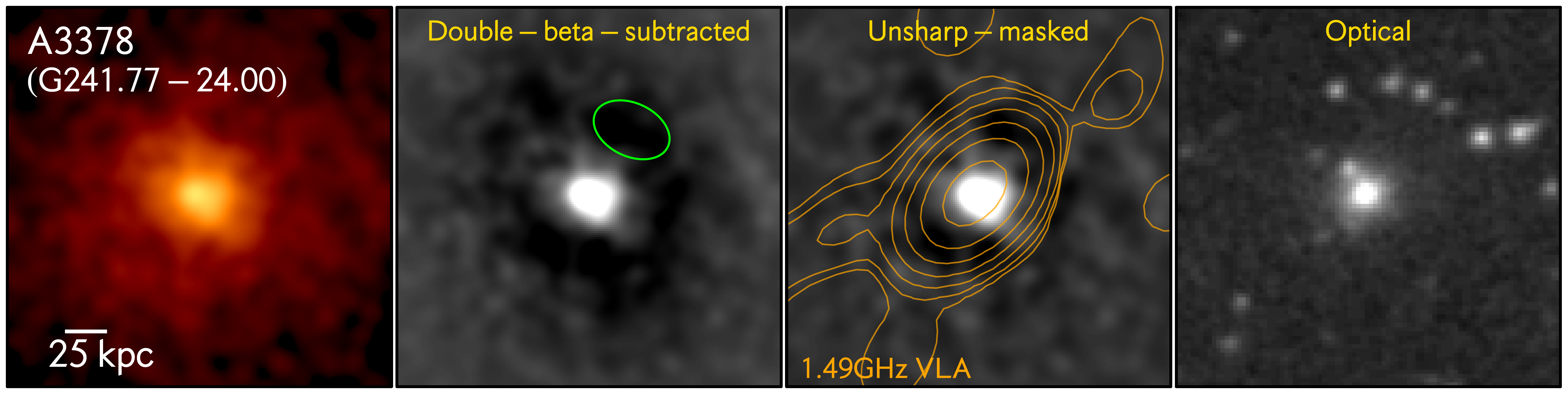}
        \includegraphics[width=\mylength\textwidth]{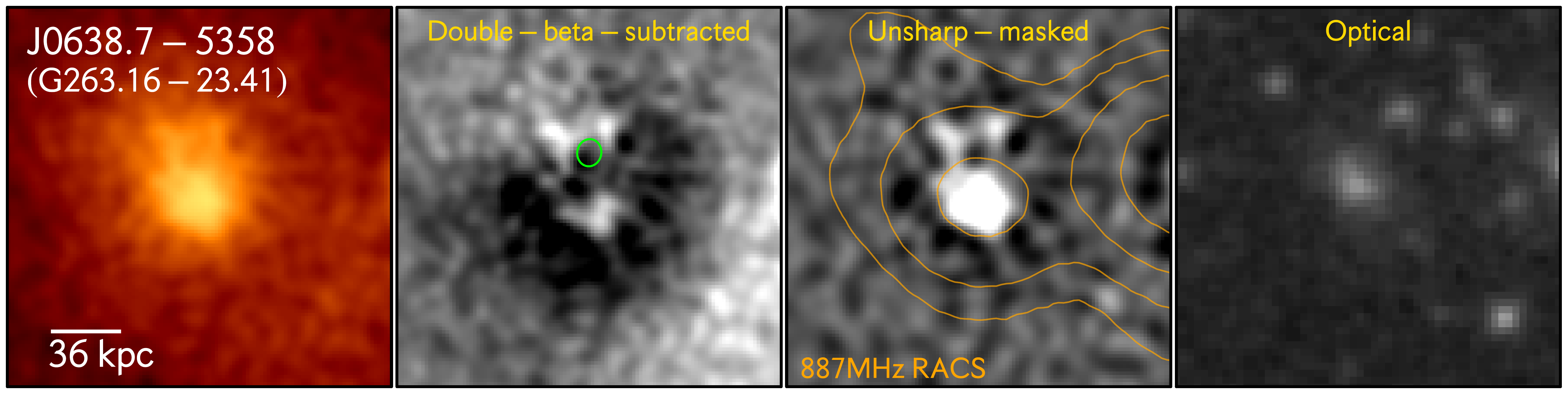}
        \includegraphics[width=\mylength\textwidth]{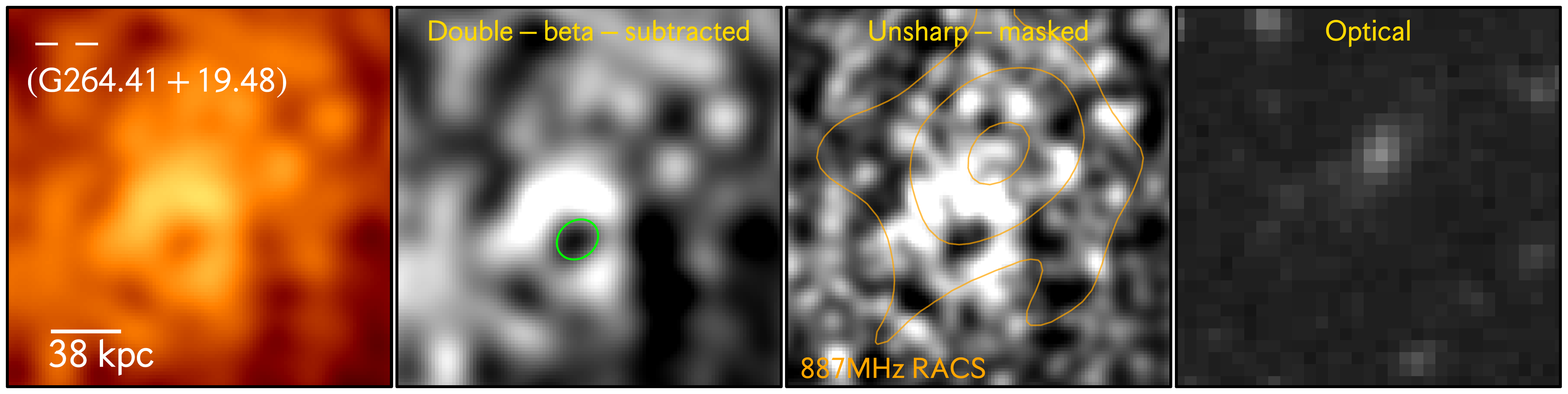}
    \caption{Continuation of Figure \ref{fig:chandra_images_P} (see text for details).}
\end{figure*}

\begin{figure*}
    \figurenum{A.2}
    \centering
        \includegraphics[width=\mylength\textwidth]{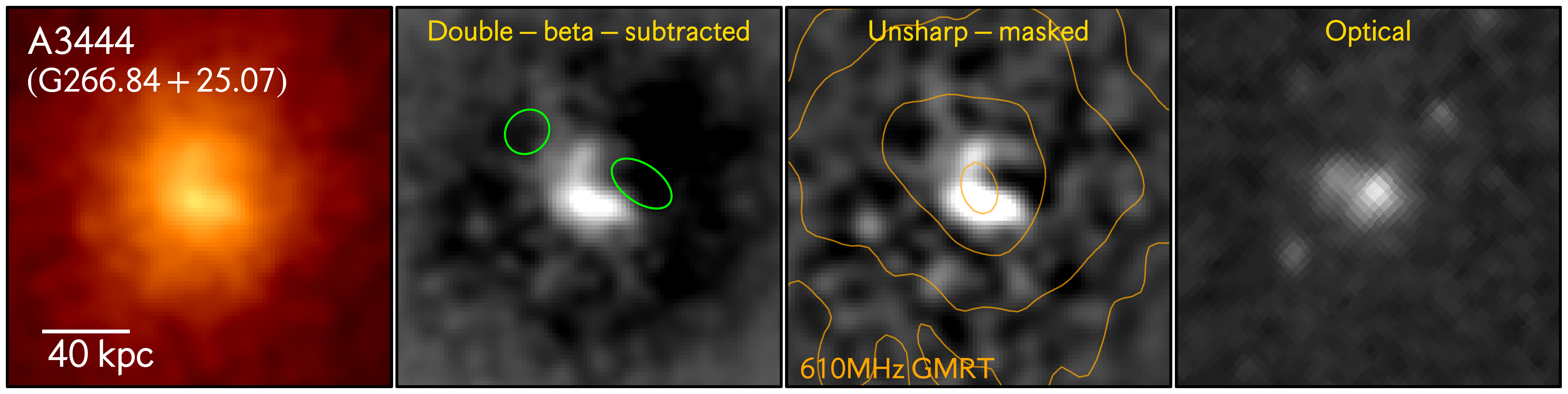}
        \includegraphics[width=\mylength\textwidth]{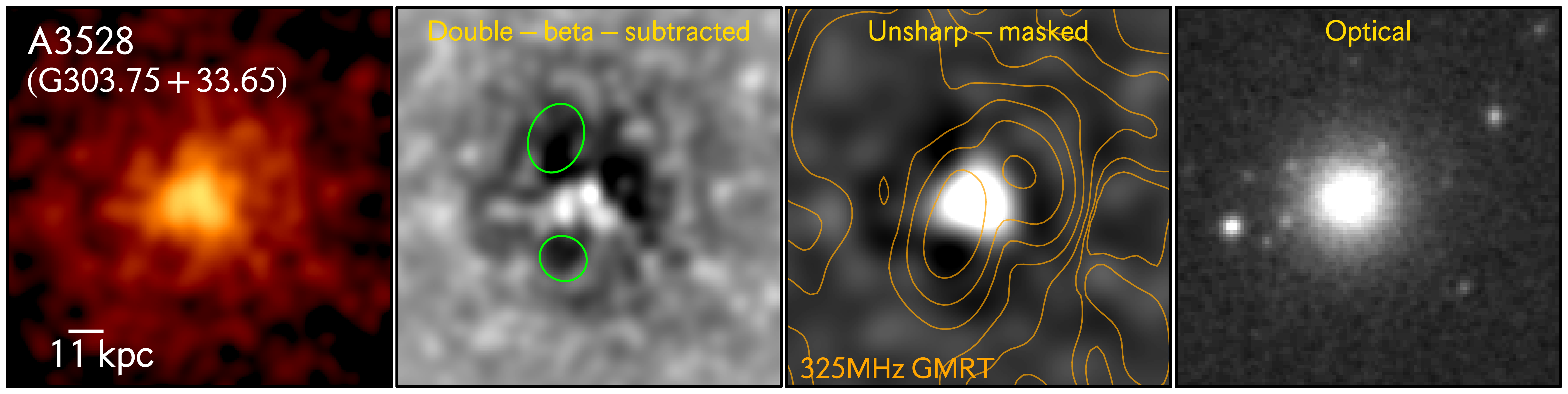}
        \includegraphics[width=\mylength\textwidth]{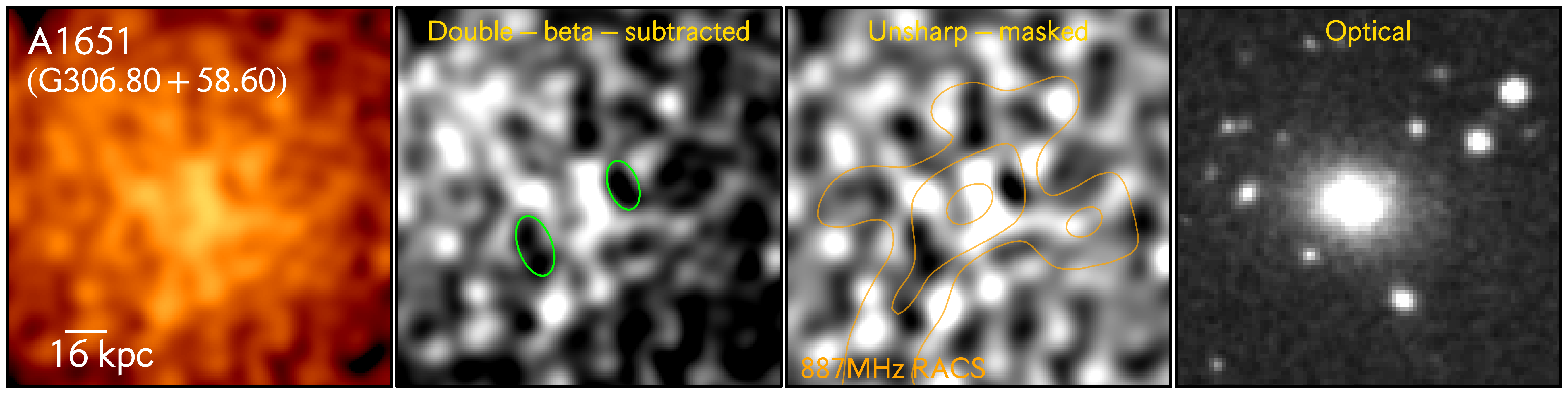}
        \includegraphics[width=\mylength\textwidth]{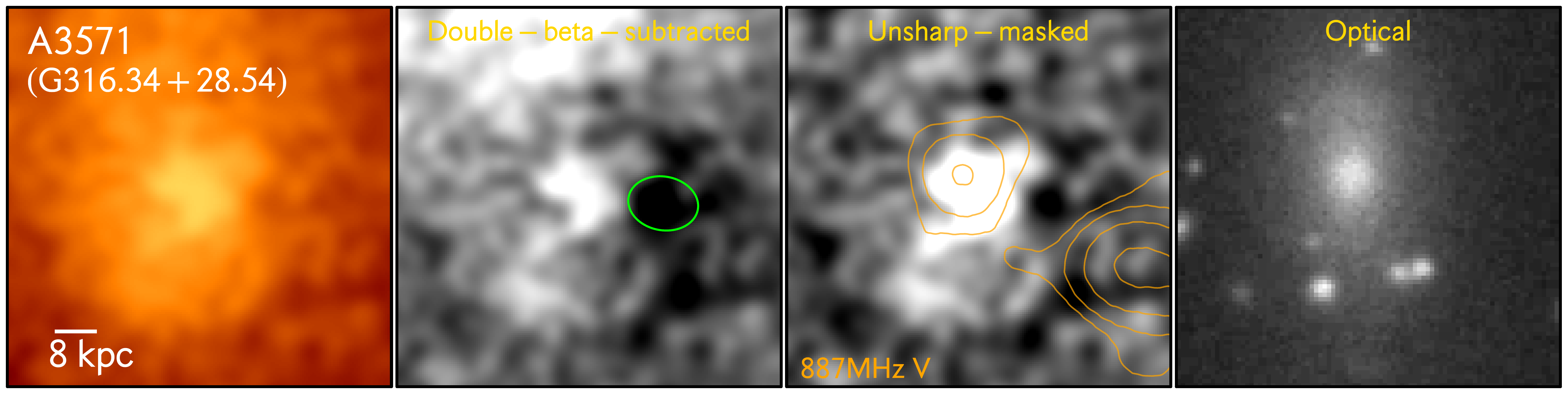}
        \includegraphics[width=\mylength\textwidth]{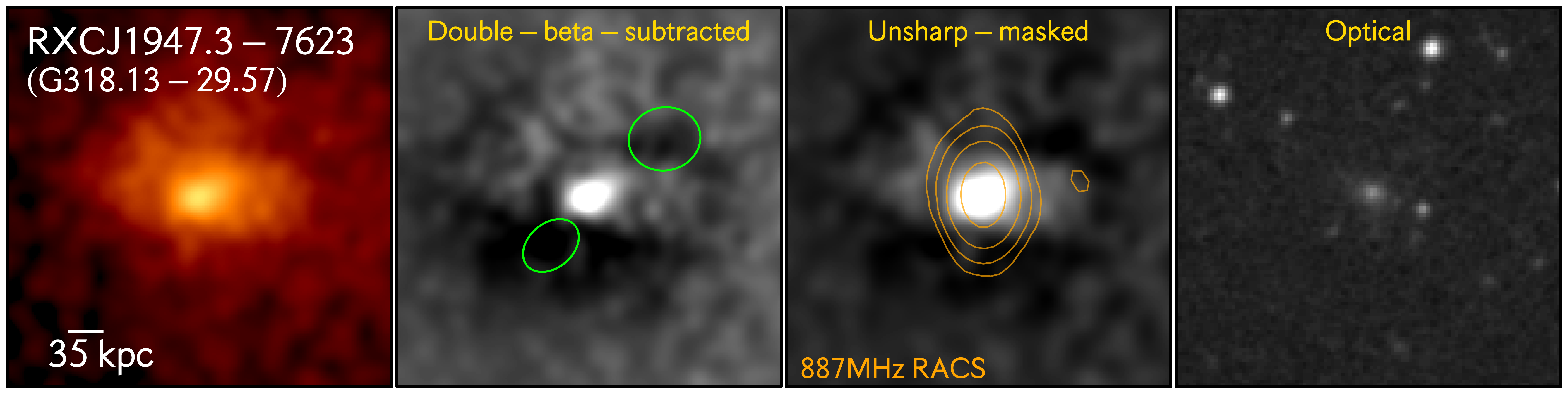}
    \caption{Continuation of Figure \ref{fig:chandra_images_P} (see text for details).}
\end{figure*}

\begin{figure*}
    \figurenum{A.2}
    \centering
        \includegraphics[width=\mylength\textwidth]{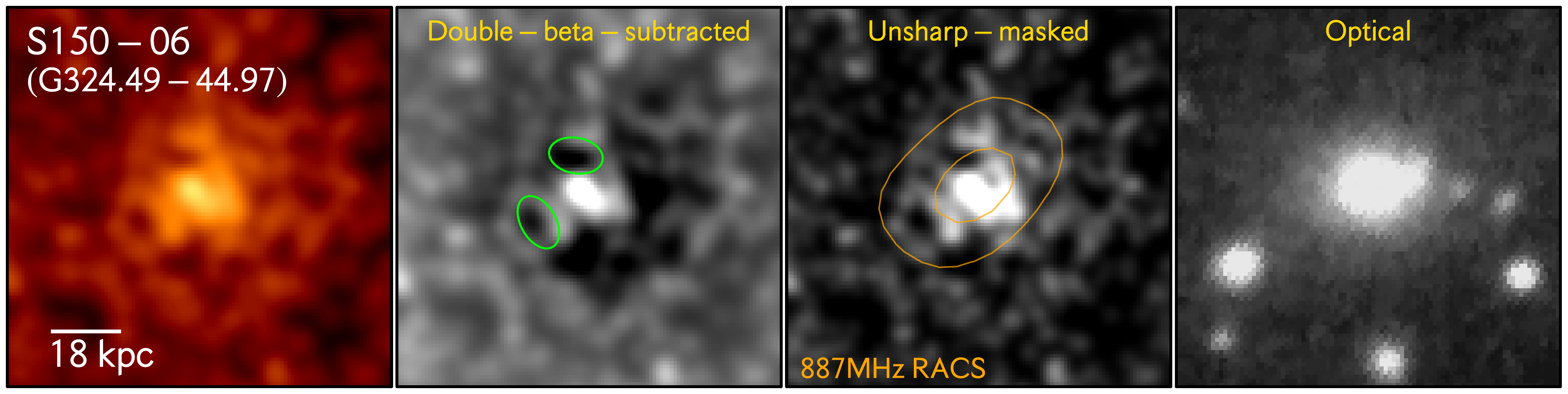}
        \includegraphics[width=\mylength\textwidth]{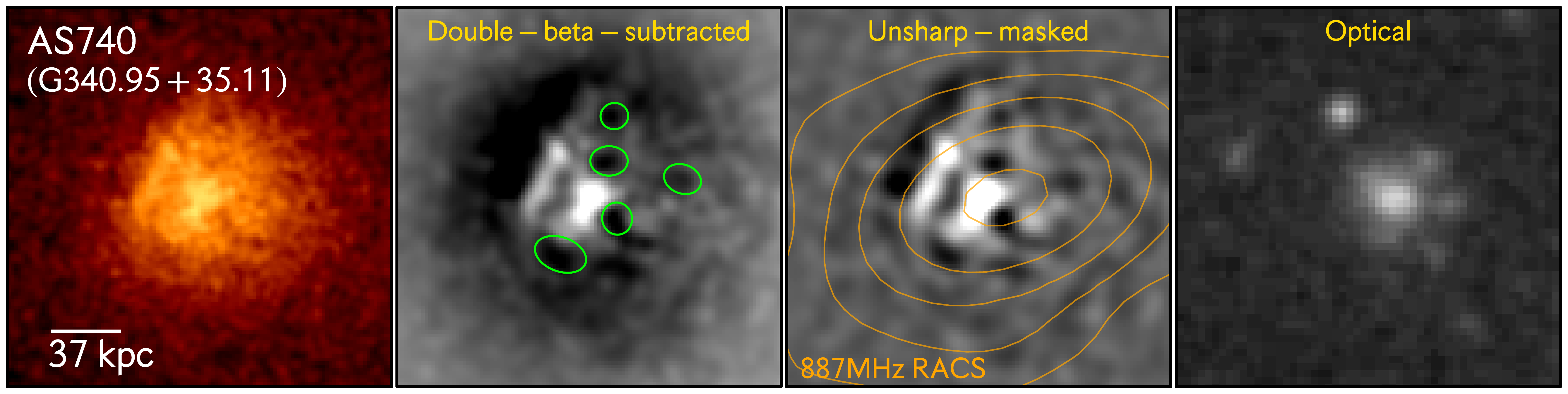}
    \caption{Continuation of Figure \ref{fig:chandra_images_P} (see text for details).}
\end{figure*}

\section{Optical line emitting gas and cavities distribution}\label{sec:app_Ha_filaments}
This section describes the distribution of the H$\alpha$ filaments detected using MUSE observations. We also discuss the distribution of the filaments as a function of the presence or lack of X-ray cavities as shown in Figure~\ref{fig:Halpha_maps} and \ref{fig:Halpha_maps2}.

\subsection{G021.09+33.25 (A2204)} G021.09+33.25 reveals several filaments with a complex distribution. In particular, the northwest and western filaments appear over the edges of the detected X-ray cavities. The eastern filament is positioned in projection beneath the giant cavity, forking into two filaments. Noteworthy the BCG appears to be interacting with a nearby galaxy.

\subsection{G033.78+77.16 (A1795)} G033.78+77.16 hosts a very elongated set of filaments arising from the center to the south of the BCG. The end of these filaments is closer to the southern cavity. Closer to the central galaxy, there are two sets of filaments located north and south of the BCG, which appears to be entrained by the radio jet.

\subsection{G073.96-27.82 (A2390)} Optical HST (F555W) observations of G073.96-27.82 reveal two merging galaxies, but appear blended as a single object on the MUSE observations. In addition, the sharp discontinuity found in the X-ray image suggests that the ICM is sloshing \citep{sonkamble15}, probably generated by the potential galaxy interaction. The northwesterly filaments in G073.96-27.82 is located in projection behind the rim of the western X-ray cavity. There are a couple of potential X-ray depressions at the Northeast and Southwest of the galaxy center that could correspond to cavities.

\subsection{G115.16-72.09 (A85)} G115.16-72.09 hosts a complex network of clumpy radial filaments extending to the S, N, NE, and NW of the BCG. According to the hot gas distribution, G115.16-72.09 shows signatures of sloshing (see \citealt{ichinohe15}). The S filament is located in projection behind the single detected cavity. The rest of the filaments and clumps are not associated with any cavity; however, some weak X-ray depressions are seen NE and NW of the BCG.

\subsection{G176.28-35.05 (2A0335+096)} The filamentary network of G176.28-35.05 is very complex, but most of the filaments appear to be in projection associated with one of the detected X-ray cavities. In particular, the northwester filament has a morphology and X-ray cavity association reminiscent of the horseshoe filament in the Perseus cluster \citep{Fabian_2008}. We note that this system is dynamically active with a cold front likely induced by sloshing motions \citep{mazzotta03}. In the same line, we find an offset between the peak of the ICM and the central BCG of 10~kpc.

\subsection{G182.44-28.29 (A478)} The N and SE set of filaments in G182.44-28.29 with ``snakes tongue'' shape are trailing behind the X-ray bubble. The SW filament with a similar structure does not associate with an X-ray depression.

\subsection{G209.56-36.49 (A496)} The NW filament appears, in projection inside of the northern cavity, while the SE filament trail one of the southern cavities. The filament position along the EW direction trails no cavities, although it may have detached from the northern cavity.

\subsection{G241.77-24.00 (A3378)} This system owns an extended clumpy H$\alpha$ nebula. The MUSE observations reveal two H$\alpha$ peaks surrounded by a diffuse emission. The northern H$\alpha$ peak is associated with the central BCG. Despite this system revealing an X-ray depression, the optical nebula is not spatially associated with the cavity.


\subsection{G244.34-32.13 (RXCJ0528.9-3927)} G244.34-32.13 reveals an extended (50~kpc) H$\alpha$ plume to the Northwest of the BCG. This source lack X-ray cavities, but the ICM distribution and the offset of 45~kpc (6$\arcsec$) between the X-ray peak and the BCG provide mild evidence for a recent interaction. {Additionally, the BCG in RXCJ0528.9-3927 (G244.34-32.13) reveals unresolved 1.4~GHz radio emission.}

\subsection{G252.96-56.05 (A3112)}
MUSE observations of G252.96-56.05 reveal an extended H$\alpha$ plume of 13~kpc of extension from NE to SW of the central galaxy. This system does not present any cavity, and the radio-jets are perpendicular to the H$\alpha$ plume. the central galaxy of A3112 (G252.96-56.05) hosts a kiloparsec radio jet with a direction towards the SE and SW of the radio core \citep{takizawa03}, and the SW X-ray tunnel may correspond to a cavity viewed almost face-on, which could produce less surface brightness contrast (see Figure~\ref{fig:cluster_nocav_halpha}).

\subsection{G256.45-65.71 (A3017)} MUSE observations reveal two potentially interacting galaxies. The H$\alpha$ emitting gas shows a double peak distribution co-spatial with the two systems, followed by an extended tail to the southeast of the main BCG. The easterly filament appears to be located close to the most E cavity. There is no optical emission-line gas associated with the W cavity. We note that this system show signatures of gas sloshing or merger \citep{pandge21}.

\subsection{G260.03-63.44 (RXCJ0232.2-4420)} The distribution of the H$\alpha$ emitting gas in G260.03-63.44 is highly compact and co-spatial with the central galaxy. The potential cavities detected on the ICM are far in projection from the H$\alpha$ emitting gas and may correspond to a past feedback activity.

\subsection{G266.84+25.07 (A3444)} G266.84+25.07 has a network of H$\alpha$ filaments extending away from the BCG to the N and NW over the central $\sim$40~kpc, which roughly matches the position of a ridge of cooling gas seen in X-ray emission. This cluster has two potential cavities located W and NW of the central BCG. We find an offset between the ICM peak and the BCG of 4~kpc indicative of sloshing motions.

\subsection{G269.51+26.42 (A1060)} A gaseous rotating disk is detected in the G269.51+26.42 (A1060) with a blueshifted structure located to the Northeast of the central galaxy. The ionized gas is co-spatial with the dust lanes detected on the optical HST observations. The H$\alpha$ emitting gas is unrelated to the potential X-ray cavities.
An unresolved CC hot corona \citep{sun09a} may likely be the explanation of the detected colder gas. However, deeper X-ray \textit{Chandra} observations are needed to confirm this hypothesis.

\subsection{G303.75+33.65 (A3528)} G303.75+33.65 is an interacting system with two central galaxies likely in the process of merging. The distribution of the H$\alpha$ emitting gas is relatively compact, with the peak at the position of the main BCG, followed by a tiny filament spreading to the southeasterly of the BCG. The secondary galaxy is cold gas-free. The potential detected X-ray cavities in G303.75+33.65 are wholly dissociated from the H$\alpha$ emitting gas.

\subsection{G304.89+45.45 (A1644)} G304.89+45.45 hosts a complex network of clumpy filaments located S and near NW of the central galaxy. In projection, the two southern networks of filaments appear to be draped over the edges of the single detected X-ray cavity. Interestingly, the optically emitting gas and the BCG are spatial coincidence with the X-ray peak, although the hot plasma distribution reveals clear evidence of sloshing (see Figure \ref{fig:Halpha_maps} and \citealt{johnson10}).

\subsection{G318.13-29.57 (RXCJ1947.3-7623)} G318.13-29.57 reveals a unresolved ionized gas distribution with an extension of 14~kpc centered on the BCG. The ionized gas is not connected with the potential X-ray cavities detected in the surrounding hot ICM.

\subsection{G340.95+35.11 (AS0780)} G340.95+35.11 reveals a compelling network of filaments, with two blobs of H$\alpha$ emitting gas located parallel on the WE plane, followed by two radial filaments to the N and three to the SW of the cluster center. With a loop distribution, a sixth filament is located W of the BCG. It is worth noticing that the main H$\alpha$ peak is spatially aligned with the central BCG. Each of the filaments is connected in projection to a potential X-ray cavity.

\section{Chandra Observations}

\startlongtable
\movetableright=-5cm
\tabletypesize{\scriptsize} 
\setlength{\tabcolsep}{2.5pt}
\begin{deluxetable*}{lccccccc}
\centering
\tablecaption{Summary of \textit{Chandra} observations}\label{tab:sample}
\tablehead{
{Cluster} & {Alt. name} &
{R.A.} &
{Dec.} & {$z$} & {Obs. ID} & {Date} & {Exp.} }
\decimalcolnumbers
\startdata
\textit{CC clusters }\\
G006.47+50.54 &            A2029 & 15 10 56.117 & +05 44 40.38 & 0.077 &                                      891 4977 6101 &                                  2000, 2004, 2004  &          107.6 \\
G008.44-56.35 &            A3854 & 22 17 45.701 & -35 43 32.55 & 0.149 &                                              15116 &                                              2013  &           11.9 \\
G021.09+33.25 &            A2204 & 16 32 46.854 & +05 34 31.61 & 0.151 &                                      499 6104 7940 &                                  2000, 2004, 2007  &           96.8 \\
G033.46-48.43 &            A2384 & 21 52 21.245 & -19 32 54.19 & 0.094 &                                               4202 &                                              2002  &           31.4 \\
G033.78+77.16 &            A1795 & 13 48 52.710 & +26 35 31.20 & 0.062 & 493 494 3666 5286 5287 &  2000, 1999, 2002, 2004  &         2480.1 \\
 &            &  &  & &              5289 5290 6159 6160 6161 & 2004, 2004, 2004, 2004    &           \\
G036.72+14.92 &               -- & 18 04 31.215 & +10 03 24.21 & 0.152 &                                              15098 &                                              2014  &            9.6 \\
G042.82+56.61 &            A2065 & 15 22 29.473 & +27 42 18.76 & 0.072 &                                          3182 7689 &                                        2002, 2007  &           54.5 \\
G044.22+48.68 &            A2142 & 15 58 21.100 & +27 13 47.87 & 0.089 &              1196 1228 5005 7692 15186 &          1999, 1999, 2005, 2007, 2014&          228.3 \\
 &            &  &  & &              16564 16565 &         2014, 2014    &           \\
G049.20+30.86 &   RXJ1720.1+2638 & 17 20 09.957 & +26 37 30.79 & 0.164 &                             304 549 1453 3224 4361 &                      2000, 1999, 1999, 2002, 2002  &           59.8 \\
G049.66-49.50 &            A2426 & 22 14 32.554 & -10 22 17.84 & 0.098 &                                              12279 &                                              2010  &            9.7 \\
G055.60+31.86 &            A2261 & 17 22 27.300 & +32 07 57.98 & 0.224 &                               550 5007 20413 21960 &                            1999, 2004, 2018, 2018  &          128.3 \\
G056.81+36.31 &            A2244 & 17 02 42.571 & +34 03 38.15 & 0.095 &                                          4179 7693 &                                        2003, 2007  &           62.1 \\
G057.92+27.64 &         ZwCl1742 & 17 44 15.426 & +32 59 31.71 & 0.076 &                                         8267 11708 &                                        2007, 2009  &           53.4 \\
G062.42-46.41 &               -- & 22 23 47.779 & -01 39 00.82 & 0.091 &                                        15107 15312 &                                        2013, 2013  &           29.2 \\
G062.92+43.70 &            A2199 & 16 28 38.232 & +39 33 03.36 & 0.030 &                    497 498 10748 10803 10804 10805 &                2000, 1999, 2009, 2009, 2009, 2009  &          158.3 \\
G067.23+67.46 &            A1914 & 14 26 00.269 & +37 49 40.52 & 0.171 &             542 3593 18252 20023 20024  &          1999, 2003, 2017, 2017, 2017  &          147.9 \\
 &            &  &  & &              20025 20026 &         2017, 2017  &           \\
G072.80-18.72 &               -- & 21 22 27.115 & +23 11 50.12 & 0.143 &                                              13379 &                                              2011  &            8.9 \\
G073.96-27.82 &             A239 & 21 53 36.797 & +17 41 43.53 & 0.233 &                                       500 501 4193 &                                  2000, 1999, 2003  &          114.0 \\
G080.99-50.90 &               -- & 23 11 33.144 & +03 38 08.17 & 0.300 &                                         3288 11730 &                                        2002, 2009  &           36.3 \\
G086.45+15.29 &               -- & 19 38 18.297 & +54 09 36.16 & 0.260 &                                              15104 &                                              2014  &           14.9 \\
G094.01+27.42 &               -- & 18 21 57.197 & +64 20 36.30 & 0.299 &       1599 2186 2310 2311 2418  &  2001, 2001, 2001, 2001, 2001 &          656.8 \\
 &                &  &  &  &        9398 9845 9846 9848 &   2001, 2008, 2008, 2008  &          656.8 \\
G096.85+52.46 &               -- & 14 52 58.061 & +58 03 00.56 & 0.318 &                                      906 7021 7713 &                                  2000, 2006, 2008  &          113.1 \\
G098.95+24.86 &               -- & 18 54 02.098 & +68 23 01.24 & 0.093 &                                              15121 &                                              2013  &           15.9 \\
G115.16-72.09 &              A85 & 00 41 50.390 & -09 18 09.53 & 0.056 &                        904 15173 15174 16263 16264 &                      2000, 2013, 2013, 2013, 2013  &          195.1 \\
G115.71+17.52 &               -- & 22 26 30.303 & +78 19 16.11 & 0.300 &                                        13383 16281 &                                        2012, 2014  &           58.4 \\
G124.21-36.48 &             A115 & 00 55 50.297 & +26 24 36.47 & 0.197 &                       3233 13458 13459 15578 15581 &                      2002, 2012, 2012, 2012, 2012  &          361.5 \\
G125.70+53.85 &               -- & 12 36 57.703 & +63 11 13.65 & 0.302 &                                         7938 15127 &                                        2007, 2013  &           44.5 \\
G139.59+24.18 &               -- & 06 21 49.156 & +74 42 05.10 & 0.300 &                                        15139 15297 &                                        2013, 2013  &           27.6 \\
G146.33-15.59 &               -- & 02 54 27.386 & +41 34 46.82 & 0.017 &                        908 11717 12016 12017 12018 &                      2000, 2009, 2009, 2009, 2009  &          181.6 \\
G164.61+46.38 &               -- & 09 38 19.645 & +52 02 58.03 & 0.342 &                                        15137 16573 &                                        2014, 2014  &           41.6 \\
G166.13+43.39 &             A773 & 09 17 53.077 & +51 43 41.38 & 0.217 &              533 3588 5006 &                                  2000, 2003, 2004  &          165.4 \\
 &                &  &  &  &        22051 22052 &    &          656.8 \\
G176.28-35.05 &       2A0335+096 & 03 38 40.698 & +09 58 03.07 & 0.035 &                                      919 7939 9792 &                                  2000, 2007, 2007  &          102.9 \\
G180.62+76.65 &            A1423 & 11 57 17.375 & +33 36 39.11 & 0.213 &                                          538 11724 &                                        2000, 2010  &           35.6 \\
G182.44-28.29 &             A478 & 04 13 25.196 & +10 27 53.66 & 0.088 &                                          1669 6102 &                                        2001, 2004  &           52.4 \\
G182.63+55.82 &               -- & 10 17 03.531 & +39 02 53.32 & 0.206 &                                           903 7704 &                                        2000, 2007  &           41.4 \\
G182.63+55.82 &               -- & 10 17 03.531 & +39 02 53.32 & 0.206 &                                           903 7704 &                                        2000, 2007  &           41.4 \\
G209.56-36.49 &             A496 & 04 33 37.913 & -13 15 42.03 & 0.033 &                                      931 3361 4976 &                                  2000, 2001, 2004  &          104.0 \\
G226.24+76.76 &            A1413 & 11 55 17.943 & +23 24 20.25 & 0.143 &                            537 1661 5002 5003 7696 &                            2000, 2001, 2004, 2007  &          136.1 \\
G228.49+53.12 &               -- & 10 25 58.011 & +12 41 08.71 & 0.143 &                                              13375 &                                                    &            8.9 \\
G241.74-30.88 & MCXCJ0532.9-3701 & 05 32 55.628 & -37 01 35.71 & 0.271 &                                              15112 &                                              2013  &           24.8 \\
G241.77-24.00 &            A3378 & 06 05 53.936 & -35 18 08.71 & 0.139 &                                              15315 &                                              2013  &            9.9 \\
G244.34-32.13 &               -- & 05 28 52.997 & -39 28 13.50 & 0.284 &                                   4994 15177 15658 &                                  2004, 2013, 2013  &          111.4 \\
G252.96-56.05 &            A3112 & 03 17 57.637 & -44 14 17.40 & 0.075 &                     2216 2516 6972 7323 7324 13135 &                2001, 2001, 2006, 2006, 2006, 2011  &          149.9 \\
G253.47-33.72 &            A3343 & 05 25 48.812 & -47 15 10.22 & 0.191 &                                              15122 &                                              2014  &           17.9 \\
G256.45-65.71 &            A3017 & 02 25 53.140 & -41 54 52.95 & 0.220 &                                        15110 17476 &                                        2013, 2015  &           28.8 \\
G257.34-22.18 &               -- & 06 37 14.638 & -48 28 18.15 & 0.203 &                                              15125 &                                              2013  &           24.8 \\
G260.03-63.44 &  RXCJ0232.2-4420 & 02 32 18.714 & -44 20 46.38 & 0.284 &                                               4993 &                                              2004  &           23.4 \\
G263.16-23.41 &               -- & 06 38 48.541 & -53 58 26.09 & 0.227 &                             9420 15176 16572 16598 &                            2008, 2014, 2014, 2014  &          108.5 \\
G263.66-22.53 &               -- & 06 45 28.586 & -54 13 43.00 & 0.164 &                                              15301 &                                              2013  &           10.0 \\
G264.41+19.48 &               -- & 10 00 01.753 & -30 16 37.46 & 0.240 &                                              15132 &                                              2014  &           30.7 \\
G266.84+25.07 &  RXCJ1023.8-2715 & 10 23 50.137 & -27 15 22.24 & 0.254 &                                               9400 &                                              2008  &           36.7 \\
G286.58-31.25 &               -- & 05 31 28.179 & -75 10 37.86 & 0.210 &                                              15115 &                                              2014  &           22.2 \\
G303.75+33.65 &            A3528 & 12 54 40.703 & -29 13 40.55 & 0.054 &                                               8268 &                                              2007  &            8.1 \\
G304.89+45.45 &            A1644 & 12 57 11.570 & -17 24 33.57 & 0.047 &                                          2206 7922 &                                              2001  &           70.2 \\
G306.68+61.06 &            A1650 & 12 58 41.433 & -01 45 43.44 & 0.085 &            4178 5822 5823 6356 6357 &    2003, 2005, 2005, 2005, 2005  &          230.1 \\
 &             &  &  &  &            6358 7242 7691  &  2005, 2006, 2007 &          \\
G306.80+58.60 &            A1651 & 12 59 22.339 & -04 11 45.31 & 0.085 &                                               4185 &                                              2003  &            9.6 \\
G313.36+61.11 &            A1689 & 13 11 29.460 & -01 20 27.68 & 0.183 &                       540 1663 5004 6930 7289 7701 &                2000, 2001, 2004, 2006, 2006, 2007  &          197.1 \\
G313.87-17.10 &               -- & 16 01 48.426 & -75 45 15.56 & 0.153 &                                              14386 &                                              2012  &            9.0 \\
G316.34+28.54 &            A3571 & 13 47 28.110 & -32 51 57.98 & 0.039 &                                               4203 &                                              2003  &           34.0 \\
G318.13-29.57 &               -- & 19 47 14.773 & -76 23 44.99 & 0.217 &                                              15102 &                                              2013  &           19.8 \\
G324.49-44.97 &               -- & 22 18 00.518 & -65 10 52.42 & 0.095 &                                              15314 &                                              2014  &            9.7 \\
G337.09-25.97 &               -- & 19 14 37.070 & -59 28 17.20 & 0.120 &                                              15135 &                                              2013  &           24.8 \\
G340.95+35.11 &            AS740 & 14 59 28.730 & -18 10 45.08 & 0.236 &                                               9428 &                                        2008, 2022  &           39.6 \\
G349.46-59.94 &               -- & 22 48 44.562 & -44 31 48.94 & 0.347 &                                   4966 18611 18818 &                                  2004, 2016, 2016  &          123.7 \\
\enddata
\tablecomments{
(1) Column lists the cluster name (the prefix PLCKESZ is omitted).\\
(2) Alternative name.\\
(3) and (4) position of the clusters.\\
(5) redshift.\\
(6) ID observation used in this work. \\
(7) Date of the OBSID\\
(8) Total exposure time in ks.}
\end{deluxetable*}

\bibliographystyle{aasjournal}

\end{document}